\documentclass[12pt]{article}
\usepackage{epsf}
\renewcommand{\thefootnote}{\fnsymbol{footnote}}

\def\g{\Gamma} 
\def\gg{{\rm I}\!\g}
\newcommand{\fp}{\Phi\Pi}  
\newcommand{\alp}{\alpha} 
\def\bsg{$b \rightarrow s \gamma$} 
\def\hgg{$H \rightarrow \gamma \gamma$} 
\newcommand{\nabh}{\hat{\nabla}}  

\newcommand{\gh}{\hat{\g}} 
  
\newcommand{\dms}{\displaystyle} 
\def\cf{{\cal F}}

\begin{document} 

\renewcommand {\theequation}{\arabic{section}.\arabic{equation}}

\begin{flushright} 
  BUTP--99/13\\
  MPI/PhT-98-90\\ 
%%%  revised version\\
  
\end{flushright} 
 
\vskip 0.5cm 
\begin{center} 
  \centerline{\LARGE\bf  Practical Algebraic Renormalization} 
  \vskip 1.0cm
  {\large
    Pietro Antonio Grassi$^{(a)}$
\footnote{pag5@nyu.edu; present address: 
    New York University, Physics Dep., New York, 10003, NY, USA.},
    Tobias Hurth$^{(a)}$\footnote{tobias.hurth@cern.ch; present address: 
      CERN, Theory Division, CH-1211 Geneve 23, Switzerland.}
    and Matthias Steinhauser$^{(b)}$\footnote{Present address: II. Institut f\"ur Theoretische Physik,  
   Universit\"at Hamburg, 22761 Hamburg, Germany.} 
    \vskip 1,7cm 
    {(a) Max-Planck-Institut f\"ur Physik,}\\  
    {Werner-Heisenberg-Institut, D-80805 Munich, Germany} 
      \\[.4em]
    {(b) Institut f\"ur Theoretische Physik,} \\
    {Universit\"at Bern, CH-3012 Bern, Switzerland}
    \vskip 0,2cm
    } 
  \vskip 0,5cm 
  \begin{minipage}{14.8cm}
    {\small 
      \begin{center}
        {\bf Abstract} 
      \end{center}
      \vskip 0,2cm 
      A practical approach is presented  which allows
      the use of a non-invariant regularization scheme
      for the computation
      of quantum corrections in perturbative quantum field theory.
      The theoretical control of algebraic renormalization  
      over non-invariant counterterms
      is translated into a practical computational method.
      We provide a detailed introduction into the handling of the
      Slavnov-Taylor and Ward-Takahashi identities in
      the Standard Model both in the conventional and the background gauge.
      Explicit examples for their practical derivation are presented.
      After a brief introduction into the Quantum Action Principle
      the conventional algebraic method which allows for
      the restoration of the functional identities is discussed.
      The main point of our approach is the optimization of this procedure
      which results in an enormous reduction of the calculational effort.
      The counterterms which have to be computed are universal in
      the sense that they are independent of the regularization scheme.
      The method
      is explicitly illustrated for two processes of phenomenological
      interest: QCD corrections to the decay of the Higgs boson 
      into two
      photons and two-loop electroweak corrections to the process
      $B \rightarrow  X_s \gamma$.
      }
  \end{minipage}
\end{center} 

\thispagestyle{empty}
\newpage

\setcounter{page}{1}

\renewcommand{\thefootnote}{\arabic{footnote}}
\setcounter{footnote}{0}

%%%%%%%%%%%%%%%%%%%%%%%%%%%%%%%%%%%%%%%%%%%%%%%%%%%%%%%%%%%%

\setcounter{equation}{0}
\section{Introduction} 
\label{sec:introduction}
A regularization method which respects all symmetries
of the Standard Model (SM)~\cite{SM} 
does not exist.  The popular and powerful
method of Dimensional Regularization~\cite{dim_reg}
is at least an invariant scheme for QCD. 
In the electroweak sector, however, the coupling to chiral fermions
introduces the well-known $\gamma_5$ problem. One also has to face
additional 
technicalities due to evanescent operators in the effective field theory
approach. 
It is well-known that in the framework of Dimensional
Regularization
only the 't~Hooft-Veltman-Breitenlohner-Maison scheme \cite{Hoo,maiso} 
for $\gamma_5$ is shown to be consistent to all orders.
The so-called naive dimensional scheme (with an anticommuting $\gamma_5$) 
does not reproduce the
chiral anomaly and is not consistent to all orders. 
For specific examples it leads to correct results at the lowest
orders in perturbation theory.
Nevertheless, it seems desirable to have a powerful practical
alternative even in the SM, at least for cross-checks, as suggested
controversies in the past suggest (see, e.g.,~\cite{contr}).
Moreover, the impressive experimental precision mainly reached at the 
electron positron  colliders LEP and SLC and the proton anti-proton 
collider TEVATRON has made it mandatory to evaluate specific two- or even 
three-loop contributions to observables where the inconsistencies of the 
well-known ``naive dimensional'' scheme are unavoidable.

Going beyond the SM,
it is well-known that Dimensional Regularization 
breaks the Ward identities of supersymmetry.
However,  one very often prefers to
keep the dimensional scheme for the practical calculations, also
beyond the SM,
in order to take advantage of already
well-developed computer tools~\cite{HarSte98}.
Thus, one needs a practical procedure to
restore
the Ward identities of supersymmetry in the final step of the
renormalization
procedure.

From the principal point of view, the calculation of higher-loop contributions
in perturbative quantum field theories is a well-understood issue. The
axioms of relativistic quantum field theory, such as causality and Poincar\'e
invariance, fix the matrix elements completely to all orders up to a limited
number
of free constants. They have to be determined by renormalization
conditions. These free constants correspond to a renormalization ambiguity for
coinciding points in the definition of time-ordered products of
operator-valued distributions~\cite{Epstein}.
The main question is whether the 
renormalization ambiguity can be fixed in such a way
that the time-ordered products
fulfill the symmetry constraints.
The  question behind this is the  compatibility of the symmetries of 
the classical Lagrangian with quantization.

Here the method of algebraic renormalization offers a complete theoretical
answer: In general, the subtraction of ultra-violet
divergences in quantum field
theories leads to non-invariant Green functions, which means that the
regularization scheme and the subsequent renormalization do not respect the
symmetries of the theory like supersymmetry or local gauge symmetries.
As we mentioned above, Dimensional Regularization preserves gauge 
symmetries (up to the $\gamma_5$ problem) but breaks supersymmetry.

The Quantum Action Principle~\cite{QAP}
tells us that the breaking terms are local at the
lowest non-vanishing order.  This fact provides a possible path for the
construction of invariant Green functions, independent of the regularization
scheme. One introduces, order by order, finite non-invariant local
counterterms which restore the symmetry relations (provided there are no
anomalies)~\cite{brs}.  
Thus, one can in principle show that in anomaly-free theories the
local renormalization ambiguity (which is not fixed by the axioms of
relativistic 
quantum field theory) can always be used  in such a way that the
perturbative S-matrix enjoys all symmetry properties of the classical
theory (for a review, see~\cite{libro}).

Although the method of algebraic renormalization is intensively
used as a tool for proving renormalizability of various models~\cite{libro},
its full value has not
yet been widely appreciated by the practitioners.  Indeed, the theoretical
understanding of algebraic renormalization
does not lead automatically to a practical advice
for higher-loop calculations.  One could even expect that such an algebraic
renormalization scheme becomes very complicated at  higher orders.
It is one of the main purposes of this paper to provide
theoretical procedures which minimize the additional
efforts for the restoration of the symmetries and to demonstrate the
efficiency of the combined method in some examples of phenomenological
interest. 

However, two obvious practical complications of algebraic 
renormalization have to be taken 
into account:

\begin{itemize}
\item[(a)]
  The constraints introduced  by the symmetry connect various Green functions.
  Thus, for the construction of the non-invariant counterterms corresponding
  to a specific Green function  one also has to compute the
  various other Green functions involved in the identities. 
\item[(b)]
  In the computation of higher-loop contributions one also has to analyze
  identities from lower orders
  which constrain the non-invariant counterterms.
\end{itemize}
These disadvantages can be significantly reduced:
\begin{itemize}
\item[(1)]
  First, one should state that many identities are not relevant if one is
  interested in one specific Green function only and if the corresponding
  breaking terms can be compensated by the other Green functions in the given
  identity alone.
\item[(2)]
  In the case of local gauge symmetries, the structure of the relevant
  identities can be considerably simplified by using the background field
  gauge~\cite{bkg}. 
  In a conventional gauge there is a large number of non-linear Slavnov-Taylor
  identities.  
  In the  background field gauge some of them get replaced by linear
  Ward-Takahashi identities like in QED.
\item[(3)]
  We have some well-known theoretical constraints~\cite{libro}:
  the Quantum Action
  Principle tells us that the breaking terms are local at the lowest
  non-vanishing order and thus are removable by counterterms if there is 
  no anomaly.   Furthermore, the algebraic consistency conditions
  heavily   constrain the structure of the breaking terms.
\item[(4)]
  Finally, the most important simplification we want to present  in this 
  paper is the following: the number of 
  breaking terms one has to calculate in addition can
  essentially be reduced to the ones which correspond to finite Green
  functions. This can be achieved by using a specific zero-momentum
  subtraction procedure.
\end{itemize}

In this paper we want to discuss these different ingredients from a practical
point of view and offer an algorithmic strategy for practical algebraic
renormalization.  As illustrating examples for our combined algebraic method
we have chosen two processes of phenomenological interest, namely the two-loop
contributions to $B \rightarrow X_s \gamma$ and to $H \rightarrow \gamma
\gamma$.
The important extensions of these techniques to supersymmetric 
examples will be presented in a forthcoming paper.

As mentioned above, the proposed procedure based on algebraic renormalization
is not restricted to a specific class of regularization schemes.
Once the structure of the  local breaking terms 
are under control, one can choose the most
practical 
regularization scheme for the specific case under consideration.

In the following we also use the method of Analytic
Regularization in one of our illustrating examples.  This choice 
is guided by the fact that this scheme enjoys the
property of mass independence like the minimal
subtraction (MS) \cite{dim_reg,Hoo} or
the modified minimal subtraction ($\overline{\rm MS}$) scheme~\cite{msbar}
of the Dimensional Regularization.

The delicate infra-red problem is another important task. As mentioned above,
the method includes zero-momentum subtractions which heavily rely on the
regularity properties of the Green functions at zero momentum~\cite{IR}.
Here we
mention the necessary modifications in massless theories. 

The paper is organized as follows: 

In Section~\ref{sec:sti} we recall the
fundamental symmetry constraints of the SM namely the
Slavnov-Taylor and the Ward-Takahashi identities. 
The main idea of this chapter is
to collect all technical ingredients which are necessary to derive the
symmetry constraints for a specific process in the SM.  

 In the first part of Section~\ref{sec:renorma} we discuss the practical 
consequences of two
further ingredients of the algebraic renormalization,
namely the Quantum Action
Principle and the Wess-Zumino consistency conditions, particularly 
within the background field method (BFM).
Then we propose our main procedure to 
remove the breaking terms in the specific symmetry identities.
The various practical steps are presented in an algorithmic form.

In Section~\ref{sec:hgg_coun} we 
illustrate our practical algebraic renormalization scheme in the two-loop
calculation of the decay $H \rightarrow \gamma\gamma$,
which is one of the promising channels
for the discovery of the Higgs boson with a mass of around 120~GeV.

In Section~\ref{sec:bsg_coun}
the analysis of the electroweak corrections to the decay $b
\rightarrow s \gamma$ is presented.

In the Appendices some
auxiliary technical and theoretical information 
used in Sections~\ref{sec:sti} and~\ref{sec:renorma}
are offered to the reader. In particular those parts of the SM
Lagrangian in the background field gauge which are absent in the literature
are given in Appendix~\ref{app:sourceterms}.
In Appendix~\ref{app:stiex} an explicit example on how in practice
the Slavnov-Taylor identities are derived is discussed.
In Appendix~\ref{app:triangula} we analyze 
the triangular structure  of the counterterms further.
This analysis allows to restore the identities in a 
step-by-step procedure.

%%%%%%%%%%%%%%%%%%%%%%%%%%%%%%%%%%%%%%%%%%%%%%%%%%%%%%%%%%%%

\setcounter{equation}{0}
\section{Slavnov-Taylor and Ward-Takahashi identities} 
\label{sec:sti}

The main tools for algebraic renormalization are the Slavnov-Taylor 
(STI) and Ward-Takahashi identities (WTI). 
In this Section it is shown how the {\it complete} 
set of identities corresponding to a specific
process are derived from their general form
and how it is possible to disentangle the contributions coming from QCD and 
electroweak radiative corrections.

At this point a word concerning the notation is in order.
A generic field is denoted by $\phi$.
$\Phi$ stands for scalar 
matter fields, i.e. Goldstone ($G^\pm$, $G^0$) and Higgs bosons ($H$).
Fermionic fields, respectively their conjugates are represented by
$\psi$ and $\bar{\psi}$.
A generic gauge boson field is denoted by $V^{\mu}_i$
and the ghost and the anti-ghost fields by $c$ and $\bar{c}$, respectively. 
The symbols $G^a_\mu$ and $c^a$ are used to denote gluon fields and the
corresponding  
ghosts in the adjoint representation of the Lie algebra $su(3)$. 
The background fields are marked with a hat in order to distinguish them from
their quantum counterparts. $Q_i$, respectively $Q_{q_i}$, 
denotes the electric charge of a quark $q_i$. 

Let us also introduce three different types of effective actions which will
be used in the following. The Green functions $\Gamma$ are 
regularized and renormalized.
The Green functions $\hat{\Gamma}$ are subtracted using
Taylor expansion (see Section~\ref{sub:remove}).
Finally, $\gg$ denotes the renormalized symmetric Green functions, which
satisfy the relevant WTIs and STIs.

A complete explanation of the conventions, quantum numbers and symmetry
transformations of the fields is provided in Appendix~\ref{app:sourceterms}.

%%%%%%%%%%%%%%%%%%%%%%%%%%%%%%%%%%%%%%%%%%%%%%%%%%%%%%%%%%%%

\subsection{Conventional gauge fixing} 
\label{sub:conventional}
 
In this Section the general 
form of the STI in the conventional `t Hooft gauge fixing is presented.
Thereby we follow the so-called Zinn-Justin formalism~\cite{zinn}.

Let us consider the Gell-Man-Low formula for
one-particle irreducible (truncated) Green functions (1PI) 
\begin{eqnarray}
  \label{gree.1}
  \gg_{\phi_1 \dots \phi_n}(x_1, \dots, x_n) &=& 
  \langle T \left( \phi_1(x_1) \dots \phi_n(x_n) \right) \rangle^{1PI}
  \nonumber\\
  &=& 
  \langle T \left( \phi^\circ_1(x_1) \dots \phi^\circ_n(x_n) \right) 
  e^{-i \int {\rm d}^4x {\cal L}_{int}(x)} \rangle^{1PI}
\,,
\end{eqnarray} 
where the superscript ``$\circ$'' recalls  the free fields.
The Fourier transformed Green functions are denoted by
$\gg_{\phi_1 \dots \phi_n}(p_1, \dots, p_n)$ where  
$p_i, \dots, p_n$ are the incoming 
momenta\footnote{Here and in the following momentum conservation  
  is assumed, i.e. $\sum^{n}_{i=1} p_i = 0$.}.
The definition of $\gg_{\phi_1 \dots \phi_n}(p_1, \dots, p_n)$
in terms of time-ordered products of free fields, $ \phi^\circ_1 \dots
\phi^\circ_n$,
and vertices of the interacting Lagrangian, ${\cal L}_{int}$, requires a
regularization and a subtraction prescription. In this Section we do
not rely on a specific scheme, but only on general features of the
renormalization theory such as the Quantum Action Principle (QAP) 
(see Section~\ref{sub:QAP}) and the Zimmermann identities~\cite{zimm}.

\begin{figure}[t]
  \begin{center}
    \begin{tabular}{cc}
      \epsfxsize=5.cm
      \leavevmode
      \epsffile[180 300 430 500]{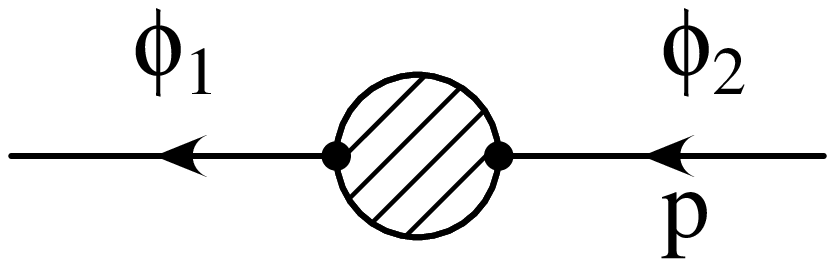}
      &
      \epsfxsize=5.cm
      \leavevmode
      \epsffile[180 300 430 500]{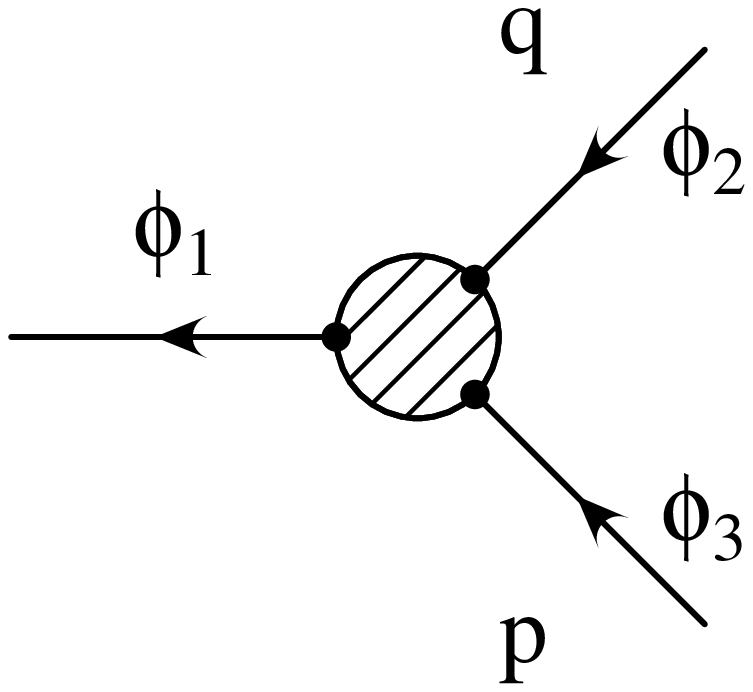}
      \\
      (a) & (b)
    \end{tabular}
    \caption[]{\label{fig:conv}
      \sf All momenta are considered as incoming. In the Green functions
      $\Gamma_{\phi_1 \dots \phi_n}$ they are assigned to the 
      corresponding fields starting from the right.
      The momentum of the most left field 
      is determined via momentum conservation. Exemplary
      $\gg^{(m)}_{\phi_1\phi_2}(p)$ 
      and $\gg^{(m)}_{\phi_1\phi_2\phi_3}(q,p)$ are pictured in (a) and (b),
      respectively.
      }
  \end{center}
\end{figure}

To handle the complete set of Green functions, it is very useful to collect
them into a generating functional
\begin{equation}  \label{gree.2}
\gg[\phi] = \sum_{n=0}^{\infty} \frac{(-i)^n}{n!}\int 
\left( \prod_{j=0}^{n} {\rm d}^4p_j \right) \delta^4 \left( \sum_k p_k \right) 
\phi_1(p_1) \dots \phi_n(p_n) 
\gg_{\phi_1 \dots \phi_n}(p_1, \dots, p_n) 
\,.
\end{equation} 
In perturbation theory $\gg_{\phi_1 \dots \phi_n}(p_1, \dots, p_n)$
is a formal power series in $\hbar$. In the following, 
we will adopt the notation $\gg^{(m)}_{\phi_1 \dots \phi_n}$ to 
indicate the $m$-loop contribution to the Green function  $\gg_{\phi_1 \dots
  \phi_n}$.
In terms of $\gg$ each single Green function of the form (\ref{gree.1})  
is obtained by means of functional derivatives
\begin{equation}  \label{gree.3}
i^n 
\left. \frac{\delta^{n} \gg[\phi]}{\delta 
\phi_1(p_1) \dots 
\delta \phi_n(p_n)} \right|_{\phi=0} =  
\gg_{\phi_1 \dots \phi_n}(p_1 \dots p_n),
\end{equation} 
where $\phi(p)$ denotes the Fourier transform of $\phi(x)$.
In Fig.~\ref{fig:conv} our conventions concerning the external momenta can be
found.
The Green functions of Eq.~(\ref{gree.1}) exhaust all the possible
amplitudes involved in the S-matrix computation, but they do not cover 
the complete set of Green functions needed for the renormalization of the
theory. Indeed, due to the non-linearity of the Becchi-Rouet-Stora-Tyutin
(BRST) transformations~\cite{brs},
the renormalization of some composite
operators (namely $s \phi_i$ where $\phi_i$ is a 
generic field of the SM
and $s$ is the BRST generator) is necessary.
This is usually done by adding the composite operators 
$s \phi_i$ coupled to BRST invariant external sources $\phi^*_i$ 
to the classical action
\begin{equation}  \label{gree.4}
{\cal L}^\prime = {\cal L}_{INV} + \sum_i \phi^*_i \, s \phi^i, 
\end{equation} 
where ${\cal L}_{INV}$ is the gauge invariant Lagrangian of the
SM (see~\cite{STII,hollik_2,krau_ew} and remarks in
Appendix~\ref{app:sourceterms}) 
and studying the renormalization of ${\cal L}^\prime$. For our
purposes we only introduce the BRST sources (also called anti-fields) 
for non-linear 
transformations as proposed by Zinn-Justin \cite{zinn}.
As a remark we mention that in the Batalin-Vilkovisky
anti-field formalism~\cite{BV} the BRST sources
are also introduced for linear BRST transformations.
The advantage is that all the gauge
fields occur on the same footing. However, they are neither necessary for 
our practical purposes nor for proving the 
renormalization of the SM.

The quantization of the theory can only be achieved by introducing a suitable 
gauge fixing ${\cal L}_{GF}$ and the corresponding Faddeev-Popov terms 
${\cal L}_{\Phi\Pi}$
\begin{equation}  \label{gree.5}
\gg_0 = \int {\rm d}^4x \left( 
{\cal L}_{INV} + \sum_i \phi^*_i \, s \phi^i + {\cal L}_{GF} +{\cal
  L}_{\Phi\Pi} \right)
\,.
\end{equation} 
Both ${\cal L}_{GF}$ and ${\cal L}_{\fp}$ break the local  gauge invariance
leaving the theory invariant under the BRST~\cite{brs} transformations. 
The BRST symmetry is crucial for proving the unitarity of the S-matrix and
the gauge independence of physical observables. Therefore it must
be implemented to all orders. For this purpose we establish the  
corresponding  STI in the functional  
form (see~\cite{aoki,hollik_2,krau_ew})
\begin{eqnarray}\label{ST} 
  {\cal S}(\gg)[\phi] & = &\int\,\,{\rm d}^4x  \Bigg[  
  \left( s_W \partial_\mu c_Z + c_W \partial_\mu c_A \right)  
  \left( s_W \frac{\delta\gg}{\delta Z_\mu} +  c_W \frac{\delta\gg}{\delta
      A_\mu} \right) 
  \nonumber\\&&\mbox{}
  + \frac{\delta \gg}{\delta W^{*,3}_\mu} 
  \left( c_W \frac{\delta\gg}{\delta Z_\mu} -  s_W \frac{\delta\gg}{\delta
      A_\mu} \right)    
  +  
  \frac{\delta \gg}{\delta W^{*,\pm}_\mu}  \frac{\delta\gg}{\delta W^\mp_\mu} +
  \frac{\delta \gg}{\delta G^{*,a}_\mu}  \frac{\delta\gg}{\delta G^a_\mu} +  
  \frac{\delta \gg}{\delta c^{*,\pm}}  \frac{\delta\gg}{\delta c^\mp}
  \nonumber\\&&\mbox{}
  +\frac{\delta \gg}{\delta c^{*,3}} 
\left(
c_W \frac{\delta\gg}{\delta c_Z} -  s_W \frac{\delta\gg}{\delta c_A}
\right)+
  \frac{\delta \gg}{\delta c^{*,a}}  \frac{\delta\gg}{\delta c^a} + 
  \frac{\delta \gg}{\delta G^{*,\pm}}  \frac{\delta\gg}{\delta G^\mp}
  +\frac{\delta \gg}{\delta G^{*,0}}  \frac{\delta\gg}{\delta G^0} 
  \nonumber\\&&\mbox{}
+  \frac{\delta \gg}{\delta H^*}  \frac{\delta\gg}{\delta H} 
  +  \sum_{I=L,Q,u,d,e} \left(  
    \frac{\delta \gg}{\delta \bar{\psi}^{*I}} \frac{\delta\gg}{\delta
      \psi^{I}} +  
    {\rm h.c.}
  \right) + \sum_{\alp=A,Z,\pm,a} b_{\alp}  \frac{\delta\gg}{\delta
    \bar{c}^{\alp}}  
  \Bigg]
  \nonumber\\
  &=& 0 
  \,,
\end{eqnarray}  
where the notation $A^\pm B^\mp = A^+ B^- + A^- B^+$ has been used.
$s_W$ and $c_W$ denote the sine and cosine of the Weinberg angle $\theta_W$
and $b_{\alp}$ are the so-called the Nakanishi-Lautrup multipliers\footnote{In
  practical calculations they can be 
  eliminated (in the case of 
  linear gauge fixing) by a  Gaussian integration.}.
The sum in the last line of Eq.~(\ref{ST}) includes the left-handed doublets
and the right-handed singlets.
For the BRST source fields no Weinberg-rotation has been introduced.
We stress that this formula represents the complete nonlinear STI to
all orders. The first two and the last 
term correspond to the linear BRST variation of the $U(1)$ abelian gauge
field and the BRST transformations of the anti-ghost fields. 
Note that the STI of the form~(\ref{ST}) contains the 
complete information of the BRST symmetry and the equation of
motion~\cite{brs,zinn}.

In the form of Eq.~(\ref{ST}) the STIs are independent from
the gauge fixing\footnote{Note that we do not have
to modify Eq.~(\ref{ST}) if the gauge fixing is changed from the
conventional `t~Hooft gauge (see Section~\ref{sub:hgg})
to the background gauge which is used in Section~\ref{sub:bsg}.
However, in order to control the dependence of the Green
functions on the background fields some new terms are conventionally
added to the STIs. They implement the equation of motion for 
the background fields 
and they are studied in the Section~\ref{sub:BFM}.}
In order to specify the gauge fixing,  we introduce the equation of motion
for the $b$ fields corresponding to the various gauge fields in the SM 
\begin{eqnarray}\label{eq:gau_fix} 
  \frac{\delta \gg}{\delta b_\alpha} &=&
  {\cal F}_\alpha({\bf V}, {\Phi}) + \xi_\alpha b_\alpha
\,,
\end{eqnarray} 
where ${\cal F}_\alpha$ ($\alpha= A,Z, W, g$) are the gauge fixing 
functions. $\xi_\alpha$ ($\alpha= A,Z, W, g$) are the corresponding gauge
parameters.  
In the case of the background gauge fixing the functions ${\cal F}_\alpha$
are explicitly given in the formula (A.2) of the Appendix.

Considering a specific process, one first has to single out the
complete set of relevant identities by using a functional derivative
(as in Eq. (\ref{gree.3})). With relevant set we mean the set of 
identities which is closed under renormalization\footnote{Up to 
additional relations which get eventually 
introduced by normalization conditions.}.
This means that the finite parts of a Green function appearing in a
given identity is fixed by other identities of the set or by renormalization
conditions. In practical calculations usually
not all identities are really necessary since they might be
automatically preserved by reasonable regularization schemes at 
lower orders.
An identity can also decouple from the others, because it only 
contains Green functions which do not influence the breaking terms 
of the other identities. The latter point will be discussed in
Section~\ref{sec:renorma}.

The most convenient procedure to deduce the complete set is the
following:  
$(i)$ consider the amplitudes involved in the physical process; $(ii)$ 
derive the identities for those amplitudes; $(iii)$ from each identity
single out the new (superficial divergent) Green functions which are not
involved in the physical process; $(iv)$ derive the identities for these 
new Green functions. The procedure stops when the new identities 
involve only new finite Green functions and no other
divergent quantities.  Finally we have to underline that 
supplementary constraints such as the Faddeev-Popov equations 
can lead to relevant identities on Green functions which 
avoid the use of a new STI.
E.g., in the case of the two-point functions no derivative w.r.t. $b_\alpha$
has to be considered. 

In order to obtain 
a meaningful expression the following two simple rules
have to taken into account:

\begin{enumerate}
\item \label{rule1}
Green functions with a positive or negative 
Faddeev-Popov ghost charge vanish as it is conserved. 
Thus, in order to extract non-zero  identities, it is necessary to 
differentiate  the expression ${\cal S}(\gg)=0$,
which carries ghost charge $+1$, 
w.r.t. one ghost field also having ghost charge $+1$.
It is also possible to differentiate w.r.t.
two ghost fields and one anti-field (carrying ghost charge $-1$).
The only exception to this rule is the case of 
anti-fields for the ghosts.
They carry two Faddeev-Popov ghost charges and, therefore, 
these charges must be compensated with three ghost fields. 
\item 
Identities for 
the Green functions are obtained by taking derivatives of the STI~(\ref{ST})
w.r.t. fields and external sources.  Clearly,
they are non-vanishing only if Lorentz invariance is respected.  
\end{enumerate}

\noindent
The  derivation of  the complete  set of non-trivial identities
is guided by the following rules:

\begin{enumerate}
\setcounter{enumi}{2}
\item \label{rule3}
If we are interested in identities involving several gauge bosons
one has to differentiate ${\cal S}(\gg)=0$ w.r.t.
the set of fields where one of 
the gauge bosons is replaced by the corresponding ghost field  $c_i$. 
The reason for this is that
the linear part of the BRST transformations of a gauge field 
is proportional to the corresponding ghost:
$s V_\mu= \partial_\mu c + \dots$.   
\item \label{rule4}
For Green functions which contain ghost fields a new
rule is needed. One ghost field must be replaced by two ghost fields. 
In fact, the BRST transformation of the ghost fields is
non-linear $s c_i = \frac{1}{2}f_{ijk} c^j c^k$ where
$f_{ijk}$ are the structure constants of the gauge group. 
In the case of ghost two-point functions
this is not necessary 
because we do not acquire any new constraints on them
from this rule (see also Appendix~\ref{app:sourceterms}). 
\item
In the identities derived with the help of rules~\ref{rule3} and~\ref{rule4}
different Green functions occur. The ones  
which still involve gauge bosons or ghosts are constrained further 
by identities which one may derive as described above.
\end{enumerate}

Since the STIs
will be our main tools in the context of algebraic
renormalization, we want to consider their derivation from Eq.~(\ref{ST})
in more detail. 
Recall that both $\gg[\phi]$ and ${\cal S}(\gg)[\phi]$ are integrated 
functionals of the fields $\phi$. 
Thus it is possible to apply the rules of functional derivatives
(see, e.g., Ref.~\cite{itzy}, Section~6-2-2).
Taking the functional derivatives of $\gg$ and setting afterwards
all fields to zero
generates a single Green function $\g_{\phi_1\dots\phi_n}$.
On the other hand, the 
functional derivatives of ${\cal S}(\gg)[\phi]$ generate a single STI
(again after setting the fields to zero after differentiation).
Note that in the expression for ${\cal S}(\gg)[\phi]$ already some
functional derivatives are present which must be 
interpreted as functionals of the form
$\frac{\delta \gg}{\delta \phi}[\phi]$.
The use of rules for taking the derivative of products
enables us to distribute the functional derivatives to the
individual expressions in ${\cal S}(\gg)[\phi]$ and  
to set all fields to zero afterwards.

From the technical point of view the only detail to be clarified is the
dependence on the space-time coordinate, respectively, the momenta
of each single STI. The presence of the integral over the
space-time in Eq.~(\ref{ST}) and the  
conservation of the momentum flow of the Green functions guarantees that no
momentum integration is left. Thus the STI can be expressed as a sum of
products of Green functions.   

An example illustrating the practical applications of the rules
collected in this Section can be found in Appendix~\ref{app:stiex}.
There we explicitly derive all  relevant STIs for a process 
involving two gauge fields and one scalar matter field. This 
general analysis covers, for instance, the
processes $H\rightarrow W^+W^-$, $H\rightarrow Z Z$ and
$H\rightarrow Z \gamma$.  
Also the identities for 
two-point functions with gauge fields and scalars are discussed
which will be 
used in our examples of Sections~\ref{sub:hgg} and~\ref{sub:bsg}.
The drastic simplifications of that analysis within the
Background Field Method (BFM)
will be discussed in Section~\ref{sub:BFM}.

%%%%%%%%%%%%%%%%%%%%%%%%%%%%%%%%%%%%%%%%%%%%%%%%%%%%%%%%%%%%

\subsection{\label{sub:hgg}Example~1: Green Functions and STIs for $H
  \rightarrow \gamma\gamma$}   

In this Section the decomposition of the S-matrix 
elements in terms of 1PI functions is described
for the process $H \rightarrow \gamma \gamma$ in two-loop approximation.
The necessary STIs which relate the finite parts of the Green functions
at the one- and two-loop level are discussed.

The decomposition of the truncated, connected off-shell Green  
functions in terms of 1PI functions is given at the two-loop level by
the following equation:
\begin{eqnarray}\label{loop.hgg.2}  
\lefteqn{G^{(\leq 2)}_{ H A_{\mu} A_{\nu} }(q_1, q_2)   
=}
\nonumber\\&&
  \gg^{(1)}_{ H A_{\mu} A_{\nu} }(q_1, q_2) + 
\gg^{(2)}_{ H A_{\mu} A_{\nu} }(q_1, q_2) + 
\gg^{(1)}_{ H H }(q_1+q_2) G^0(q_1+q_2)
\gg^{(1)}_{ H A_{\mu} A_{\nu} }(q_1,q_2)  
\nonumber \\  
&&+  
\gg^{(1)}_{ H A_{\mu} A_{\rho} }(q_1,q_2) 
G^0_{\rho\sigma}(q_2) \gg^{(1)}_{A_{\sigma} A_{\nu}}(q_2) + 
\gg^{(1)}_{A_{\mu} A_{\rho}}(q_1) 
G^0_{\rho\sigma}(q_1) \gg^{(1)}_{ H A_{\sigma} A_{\nu}}(q_1,q_2)
\,,
\end{eqnarray}
where the tree-level propagators for the photon and the Higgs boson
are given by
\begin{eqnarray}
  G^0_{\rho\sigma}(k) =  
  -i \left( \frac{g_{\rho \sigma} + (\xi -1) 
      \frac{k_\rho k_\sigma}{k^2}}{k^2 + i \epsilon}  \right)
    \quad\mbox{and}\quad G^0(k) = \frac{i}{k^2
      - m^2_H}
    \,,
  \end{eqnarray}
respectively.
$\gg^{(1)}_{A_{\sigma} A_{\rho}}(q)$ is the photon self-energy at one-loop
order, 
$\gg^{(1)}_{ H H }(q)$ the self-energy of the Higgs boson  and
$\gg^{(1)}_{ H A_{\mu} A_{\nu} }(q_1, q_2)$ and
$\gg^{(2)}_{H A_{\mu} A_{\nu} }(q_1, q_2)$  
are the one- and two-loop corrections to the 
$H\gamma\gamma$ vertex. $q_1$ and $q_2$ denote the in-going
momenta of the two photons.

\begin{figure}[ht]
   \begin{center}
     \begin{tabular}{ccc}
       \epsfxsize=5.cm
       \leavevmode
       \epsffile[180 300 430 480]{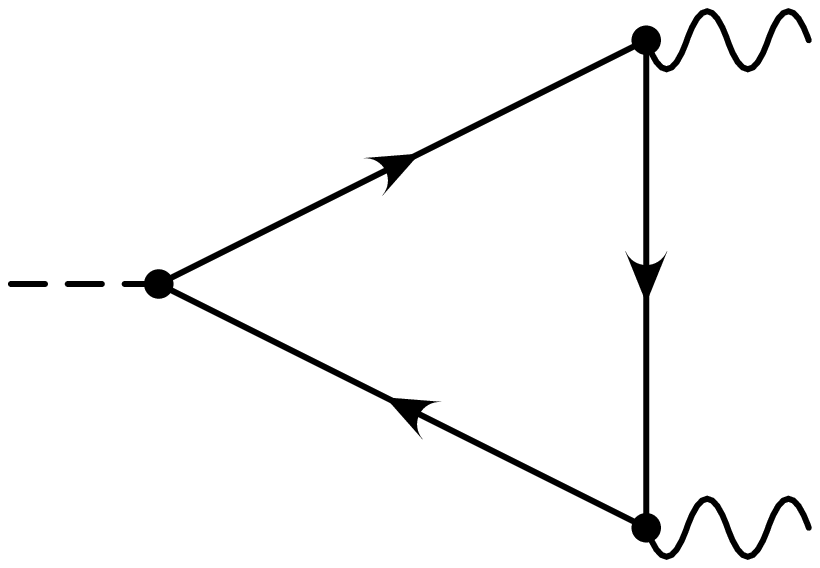}
       &
       \epsfxsize=5.cm
       \leavevmode
       \epsffile[180 300 430 480]{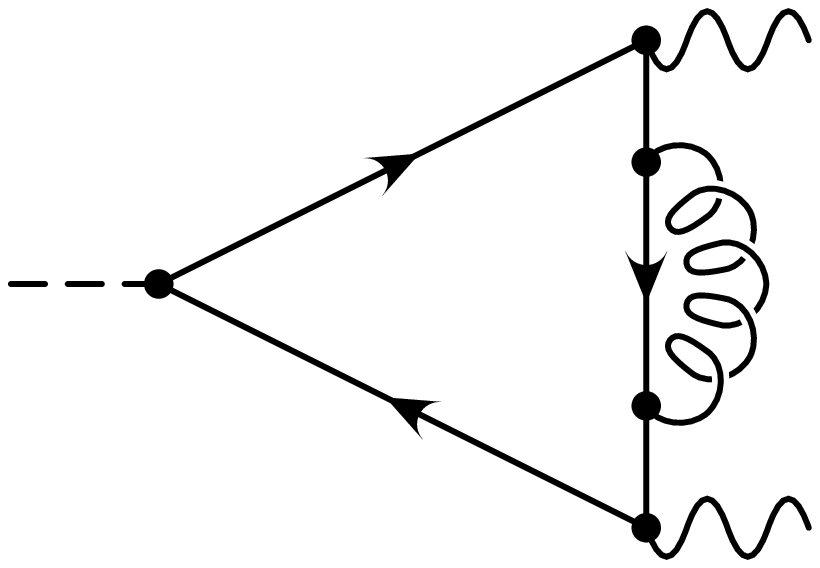}
       &
       \epsfxsize=5.cm
       \leavevmode
       \epsffile[180 300 430 480]{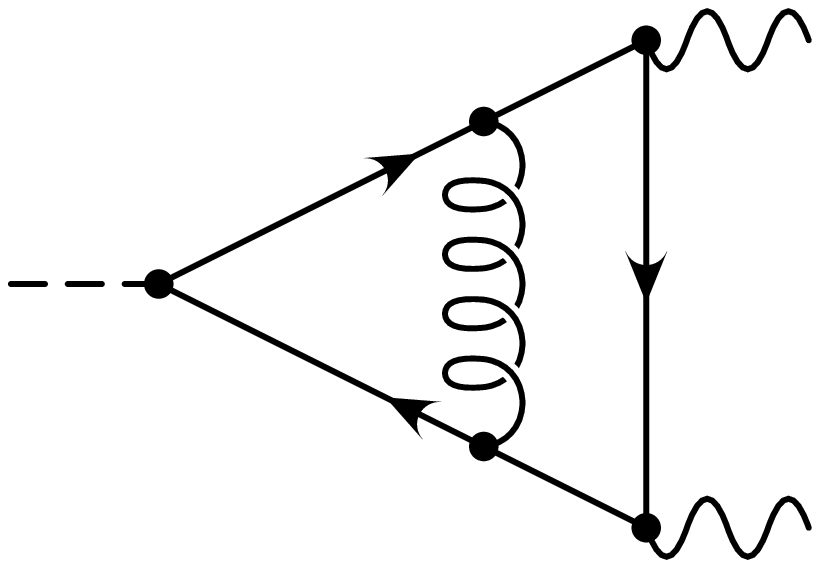}
     \end{tabular}
     \caption[]{\label{fig:hgg} \sf One- and some two-loop diagrams
       contribution to 
       $H\to\gamma\gamma$. The dashed lines correspond to the Higgs boson, the
       wavy lines to the photons, the curly ones to the gluons and the
       straight ones to the quarks.}
   \end{center}
  \end{figure}

In this calculation at two-loop order only QCD corrections are considered.
Thus it is convenient to decompose the one-loop vertex corrections into
a fermionic and a bosonic part
\begin{eqnarray}
\gg^{(1)}_{H A_{\mu} A_{\nu}}(q_1, q_2) &=&  
\gg^{(1), ferm}_{H A_{\mu} A_{\nu}}(q_1, q_2)  
+\gg^{(1), bos}_{H A_{\mu} A_{\nu}}(q_1, q_2)
\,.
\end{eqnarray}
Furthermore the two-loop terms are split into QCD and electroweak corrections:
\begin{eqnarray}
\gg^{(2)}_{H A_{\mu} A_{\nu}}(q_1, q_2) &=&  
\gg^{(2), QCD}_{H A_{\mu} A_{\nu}}(q_1, q_2)  
+\gg^{(2), ew}_{H A_{\mu} A_{\nu}}(q_1, q_2)
\,.
\end{eqnarray}
Actually, since 
at two-loop level only QCD corrections are considered, the terms 
involving the photon or Higgs boson self-energy in Eq.~(\ref{loop.hgg.2}) 
vanish. Their contribution would be of the same order as the two-loop
electroweak corrections to the genuine vertex.
 In Fig.~\ref{fig:hgg} some sample diagrams of the remaining contributions are
 pictured.
 
The physical amplitude is calculated via a projection on the physical states 
\begin{equation}\label{loop.hgg.1}
{\cal M}^{(\leq 2)}_{H\rightarrow \gamma\gamma} = \left.  
G^{(\leq 2)}_{ H A_{\mu} A_{\nu} }(q_1, q_2) \epsilon^{\mu}(q_1) 
\epsilon^{\nu}(q_2) \right|_{q_1^2 = q_2^2 = 0, (q_1 + q_2)^2 = M^2_H} 
\,,
\end{equation}
where $M_H$ is the Higgs boson mass and
$\epsilon^{\mu}(q)$ denotes the polarization vector of the photon with
momentum $q$.

The mass shell projection 
of the two-loop amplitude can be correctly performed only if the  
self-energy of the photon, $\gg^{(1)}_{A_{\sigma} A_{\rho}}(q)$,   
satisfies the well-known transversality 
condition $q^{\nu} \gg^{(1)}_{A_{\mu} A_{\nu}}(q) = 0$.
However, in a non-symmetric regularization scheme this
property is in general not valid any longer.
It has to be reestablished as will be explained in Section~\ref{sec:renorma}.

In order to obtain the complete set of one-loop 
counterterms, we observe
that the regularized two-loop Green function  
$\g^{(2)}_{ H A_{\mu} A_{\nu}}(q_1,q_2)$  
contains three different 
sub-divergences which require proper subtraction.
Furthermore, since in general the
regularization procedure breaks the symmetry, we are forced to 
introduce the following general counterterm Lagrangian 
\begin{eqnarray}\label{lag.hgg.3} 
\hspace{-.5cm}
{\cal L}^{CT}_{H\rightarrow \gamma\gamma} &=&
\sum_i  
Z_1\, e \, Q_i A_{\mu} \bar{q}_{i} \gamma^{\mu} q_{i} +  
Z_2 \bar{q}_{i} \not\!\partial q_{i} + 
Z_{m_i} Z_2 m_i \bar{q}_i q_i +  
Z_{Y_i} Z_2 Y_i H  \bar{q}_i q_i 
\,,
\end{eqnarray} 
where $H, A_{\mu}, \bar{q}_i$ and $q_i$ are the Higgs boson, the photon and
the fermion fields, respectively.
The parameters $e, m_i$ and $Y_i$ are the gauge coupling, 
the masses and the Yukawa couplings of the fermions.
In a symmetric 
regularization scheme the free parameters $Z_1, Z_2, Z_{m_i}$ and $Z_{Y_i}$ 
are related by means of QED-WTI\footnote{$Z_{m_i}$   
is not related to $Z_{Y_i}$ by a STI or WTI. However, 
in the general electroweak case two out of the three
parameters $Z_{m_i}$, $Z_{Y_i}$ and $v$, the vacuum expectation value, can be
chosen  
independently.}.
In a non-invariant
regularization scheme this is not true any longer and these coefficients  
have to be fixed separately.
For example, in Analytic
Regularization~\cite{Spe}, which we will use for the practical computation, the
renormalization constants are Laurent-expanded in powers of the regulators.
In our case it turns out that it is enough to introduce only one, $\lambda$
\cite{breite}.  
Then the renormalization constants read
$Z_{\alp} = \sum_{n\geq0} Z^{(n)}_\alp/\lambda^n$ ($\alp=1,2,m_i,Y_i$).
The pole parts are 
removed by means of the minimal subtraction scheme and the finite 
parts, $Z^{(0)}_\alp$, are fixed by the STIs.

In order to fix the one-loop counterterms explicitly, we have to consider the 
Green functions $\gg^{(1),QCD}_{A_\mu \bar{q}_i q_i}(\bar{p},p),  
\gg^{(1),QCD}_{H \bar{q}_i q_i}(\bar{p},p)$ and $\gg^{(1),QCD}_{\bar{q}_i
  q_i}(p)$
which arise as sub-diagrams of the two-loop graphs.
The fermion self-energy  
$ \gg^{(1),QCD}_{\bar{q}_i q_i}(p)$ contains the two independent 
parameters $Z_2$ and $Z_{m_i}$ which are not constrained by any 
STI. They can be fixed as usual either 
minimally, i.e. $Z^{(0)}_2=Z^{(0)}_{m_i}=0$, or by imposing
on-shell renormalization conditions
\begin{equation}\label{on.hgg.7} 
\gg^{(1),QCD}_{\bar{q}_i q_i}(\not\! p = m_i) = 0,~~~ 
\left. \frac{\partial}{\partial \!\!\! \not p}   
\gg^{(1),QCD}_{\bar{q}_i q_i}(p)\right|_{\not\, p = m_i} = 1.  
\end{equation} 
The parameter $Z_{Y_i}$ is related to the mass  
renormalization constant, $Z_{m_i}$. In our specific example,  
where fermion mixing is absent, both parameters can be identified.
$Z_1$ has to be fixed in terms of the STI which  
relates the vertex $\gg^{(1),QCD}_{A_\mu \bar{q}_i q_i}(\bar{p},p)$ to 
the fermion self-energy. By differentiating the identity w.r.t. the  
photon ghost field $c_A$ and the fermion fields  
$\bar{q}_i$ and $q_i$, one immediately gets
\begin{eqnarray}
  \label{wti.hgg.8} 
  \lefteqn{\frac{\delta^3 {\cal S}(\gg)^{(1),QCD}}{\delta c_A(-p -\bar{p}) 
      \delta \bar{q}_i(\bar{p}) \delta {q}_i({p})}\Bigg|_{\phi=0} =}
  \nonumber\\&&\mbox{}
  i (\bar{p} + p)^{\mu} \gg^{(1),QCD}_{A_\mu \bar{q}_i q_i}(\bar{p},p) + i e
  Q_i   
  \left(  
    \gg^{(1),QCD}_{\bar{q}_i q_i}({p}) - \gg^{(1),QCD}_{\bar{q}_i
      q_i}(-\bar{p})  
  \right)
  \nonumber\\
  &=& 0
  \,,
\end{eqnarray}
which is equivalent to the simple WTI in QED.
Note that this equation is only true as exclusively QCD corrections
are considered at two-loop order.

After the one-loop counterterms are fixed, let us now focus on the 
Green function $\gg^{(i)}_{ H A_{\mu} A_{\nu} }(q_1, q_2)$ ($i=1,2$)
which is our prime interest.
We again have to make sure that the finite parts of the counterterms 
are correctly fixed according to the STI. In fact, since the process
$H\to\gamma\gamma$ 
has no tree-level contribution there is no free overall parameter which 
fixes the finite parts by using renormalization conditions.
 
According to rule~3 of the previous subsection, 
we consider the derivative of ${\cal S}(\gg)=0$
w.r.t. the photon ghost field
$c_A$, one photon $A_{\nu}$ and the Higgs boson $H$.
As a result we obtain the identity which involves among others also the  
Green functions $\gg_{ H A_{\mu} A_{\nu} }(q_1, q_2)$:
\begin{eqnarray}
  \label{ide.hgg.4} 
  \lefteqn{\frac{\delta^{3} {\cal S}(\gg)}{\delta c_A(q_1)  
      \delta A_{\nu}(q_2) \delta H(p)}\bigg|_{\phi=0} = }
  \nonumber\\&&\mbox{} 
   \left( -i c^2_W q_1^\rho -s_W  \gg_{c_A W^{*,3}_\rho}(-q_1) \right) 
  \gg_{A_\rho A_\nu H}(q_2,p)
  \nonumber\\&&\mbox{} 
  + \left( -i c_W s_W q_1^\rho + c_W  \gg_{c_A W^{*,3}_\rho}(-q_1) \right) 
  \gg_{Z_\rho A_\nu H}(q_2,p)
  \nonumber\\&&\mbox{} 
  + \gg_{c_A W^{*,3}_\rho H}(q_2,p) 
  \left( c_W \gg_{Z_\rho A_\nu}(q_2)  -s_W \gg_{A_\rho A_\nu}(q_2) \right)
  \nonumber\\&&\mbox{} 
  + \gg_{c_A W^{*,3}_\rho A_\nu}(p, q_2) 
  \left( c_W \gg_{Z_\rho H}(p)  -s_W \gg_{A_\rho H}(p) \right)
  \nonumber\\&&\mbox{}+  
  \gg_{c_A G^{*,0}}(- q_1)   \gg_{G^0 A_\nu H} (q_2,p)+ 
  \gg_{c_A H G^{*,0}} (p, q_2)\gg_{G^0 A_\nu} (q_2) 
  \nonumber\\&&\mbox{}+     
  \gg_{c_A G^{*,0} A_\nu}(p, q_2) \gg_{G^0 H}(p) + 
  \gg_{c_A H^*}(-q_1)   \gg_{H A_\nu H}(q_2,p) 
  \nonumber\\&&\mbox{}+   
  \gg_{c_A H H^*}(p,q_2) \gg_{H A_\nu}(q_2)  + 
  \gg_{c_A H^* A_\nu}(p,q_2) \gg_{H H}(p)
  \nonumber\\
  &=&0
  \,.
\end{eqnarray}
Actually this equation constitutes a special case of the identity~(\ref{e_4})
derived in a more general context.

In the following we demonstrate how this equation simplifies for the special
kind of corrections we are interested in.
In order to disentangle consistently the QCD 
corrections from the Green functions appearing in the STI at a given order
one can take the derivative of the latter w.r.t. the parameters
of the $SU(3)$ colour group, namely $C_A$ and $C_F$.
Furthermore, we can disentangle the contributions 
of the fermion loop from the contribution of the bosonic corrections  
as the coupling of ghost fields (as well as the external BRST sources
$W^{*,3}_\rho$, $G^{*,0}$ and $H^*$) to the fermion lines occurs for the first
time through two-loop electroweak interactions.
Note that the Green functions $\gg_{Z_\rho H}$,  
$\gg_{H A_\nu}$, $\gg_{G^0 A_\nu H}$ and 
$\gg_{G^0 H}$ 
vanish at tree-, one- and  
two-loop level as they violate CP invariance. 
It is well-known that the CP symmetry is violated in the SM  
only through the Cabibbo-Kobayashi-Maskawa (CKM) matrix. Thus CP violation  
manifests itself in the scalar sector starting at the three-loop order.
The last term in (\ref{ide.hgg.4}) vanishes if one restricts the analysis
to fermionic contributions and their QCD corrections.
Taking these simplifications into account we finally get for 
the first term of the  
r.h.s. of Eq.~(\ref{ide.hgg.4}) expanded up to two loops 
\begin{eqnarray}
\label{ide.hgg.5} 
\left( \gg_{c_A W^{*,3}_\rho} \gg_{A_\rho A_\nu H}\right)^{(\leq 2), ferm} 
&=&  
 \gg^{(0)}_{c_A W^{*,3}_\rho}(-q_1) \gg^{(1), ferm}_{A_\rho A_\nu H}(q_2,p) +  
 \gg^{(1)}_{c_A W^{*,3}_\rho}(-q_1) \gg^{(1), ferm}_{A_\rho A_\nu H}(q_2,p) 
 \nonumber\\&&\mbox{} 
 +\gg^{(0)}_{c_A W^{*,3}_\rho}(-q_1) \gg^{(2), QCD}_{A_\rho A_\nu H}(q_2,p)  
\,.
\end{eqnarray}
Note that the three-point function $\gg^{(0)}_{A_\rho A_\nu H}(q_2,p)$ is  
absent at tree level. Further simplifications occur through the observation
that the second term on the r.h.s. of
Eq.~(\ref{ide.hgg.5}) does not give any 
contribution since there is no room for QCD corrections.
Please note that the one-loop Green function  
$\gg^{(1)}_{c_A W^{*,3}_\rho}$ does not contain any fermionic loop.
In the same line of reasoning all terms except the ones in the first two lines
of Eq.~(\ref{ide.hgg.4}) drop out.

From the Lagrangian one obtains
$ \gg^{(0)}_{c_A W^{*,3}_\rho}(-q_1) = i s_W q_1^{\mu}$.
This in combination with the accordingly simplified
remaining terms of Eq.~(\ref{ide.hgg.4})
finally lead us to the following STI
\begin{equation}\label{ide.hgg.6}  
- i q_1^{\rho} \gg^{(i), ferm}_{A_\rho A_\nu H}(q_2,p) = 0  
\,,
~~~~ i=1,2 
\,.
\end{equation}
This identity 
must be fulfilled at one- and two-loop order for the specific
corrections we are interested in.

In Section~\ref{sec:hgg_coun} the breaking
terms for these identities will be provided.
Furthermore we will compute the  
amplitude in the analytical regularization scheme and we will show how the
algebraic renormalization works for this two-loop example.

%%%%%%%%%%%%%%%%%%%%%%%%%%%%%%%%%%%%%%%%%%%%%%%%%%%%%%%%%%%%

\subsection{Background gauge fixing} 
\label{sub:BFM}
 
As is well known the BFM~\cite{bkg}
allows to derive the S-matrix elements in terms of Green functions with  
external background fields with the exception of fermion fields. 
The main simplification in the BFM results from the fact that 
the theory with background
fields possesses two different invariances: the BRST symmetry, which 
  involves quantum fields and ghosts, and the background gauge
  invariance. The latter provides several simplifications in the
  computation of physical amplitudes and in the renormalization procedure 
{due to its linearity.}
The first systematic application of BFM in the SM
for {\it invariant} regularizations at the one-loop 
level was presented in~\cite{msbkg}. 
{Our considerations regarding 
BFM, however, 
also apply in the case of {\it noninvariant} regularizations and 
also beyond the one-loop level. We focus on the practical aspects  
of the BFM  which are relevant for higher-loop calculations. 
Further details on the theoretical 
advantages of the BFM can be found  in~\cite{msbkg,papa,grassi}.}

{ The main difference between the STIs~(\ref{ST})
and the WTIs for the background gauge invariance is due to the
linearity of the latter. Linearity means that the WTIs are linear in the
functional $\gg$ and therefore they relate Green functions of the same
orders while for the STIs there is an 
interplay between higher and lower order radiative corrections.}

To renormalize properly the SM in the background gauge, one 
needs to implement the equations of motion for the background fields at the 
quantum level. 
The most efficient way to this end is to extend the BRST symmetry to the 
background fields
\begin{eqnarray}
\label{new_1}
&& s \hat{W}^{3}_{\mu} = \Omega^{3}_{\mu},~~~~s \Omega^{3}_{\mu} = 0,~~~~ 
s \hat{G}^{0}  = \Omega^{0},~~~~s \Omega^{0} = 0, \nonumber \\
&&s \hat{W}^{\pm}_{\mu} = \Omega^{\pm}_{\mu},~~~~s \Omega^{\pm}_{\mu} = 0,
~~~s \hat{G}^{\pm}  = \Omega^{\pm},~~~s \Omega^{\pm} = 0, \\
&&s\hat{G}_{\mu}^{a}  = \Omega^{a}_{\mu},~~~~~ s \Omega^{a}_{\mu} = 0,~~~~   
s \hat{H}  = \Omega^{H},~~~~~ s \Omega^{H} = 0, \nonumber 
\end{eqnarray}
where $ \Omega^{\pm}_\mu,  \Omega^{3}_\mu$ and $\Omega ^a_\mu$ are (classical) vector fields with 
the same quantum numbers as the gauge bosons $W,Z$ and $G^a_\mu$, but ghost charge $-1$ (like an anti-ghost field). 
$\Omega^{\pm},  \Omega^{0}$ and $\Omega ^H$ are scalar fields with ghost number $-1$; in the following we will 
denote by $\Omega$ the complete set of these fields. 

In the following we will denote with $\gg'$ the effective action which 
depend on the  $\Omega$ (and correspondently the ${\cal S}'_{\gg}$ its Slavnov-Taylor operator) and 
with $\gg =  \left. \gg'\right|_{\Omega=0}$, i.e. 
by setting $\Omega$ to zero. 

Therefore one has  to modify correspondingly the STI
\begin{eqnarray}\label{ST_bfm} 
  {\cal S}'(\gg')[\phi] & = & {\cal S}(\gg')[\phi] + 
\int\,\,{\rm d}^4x  \Bigg\{ \Omega^{3}_\mu 
  \left[ c_W  \left(  \frac{\delta\gg'}{\delta \hat{Z}_\mu} -  \frac{\delta\gg'}{\delta Z_\mu}\right) 
 -  s_W \left( \frac{\delta\gg'}{\delta \hat{A}_\mu}      -  \frac{\delta\gg'}{\delta A_\mu} \right) \right]
\nonumber\\&&\mbox{}
+ \Omega^{\pm}_\mu \left(  \frac{\delta\gg'}{\delta \hat{W}^\mp_\mu} - \frac{\delta\gg'}{\delta W^\mp_\mu}\right) 
+ \Omega ^a_\mu  \left(  \frac{\delta\gg'}{\delta \hat{G}^a_\mu} - \frac{\delta\gg'}{\delta G^a_\mu}\right) 
\nonumber\\&&\mbox{}
+   \Omega^{\pm} \left( \frac{\delta\gg'}{\delta \hat{G}^\mp} - \frac{\delta\gg'}{\delta G^\mp}\right) 
  +  \Omega^0  \left( \frac{\delta\gg'}{\delta \hat{G}^0} -  \frac{\delta\gg'}{\delta {G}^0}\right) 
 % \nonumber\\&&\mbox{}
+  \Omega^H  \left( \frac{\delta\gg'}{\delta \hat{H}} - \frac{\delta\gg'}{\delta H}\right) 
 \Bigg\} 
  \nonumber\\
  &=& 0 
  \,.
\end{eqnarray}  
Here $ {\cal S}(\gg')[\phi]$ are the STIs given in Eq.~(\ref{ST}). 
Notice that also the scalar fields $G^\pm$, $G^0$ and $H$ are paired 
with their own background fields, 
$\hat{G}^\pm, \hat{G}^0$ and $\hat{H}$ in order to 
extend the `t~Hooft 
gauge fixing to a background gauge invariant one described in 
Appendix~\ref{app:sourceterms}. 

To study how the effective action $\gg'$ depends on  $\hat W^\pm_\mu$, for instance,  
one has to derive the STI (\ref{ST_bfm}) with respect to the fields $\Omega^\pm_\mu$.   
After setting $\Omega^{\pm}_\mu=0$ one obtains
\begin{equation}
  \label{eq:new_1}
  \left(  \frac{\delta\gg}{\delta \hat{W}^\mp_\mu} - \frac{\delta\gg}{\delta W^\mp_\mu}\right)  = 
{\cal S}_{\gg} \left( \frac{\delta\gg'}{\delta {\Omega}^\mp_\mu}
  \right)_{\Omega=0} \,,
\end{equation}
where ${\cal S}_{\gg}$ is the linearized (conventional) Slavnov-Taylor 
operator of Eq.~(A.8). These equations describe the relations between the quantum and
the background fields and they supplement the STI and 
the WTI:
\begin{equation}
  \label{eq:an_1}
{\cal S}\left( \gg \right) = 0, ~~~~{\cal W}_{(\lambda)} (\gg) = 0 
\end{equation}
where
${\cal W}_{(\lambda)}$ is the WTI operator of the background gauge invariance (cf. Eq. (\ref{WTI})). 

The space of counterterms and of possible  breaking terms to the STI
is enlarged  
by those monomials which contain the background fields $\hat{\phi}$ and the fields $\Omega$ in addition 
to the conventional fields $\phi$ and anti-fields $\phi^*$. 
This requires a new analysis. 

In the calculation of the necessary 
counterterms, there are two main approaches. They only differ by the ordering 
in which the three  equations in (2.21) and (2.22) are used. 
\begin{enumerate}

\item 
We first use the WTI for the background gauge invariance and a subset of 
the conventional STIs (\ref{ST}). 
{The only missing parts are  counterterms  which relate
the two-point functions of the 
background field to the two-point functions of the quantum fields and for the two-point 
functions $\Gamma_{\phi^* \Omega}$, where $\phi^*$ is a generic anti-field}. 
To fix these last counterterms one has to use the extended 
STI (\ref{ST_bfm}). 

Thus, to fix for example the counterterms for the two-point function
for the quantum field $W^\pm_\mu$,  
one has to derive the STI (\ref{ST_bfm}) with respect to
$\Omega^\pm_\mu, \hat W^\mp_\nu$ and with  
respect to  $\Omega^\pm_\mu, W^\mp_\nu$:
\begin{eqnarray}
  \label{new_2}
  \gg_{\hat{W}^{+} {W}^{-}}(p) 
&=&  \gg_{W^{+} W^{-}}(p) + \gg_{\Omega^{+} W^{*,-}}(p)  \gg_{W^{+} W^{-}}(p) \,,\nonumber \\
\gg_{\hat{W}^{+} \hat{W}^{-}}(p) 
&=&  \gg_{W^{+} \hat{W}^{-}}(p) + \gg_{\Omega^{+} W^{*,-}}(p)  \gg_{W^{+} \hat{W}^{-}}(p)  \,.
\end{eqnarray}
These equations fix completely  the two-point functions
$\gg_{\hat{W}^{+} {W}^{-}}(p),  
\gg_{\hat{W}^{+} \hat{W}^{-}}(p)$ in terms of the quantum one $
\gg_{W^{+} W^{-}}(p)$. 
Finally, in order to fix the counterterms for the two-point functions
$\Gamma_{\phi^* \Omega}$  
(a complete discussion has been given in \cite{grassi}), one has to
consider the derivative of (\ref{eq:new_1}) with respect to  
the anti-field $\phi^*$ and one ghost $c$. This conclude the algebraic
renormalization program in the case of the BFM.
  
\item 
{The second alternative approach exploits completely the use of the extended 
STI, namely equations of the type (\ref{eq:new_1}): 
Thus, one first computes all possible counterterms to restore the WTI, 
then one fixes the remaining counterterms 
by considering the functional derivative of the extended STI (\ref{ST_bfm})
with respect to $\Omega$, the background fields $\hat\phi$ and the 
quantum fields $\phi$. However, as in the former approach this cannot
exhaust completely the  
algebraic renormalization program: 
One shows that one still needs a reduced set of STIs in addition which
can be derived from the conventional  
STI (\ref{ST}). }{In particular, besides the Eqs.  (\ref{eq:new_1}),
one needs the STI to fix the anti-field  
part of the action. This guarantees that the BRST transformation are
preserved to all orders.} 
\end{enumerate}

In a practical analysis of a physical process where 
one has not to compute the same 
Green functions with the quantum fields replaced by the corresponding 
background fields nor vice versa, the first approach is favourable ---
as in the two phenomenological examples discussed in this paper.
One can disregard 
all equations of the type (\ref{eq:new_1}). This  simplifies 
the analysis significantly. 
A phenomenological 
example using the second approach  will be discussed in a forthcoming 
publication. In Section~\ref{sub:consi} we discuss the analysis 
of the breaking terms in both methods within the BFM.

In order to single out the relevant WTI, we have to take into
account the rules stated in  Section~\ref{sub:conventional}.
However, in the case of the background fields,
the role of the ghost particles is played by the parameter of the
infinitesimal background gauge transformations
(see Appendix~\ref{app:sourceterms}).
To each generator of the gauge group 
$SU_C(3) \times SU_I(2) \times U_Y(1)$ we consider the corresponding
local infinitesimal parameters. They are denoted by
$\lambda_A(x), \lambda_Z(x)$ and $\lambda_\pm(x)$
for the electroweak part and $\lambda_a(x)$ for the QCD sector.
Thus, the functional WTI for the effective action $\gg'$ reads:
\begin{eqnarray}
  \label{WTI}
  {\cal W}'_{(\lambda)}(\gg') &=& \sum_{\phi} \int {\rm d}^4x
  \left(\delta_{\lambda(x)}\phi\right)
  \frac{\delta 
    \gg'}{\delta \phi(x)} = \sum_{\alp = A,Z\pm, a} 
  \int {\rm d}^4x \lambda_\alp(x) {\cal W}'_{\alpha}(x) (\gg') \,\,=\,\, 0
\,,   
\end{eqnarray}
where the variations $\delta_{\lambda}\phi(x)$ are explicitly given in
Appendix~\ref{app:sourceterms}
(see Eqs.~(\ref{BKG_tranf})--(\ref{BKG_trans_ferm})).
The sum runs over all possible fields and anti-fields.
 ${\cal W}'_{\alpha}(\lambda) ({\cal W}'_{\alpha}(x))$ is 
called Ward-Takahashi operator and acts on the 
functional $\gg'[\phi]$. An explicit expression is given in \cite{msbkg}.

Concerning the rules of Section~\ref{sub:conventional}, 
only two slight modifications of the  
rules~\ref{rule3} and~\ref{rule4} are necessary:
\begin{itemize}
\item[$3^\prime$.]\label{rule3p}
  One has to  differentiate the general WTI (\ref{WTI})
  w.r.t. the infinitesimal parameters $\lambda_V$ in order to get
  constraints on the Green functions  
  involving  the corresponding background gauge fields $\hat{V}$.
\item[$4^\prime$.]\label{rule4p}
  To derive constraints on Green functions involving one
  ghost and one anti-field plus other  
  quantum fields one either can derive a corresponding STI with
  rule~\ref{rule4} (i.e. differentiate Eq.~(\ref{ST}) w.r.t.
  two ghost fields) or one can derive a linear WTI by  
  differentiating w.r.t. one ghost field and one infinitesimal
  parameter $\lambda_\alpha$.
  We prefer to use the second version since linear WTIs are simpler to
  handle within our specific subtraction method
  (cf. Section~\ref{sub:remove}).
  Note that this choice implies some assumptions on
  the wave function renormalization for multiplets of fields  
  as will be explained in more detail in the example on \bsg~(cf.
  Section~\ref{sec:bsg_coun}).
\end{itemize}

In connection to these modifications a remark is in order.
If we consider a Green function with one gauge field, one ghost field
and one anti-field (e.g., $\gg_{c_A W^{*,+}_\mu W^{-}_\nu}$)
we have to differentiate
the WTI~(\ref{WTI}) w.r.t. the ghost field, the anti-field 
and the infinitesimal parameter (which in the example is $\lambda_-$)
associated to the gauge field. This provides an identity which fixes the
considered Green function. In the case that no gauge field is involved 
(e.g., $\gg_{c_A \bar{q}^{*} q'}$)
one has to consider the background variation of the ghost field,
of the anti-fields and of the 
quantum field (which in the example is $q'$).
Thus the WTI has to be differentiated w.r.t. 
$\delta_{\lambda} c_A, \bar{q}^{*}$ and $q'$,
and $c_A, \delta_{\lambda} \bar{q}^{*}$ and $q'$,
and $c_A, \bar{q}^{*}$ and $\delta_{\lambda} q'$.
Some of the resulting WTIs coincide\footnote{
This is a consequence of the consistency conditions 
to be discussed in Section~\ref{sub:consi}.}. 
One has to select the independent ones, but this can 
be easily done by inspection of the WTIs themselves.

At this point let us consider 
the example discussed in 
Appendix~\ref{app:stiex}
the amplitude involving two gauge fields $V^\mu_i$ 
and one scalar field $\Phi$ in the context of the BFM.
Thus we are able to compare the two approaches and to
underline the differences. 

In the framework of the BFM the two gauge fields, $V^{\mu}_1$ and
$V^{\nu}_2$, and the scalar field $\Phi$ are replaced by 
their counterparts $\hat{V}^{\mu}_1$, $\hat{V}^{\nu}_2$
and $\hat{\Phi}$, respectively.
As in the conventional gauge fixing the amplitude 
is built up by irreducible Green functions which in this case read
$\gg_{\hat{V}^{\mu}_i \hat{V}^{\nu}_j}$ $(i,j=1,2)$,
$\gg_{\hat{\Phi} \hat{\Phi}}$ and
$\gg_{\hat{V}^{\mu}_1 \hat{V}^{\nu}_2 \hat{\Phi}}$.
In the following we will denote irreducible Green functions
where only external background fields are involved as
background Green functions\footnote{Notice that in the following 
equations where we consider  single components of WTIs or STIs, or 
even some specific Green functions obtained  from $\gg'$ we can 
avoid the prime.}.

Let us in a first step consider
the two-point function $\gg_{\hat{V}^i_{\mu} \hat{V}^j_{\nu}}(p)$. We
get the following identities using~(\ref{WTI}) in combination
with Eqs.~(\ref{BKG_tranf})--(\ref{BKG_trans_ferm}):
\begin{eqnarray}
  \label{bfm_2}
  \left. \frac{\delta ^2 {\cal W}_{(\lambda)}(\gg) }{\delta \lambda_V^i(-p)
      \delta  \hat{V}^j_{\nu}(p)}
  \right|_{\phi=0} &=& i p^\mu \gg_{\hat{V}^i_{\mu} \hat{V}^j_{\nu}}(p) + 
  \sum_{\Phi'} M_{i, \Phi'} \gg_{\hat{\Phi}' \hat{V}^j_{\nu}}(p)
\,\,=\,\, 0 
\,.
\end{eqnarray}
The sum runs over all Goldstone fields $G^0$ and $G^\pm$
with masses $M_{\pm, G^\pm} = \pm i M_W$,  $M_{Z, G^0} = - M_Z$ 
and zero for all the other combinations.
In the following the summation sign will be omitted.
A comparison with the corresponding identity in the conventional
formalism, Eq.~(\ref{e_1}), shows that Eq.~(\ref{bfm_2})
is linear in the Green functions.
However, it requires the renormalization of the mixed two-point
functions $\gg_{\hat{\Phi}' \hat{V}^j_{\nu}}(p)$ which can be studied
with the help of rule~$3^\prime$:
\begin{eqnarray}
  \label{bfm_3.0}
  \left. \frac{\delta ^2 {\cal W}_{(\lambda)}(\gg) }{\delta \lambda_V^i(-p)
      \delta  \hat{\Phi}(p)}
  \right|_{\phi=0} &=& i p^\mu \gg_{\hat{V}^i_{\mu} \hat{\Phi}}(p) + 
   M_{i, \Phi'} \gg_{\hat{\Phi}' \hat{\Phi}}(p) \,\,=\,\, 0
\,.  
\end{eqnarray}

Let us now come to the three-point function
$\gg_{\hat{V}^{\mu}_1 \hat{V}^{\nu}_2 \hat{\Phi}}(p_1, p_2)$.
From Eq.~(\ref{WTI}) we get
\begin{eqnarray}
  \label{bfm_3}
  \lefteqn{\left. \frac{\delta ^3 {\cal W}_{(\lambda)}(\gg) }{\delta
        \lambda_V^i(-p_1 -p_2)
        \delta  \hat{V}^j_{\nu}(p_1) \delta \hat{\Phi}(p_2)}
    \right|_{\phi=0} =}
  \nonumber\\&&\mbox{}
  i \left( p_1 + p_2 \right)^\mu 
  \gg_{\hat{V}^i_{\mu} \hat{V}^j_{\nu} \hat{\Phi}}(p_1,p_2)
  +  M_{i, \Phi'} \gg_{\hat{\Phi}' \hat{V}^j_{\nu} \hat{\Phi}}(p_1,p_2)  + 
  f_{ijk} \gg_{\hat{V}^k_{\nu} \hat{\Phi}}(p_2) 
  + t_{i, \Phi \Phi'} \gg_{ \hat{\Phi'} \hat{V}^j_{\nu}}(p_1)
  \nonumber\\
  &=&0
  \,, 
\end{eqnarray}
where $f_{ijk}$ and $t_{i, \Phi \Phi'}$ represent the structure constants,
respectively, the generators of the gauge group in the representation for
scalar fields. We refrain from listing them explicitly. 
In the above identity the two-point functions are already
known and only the function 
$\gg_{\hat{\Phi}' \hat{V}^j_{\nu} \hat{\Phi}}(p_1,p_2)$
is new. It is fixed by the WTI
\begin{eqnarray}
  \label{bfm_4}
  \lefteqn{\left. \frac{\delta ^3 {\cal W}_{(\lambda)}(\gg) }{\delta
        \lambda_V^i(-p_1 -p_2)
        \delta  \hat{\Phi}(p_1) \delta \hat{\Phi}'(p_2)}
    \right|_{\phi=0} =}
  \nonumber\\&&\mbox{}
  i \left( p_1 + p_2 \right)^\mu 
  \gg_{\hat{V}^i_{\mu} \hat{\Phi}' \hat{\Phi}}(p_1,p_2)
  + M_{i, \Phi''} \gg_{\hat{\Phi}'' \hat{\Phi} \hat{\Phi}'}(p_1,p_2)  +
  t_{i, \Phi' \Phi''}   \gg_{\hat{\Phi}'' \hat{\Phi}}(p_1) 
  + t_{i, \Phi \Phi''}  \gg_{\hat{\Phi}' \hat{\Phi}''}(-p_2)
  \nonumber\\
  &=& 0
\,,
\end{eqnarray}
where again 
the sum over $\Phi''$ takes the values $G^0$ and $G^\pm$.
  
The four equations~(\ref{bfm_2}), (\ref{bfm_3.0}), (\ref{bfm_3})
and (\ref{bfm_4}) 
already form the complete set of identities needed for the
computation of the amplitude.
In fact, all identities are linear in
$\gg$ and therefore they keep the same form to all orders. This also
implies that the coefficients $f_{ijk}, t_{i, \Phi \Phi'}$ and $M_{i, \Phi}$
are not 
renormalized. Their renormalization is fixed from the renormalization
conditions.
Furthermore no Green function involving ghosts or anti-fields occur.
Let us mention that instead of the four identities derived above
roughly ten mostly non-linear STIs have to be analyzed
in the conventional gauge fixing. 
Thus, in this case the BFM is obviously superior as compared to the 
conventional gauge fixing.

{At this point a practical remark is in order: the obvious advantage of the 
BFM due the linearity of the WTIs is only valid at the highest order of 
the computation. In lower orders, i.e. in sub-diagrams, also quantum 
field Green functions are involved which makes it necessary to use all three
types of identities, namely (\ref{ST_bfm}), (\ref{WTI}) and (\ref{eq:new_1}), in general. 
Thus, in specific examples, the BFM 
may introduce more complications in the sub-diagrams in comparison
with a conventional gauge fixing such that  
the advantages of the BFM at the highest loop level could get 
partly compensated. However, in the example of the two-loop corrections
to \bsg, to be discussed in the next section, we will show that 
the analysis of the sub-diagrams within the BFM is still favourable
compared with the analysis within a conventional gauge.} 

{Let us finish this section with two practical remarks about the gauge fixing
and renormalization conditions within the BFM.}

In order to evaluate the S-matrix elements
in the BFM a gauge fixing has to be chosen for the background gauge fields. 
However, this choice is completely independent from the gauge fixing    
used for the internal gauge fields. This allows for a more   
convenient choice oriented on the physical process.
For instance, the BFM Green functions with external unphysical scalar   
bosons ($G^{0}$ and $G^{\pm}$), with external ghost fields  
as well as longitudinal gauge bosons can be neglected
in the decomposition of S-matrix elements in terms of 1PI parts.
This can be achieved via the use of the unitary gauge fixing for the BFM
propagators (see~\cite{msbkg} for explicit examples).
The procedure can be implemented both for the electroweak and the QCD sector
of the SM.

Besides the symmetries of the BFM one has to impose some  
renormalization conditions in order to unambiguously fix the finite parts of 
Green functions. It appears very useful to
implement them in terms of background Green functions~\cite{grassi}. 
The relation between
the renormalization conditions for the background Green functions and those
for the quantum fields are considered in~\cite{grassi}.

%%%%%%%%%%%%%%%%%%%%%%%%%%%%%%%%%%%%%%%%%%%

\subsection{Example 2: Green Functions, STIs and WTIs for the \bsg}      
\label{sub:bsg}

In this section, we  briefly describe the decomposition of the S-matrix 
elements for the process \bsg~in terms of 1PI functions
at two-loop approximation.
Simplifications concerning the renormalization procedure
are discussed in the context of the BFM. We explicitly derive 
all WTIs and STIs constraining the counterterms at the one-
and two-loop level  for this specific process . 
 
The decomposition of the (truncated) connected BFM Green
functions in terms of 1PI functions can be split into two 
contributions.
The first one, $G^{(2),vertex}_{\hat{A}_{\mu} b \bar{s}} (p_{b}, p_{s})$,
contains the flavour changing neutral current (FCNC) vertex corrections
whereas the second one, 
$G^{(2),s.e.}_{\hat{A}_{\mu} b \bar{s}} (p_{b}, p_{s})$, the 
FCNC self-energies: 
\begin{eqnarray}
  \label{loop.1}  
  G^{(2)}_{\hat{A}_{\mu} b \bar{s}} (p_{b}, p_{s})  
  &=&  G^{(2),vertex}_{\hat{A}_{\mu} b \bar{s}} (p_{b}, p_{s})  +  
  G^{(2),s.e.}_{\hat{A}_{\mu} b \bar{s}} (p_{b}, p_{s})  
  \,.
\end{eqnarray}
Remember that for the background fields we have chosen to use
the unitary gauge in order to avoid external unphysical particles.
Then the two contributions are given by:
\begin{eqnarray}
  \label{loop.2}  
  G^{(2),vertex}_{\hat{A}_{\mu} b \bar{s}} (p_{b}, p_{s}) & = & 
  \gg^{(2)}_{\hat{A}_{\mu} b \bar{s}} 
  +  
  \gg^{(1)}_{\hat{A}_{\mu} \hat{A}_{\sigma}} 
  G^{(0)}_{\hat{A}_{\sigma} \hat{A}_{\nu}}  
  \gg^{(1)}_{\hat{A}_{\nu} b \bar{s}}  
  + 
  \gg^{(1)}_{b \bar{b}}  G^{(0)}_{b \bar{b}} 
  \gg^{(1)}_{\hat{A}_{\mu} b \bar{s}}  
  +  
  \gg^{(1)}_{\hat{A}_{\mu} b \bar{s}}  
  G^{(0)}_{s \bar{s}} \gg^{(1)}_{s \bar{s}}
  \nonumber\\&&\mbox{} 
  +  
  \gg^{(1)}_{\hat{A}_{\mu} \hat{H}}  
  G^{(0)}_{\hat{H} \hat{H}} \gg^{(1)}_{\hat{H} b \bar{s}} 
  +  
  \gg^{(1)}_{\hat{A}_{\mu} \hat{Z}_{\sigma}}  
  G^{(0)}_{\hat{Z}_{\sigma} \hat{Z}_{\nu}}  
  \gg^{(1)}_{\hat{Z}_{\nu} b \bar{s}}   
  \,,
  \\
  \label{loop.3}  
  G^{(2),s.e.}_{\hat{A}_{\mu} b \bar{s}} (p_{b}, p_{s}) & = & 
  \gg^{(0)}_{\hat{A}_{\mu} b \bar{b}}  
  G^{(0)}_{b \bar{b}} \gg^{(2)}_{b \bar{s}}  
  +  
  \gg^{(2)}_{b \bar{s}}  G^{(0)}_{s \bar{s}}  
  \gg^{(0)}_{\hat{A}_{\mu} s \bar{s}} 
  +  
  \gg^{(1)}_{\hat{A}_{\mu} b \bar{b}}  
  G^{(0)}_{b \bar{b}} \gg^{(1)}_{b \bar{s}}   
  +  
  \gg^{(1)}_{b \bar{s}}  G^{(0)}_{s \bar{s}}   
  \gg^{(1)}_{\hat{A}_{\mu} s \bar{s}}   
  \nonumber\\&&\mbox{} 
  +  
  \gg^{(1)}_{\hat{A}_{\mu} \hat{H}}   
  \gg^{(1)}_{b \bar{s}}  G^{(0)}_{s \bar{s}} 
  \gg^{(0)}_{\hat{H} s \bar{s}}  
  +  
  \gg^{(1)}_{\hat{A}_{\mu} \hat{H}}  
  \gg^{(0)}_{\hat{H} b \bar{b}}  
  G^{(0)}_{b \bar{b}} \gg^{(1)}_{b \bar{s}}  
  \nonumber\\&&\mbox{} 
  +
  \gg^{(1)}_{\hat{A}_{\mu} \hat{Z}_{\nu}} G^{(0)}_{b \bar{b}}  
  \gg^{(1)}_{b \bar{s}} \gg^{(0)}_{\hat{Z}_{\nu} s \bar{s}}  
  + 
  \gg^{(1)}_{\hat{A}_{\mu} \hat{Z}_{\nu}}  
  \gg^{(0)}_{\hat{Z}_{\nu} b \bar{b}}  
  G^{(0)}_{b \bar{b}} \gg^{(1)}_{b \bar{s}}  
\,.
\end{eqnarray} 
We recall that  $G^{(0)}_{i j}$ denotes the tree-level propagators and 
$\gg$ the irreducible Green functions.
After projection on the physical states,
the contributions from the $\gamma-Z$ and $\gamma-H$ mixings
vanish because of the WTI
\begin{eqnarray} 
\left. \frac{\delta^2{\cal W}_{(\lambda)}(\gg)}{\delta \lambda_A(-p) \delta
    \hat{H}(p)} \right|_{\phi=0}  
&=& i p_\mu \gg_{\hat{A}_\mu \hat{H}}(p) \,\,=\,\,0
\,,
\end{eqnarray}
where an analogous equation holds for $\gg_{\hat{A}_\mu \hat{Z}_\nu}$. 
Thus the second line in Eq.~(\ref{loop.2}) and the second and third lines of
Eq.~(\ref{loop.3}) drop out from the amplitudes.

Let us in a first step consider the two-loop contribution to the
\bsg~amplitude and derive the corresponding WTI.
According to our rules we have to
replace the photon field, $A_{\mu}$, by the corresponding infinitesimal
parameter of\footnote{Here the index $Q$ reminds that the abelian group of QED
  is meant.}
$U_Q(1)$, $\lambda_A$. Furthermore we have to
take the derivative of Eq.~(\ref{WTI}) w.r.t. $\lambda_A$, $b$ and $s$.
Using the formulae~(\ref{BKG_tranf})--(\ref{BKG_trans_ferm}) we 
obtain
\begin{eqnarray}
  \label{bsg1}
  \lefteqn{\left. \frac{\delta^3 {\cal W}_{(\lambda)}(\gg)^{(2)}}
    {\delta \lambda_A(-p_s - p_b) \delta \bar{s}(p_s)  \delta b(p_b)}
  \right|_{\phi=0} =}
\nonumber\\&&\mbox{} 
  i (p_{s} + p_{b})^{\mu}    
  \gg^{(2)}_{\hat{A}_{\mu} \bar{s} b}(p_{s},p_{b}) + i e Q_{d} \left(    
    \gg^{(2)}_{\bar{s} b}(p_{b})   
    - \gg^{(2)}_{\bar{s} b}(-p_{s})  \right)
  \nonumber\\
  &=& 0
  \,,
\end{eqnarray}
where $Q_{d}$ is the charge of the down-type quarks
and $p_b$ and $p_s$ are the in-going momenta
of the quark lines.
This identity (for a one-loop analysis see also~\cite{ferrari,barroso})
can be used
to fix the overall counterterms defined through

\begin{equation} \label{twocnt} 
{\cal L}^{(2),CT}_{b\rightarrow s \gamma} =   
Z^{(2)}_{L,1} \hat{A}_{\lambda} \bar{s} \gamma^{\lambda} P_L b + 
Z^{(2)}_{L,2} \bar{s} \not\!\partial P_L b  +
Z^{(2)}_{R,1} \hat{A}_{\lambda} \bar{s} \gamma^{\lambda} P_R b + 
Z^{(2)}_2      \bar{s} \not\!\partial P_R b 
+  {\rm h.c.} 
\,,
\end{equation} 
where the $Z$ factors, in general, contain finite and
divergent contributions. Clearly the same Lagrangian also holds at one-loop
order.
$P_{L/R}=(1\mp\gamma_5)/2$ are the projectors on the left- and
right-handed components.
Note that for an invariant
regularization scheme no counterterm at all is needed for the Green function 
$\gg^{(2)}_{\hat{A}_{\mu} b \bar{s}}(p_{b},p_{s})$. 
However,  if the regularization scheme breaks the identity~(\ref{bsg1}), it
can be restored with the help of non-invariant counterterms
in~(\ref{twocnt}).

 \begin{figure}[ht]
   \begin{center}
     \begin{tabular}{cc}
       \epsfxsize=5.cm
       \leavevmode
       \epsffile[180 280 440 420]{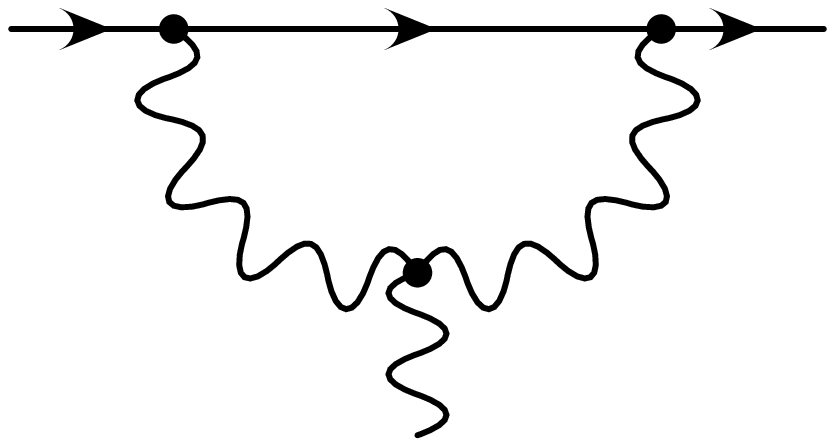}
       &
       \epsfxsize=5.cm
       \leavevmode
       \epsffile[180 280 440 420]{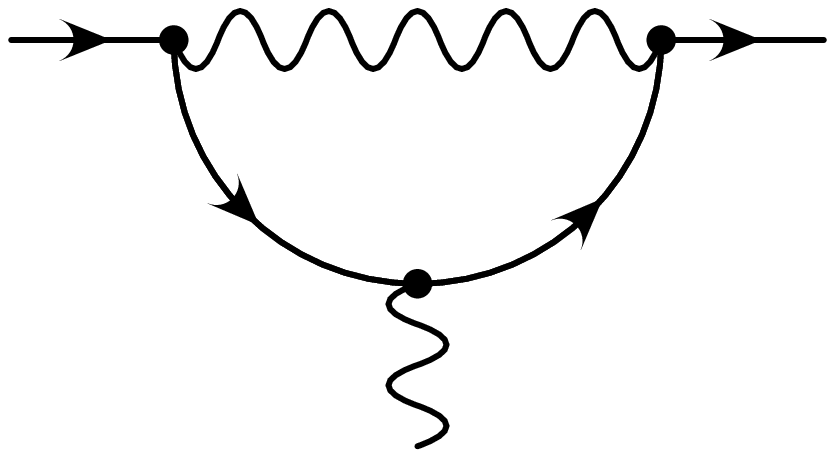}
     \end{tabular}
     \caption[]{\label{fig:bsg} \sf One-loop diagrams contributing to $b\to
       s\gamma$. In order to obtain the two-loop diagrams one-loop corrections
       to each vertex and internal propagator have to be considered.
       The two external fermion lines represent the $b$, respectively, the $s$
       quark and the external wiggled line the background photon,
       $\hat{A}_\mu$.}
    \end{center}
   \end{figure}

Let us now have a closer look to the {\it sub-divergences}. From the  
topological structure of the two-loop diagrams
with only external background gauge fields,
which are contained in $\gg^{(2)}_{\hat{A}_{\mu} b \bar{s}}$
(cf. Fig.~\ref{fig:bsg}),
it is evident that the 1PI three- and four-point Green functions
with external quantum, respectively, ghost
fields do not appear at one-loop order:
\begin{equation}\label{ta:vert}   
  \gg^{(1)}_{V^{a}_{\mu} V^{b}_{\nu} V^{c}_{\rho}}\,,\qquad   
  \gg^{(1)}_{V^{a}_{\mu} V^{b}_{\nu} V^{c}_{\rho} V^{d}_{\sigma}}\,,\qquad
  \gg^{(1)}_{V^{a}_{\mu} \bar{c}^{b} c^{c}}
\,.
\end{equation}
Here $V^{a}_\mu$ are the vector quantum fields and $c^{a}$ are ghost fields.  
Also the Green functions where the vector fields are replaced by scalar fields
do not contribute.
Actually the renormalization of sub-divergences
with more then two quantum 
fields enter the calculation only at three-loop order. 
Thus, we only have to consider Green functions
which are either  sub-diagrams of $\gg^{(2)}_{\hat{A}_{\mu} b \bar{s}}$
or diagrams occurring in~(\ref{loop.2}) and ~(\ref{loop.3}) respectively.
Concerning the three-point diagrams the following functions have to be
taken into account:
\begin{eqnarray}
  \label{gr:bsga}
  \begin{array}{llll}   
    \gg^{(1)}_{\hat{A}_{\mu} \bar{s} b}(p_{s},p_{b})\,,
    &
    \gg^{(1)}_{\hat{A}_{\mu} \bar{q}_{2} q_1}(p_{2},p_{1})\,,
    &
    \gg^{(1)}_{\hat{A}_{\mu} W^{+}_{\nu} W^{-}_{\rho}} (p_{+},p_{-})\,,
    & 
    \gg^{(1)}_{\hat{A}_{\mu} G^{+} G^{-}} (p_{+},p_{-})\,,
    \\
    \\   
    \gg^{(1)}_{\hat{A}_{\mu} G^{+} W^{-}_{\nu}} (p_{+},p_{-})\,,
    &
    \gg^{(1)}_{\hat{A}_{\mu} W^{+}_{\nu} G^{-}} (p_{+},p_{-})\,,
    & 
    \gg^{(1)}_{W^{+}_{\mu} \bar{q}_{2} q_1}(p_{2},p_{1})\,,
    & 
    \gg^{(1)}_{G^{+} \bar{q}_{2} q_1}(p_{2},p_{1})\,.
  \end{array} 
\end{eqnarray}   
The two-point functions
\begin{eqnarray}\label{gr:bsgb}
  \begin{array}{lll}   
    \gg^{(1)}_{\bar{s} b}(p_{s},p_{b})\,,
    &
    \gg^{(1)}_{\bar{q}_{2} q_1}(p_{2},p_{1})\,,
    &
    \gg^{(1)}_{W^{+}_{\nu} W^{-}_{\rho}} (p_{+},p_{-})\,,
    \\
    \\
    \gg^{(1)}_{G^{+} G^{-}} (p_{+},p_{-})\,,
    &
    \gg^{(1)}_{G^{+} W^{-}_{\mu}} (p_{+},p_{-})\,,
    &
    \gg^{(1)}_{W^{+}_{\mu} G^{-}} (p_{+},p_{-})\,,
  \end{array}
\end{eqnarray}   
appear as self-energies in  the two-loop graphs.
In Eqs.~(\ref{gr:bsga}) and ~(\ref{gr:bsgb}) 
$q_1$ and $q_2$ are two generic quark fields.
Notice that the Green functions 
$\gg_{\hat{A}_{\mu} G^{+} W^{-}_{\nu}}$ and
$\gg_{\hat{A}_{\mu} W^{+}_{\nu} G^{-}}$ are absent at
tree level. This is
a consequence of the choice for the gauge fixing and the 
background gauge invariance.
At one-loop level, however, contributions may appear
as soon as a non-invariant regularization scheme is used.

By inspection of Eqs.~(\ref{gr:bsga}) and~(\ref{gr:bsgb}), 
we see that the same Green functions with the quantum fields 
replaced by the corresponding 
background fields never occur. % nor vice versa. 
This implies that  the STI identities 
(\ref{eq:new_1}) relating background and quantum fields are irrelevant for the 
practical analysis of this specific process and we can restrict ourselves
to the conventional STI and the WTI. In Section~\ref{sub:consi} we present a 
general discussion on this point.

Thus, let us discuss the STIs and WTIs which constrain
the relevant sub-diagrams.
Using our rules for the BFM  we are able to derive the complete set of
identities. For the three-point functions involving background photon field 
WTIs are used whereas only a reduced set of STIs for three-point functions 
and two-point functions are indeed necessary. 

We start with the WTIs. From~(\ref{WTI}) one gets:
\begin{eqnarray}\label{wti.1} 
\lefteqn{\left. \frac{\delta^3 {\cal W}_{(\lambda)}(\gg)^{(1)}}
{\delta \lambda_A(-p_1 - p_2) \delta \bar{q}_2(p_2)  \delta q_1(p_1)}
\right|_{\phi=0} = }
 \nonumber \\&&\mbox{}
 i \left(p_{2} + p_{{1}}\right)^\mu \gg^{(1)}_{\hat{A}_{\mu}  
\bar{q}_2 q_1}(p_{2},p_{{1}})  
+ i e Q_{q} \left(  \gg^{(1)}_{\bar{q}_2 q_1}(p_{1})  -  
\gg^{(1)}_{\bar{q}_2 q_1}(-p_2) \right)
\nonumber\\&=&0
\,,
\end{eqnarray} 
which is the analogue equation to~(\ref{bsg1}). Here, however,
$q_1$ and $q_2$ refer to any type of quark fields.
As already mentioned above, an advantage of 
the BFM is the linearity of the identities w.r.t. $\gg$.
This allows to disentangle easily the fermionic corrections from
the bosonic ones. In fact, in the same way as for the $H\rightarrow
\gamma\gamma$, we can select the independent contributions and
introduce the corresponding counterterms. Notice furthermore that 
no ghosts are involved. In the case of conventional gauge fixing
the identity~(\ref{wti.1}) would be replaced by a STI obtained by
differentiating w.r.t. $c_A, q_1$ and $\bar{q}_2$. Already at one-loop order
this identity would require the computation of Green functions involving
ghost particles and off-shell quark fields. 

The other Green functions of Eq.~(\ref{gr:bsga}) involving background fields
are constrained by the following WTI
\begin{eqnarray}\label{wti.2} 
\lefteqn{\left. \frac{\delta^3 {\cal W}_{(\lambda)}(\gg)^{(1)}}
{\delta \lambda_A(-p_+ - p_-) \delta W^+_\rho(p_+)  \delta W^-_\sigma(p_-)}
\right|_{\phi=0} =}
\nonumber\\&&\mbox{}
  i \left(p_{+} + p_{-}\right)^\mu \gg^{(1)}_{\hat{A}_{\mu} W^{+}_{\rho} 
  W^{-}_{\sigma}} (p_{+},p_{-}) - i e \left(  \gg^{(1)}_{W^{+}_{\rho} 
  W^{-}_{\sigma}} (p_{-}) - \gg^{(1)}_{W^{+}_{\rho} W^{-}_{\sigma}} (-p_{+}) 
\right)
\nonumber\\&=&0
\,,
\label{wti.3}\\
\lefteqn{\left. \frac{\delta^3 {\cal W}_{(\lambda)}(\gg)^{(1)}}
{\delta \lambda_A(-p_+ - p_-) \delta G^+(p_+)  \delta G^-(p_-)}
\right|_{\phi=0} =}
 \nonumber\\&&\mbox{}
i \left( p_{+} + p_{-} \right)^\mu \gg^{(1)}_{\hat{A}_{\mu} G^{+} 
  G^{-}} (p_{+},p_{-}) - i e \left( \gg^{(1)}_{G^{+} G^{-}} 
(p_{-}) - \gg^{(1)}_{G^{+} G^{-}}(-p_{+}) \right)
\nonumber\\&=&0
\,,
\\
\label{wti.4}
\lefteqn{\left. \frac{\delta^3 {\cal W}_{(\lambda)}(\gg)^{(1)}}
{\delta \lambda_A(-p_+ - p_-) \delta W^+_\rho(p_+)  \delta G^-(p_-)}
\right|_{\phi=0} =}
\nonumber \\&&\mbox{}
 i \left( p_{+} + p_{-} \right)^\mu \gg^{(1)}_{\hat{A}_{\mu} 
  W^{+}_{\rho} G^{-} } (p_{+},p_{-})  - i e \left( \gg^{(1)}_{
  W^{+}_{\rho} G^{-}} (p_{-}) - \gg^{(1)}_{W^{+}_{\rho} G^{-}} (-p_{+}) 
\right) 
\nonumber\\&=&0
\,,
\end{eqnarray} 
and their hermitian counterparts. 

According to our third rule of Section~\ref{sub:conventional}
for the Green functions 
$\gg_{\hat{A}_\mu W^+_\nu W^-_\rho}(p_+, p_-)$,
$\gg_{\hat{A}_\mu W_\nu^+ G^-}(p_+, p_-)$,
$\gg_{\hat{A}_\mu G^+ W_\rho^-}(p_+, p_-)$
and
$\gg_{\hat{A}_\nu G^+ G^-}(p_+, p_-)$ of the above equations 
one has to derive new STIs. For instance, one gets 
\begin{eqnarray}
  \label{new_eq}
\lefteqn{\frac{\delta^{2} {\cal S}(\gg)}{\delta \hat{A}_\mu(-p_+-p_-) \delta
   c^{+}(p_+)  \delta W^{-}_{\rho}(p_-)}\bigg|_{\phi=0} = }
\nonumber \\&&\mbox{}
  \gg_{\hat{A}_\mu c^{+} W^{*,-}_{\nu}}(p_+,p_-)  \gg_{W^{+}_{\nu}
    W^{-}_{\rho}}(p_-) +   \gg_{\hat{A}_\mu c^{+} G^{*,-}}(p_+,p_-)  
  \gg_{G^{+} W^{-}_{\rho}}(p_-)   
\nonumber \\&&\mbox{}
+ \gg_{c^{+} W^{*,-}_{\nu}}(-p_+)  \gg_{\hat{A}_\mu W^{+}_{\nu}
    W^{-}_{\rho}}(p_+,p_-) +   \gg_{c^{+} G^{*,-}}(-p_+)  
  \gg_{\hat{A}_\mu G^{+}
    W^{-}_{\rho}}(p_+,p_-) 
\nonumber\\&=&0
\,,
\end{eqnarray}
where the new Green functions 
$\gg_{\hat{A}_\mu c^{+} W^{*,-}_{\nu}}(p_+,p_-)$ and 
$\gg_{\hat{A}_\mu c^{+} G^{*,-}}(p_+,p_-)$ 
emerge. Clearly, also the STIs~(\ref{new_eq}) can be spoiled by the 
radiative corrections 
and therefore it must be restored by suitable counterterms. However,
it  turns 
out that due 
to the  consistency conditions (see Section~\ref{sub:consi})
such STIs deliver no independent constraints. They are 
automatically preserved if the STIs for the two-point functions (discussed
below (\ref{exap})),  
the WTIs (\ref{wti.2})--(\ref{wti.4}),
and  the WTIs involving the new Green functions 
$\gg_{\hat{A}_\mu c^{+} W^{*,-}_{\nu}}(p_+,p_-)$ and
$\gg_{\hat{A}_\mu c^{+} G^{*,-}}(p_+,p_-)$ 
have been restored. 
These new WTIs are obtained by differentiating the WTI~(\ref{WTI}) w.r.t. 
$\lambda_A \,, c^{+}(p_+) \,$ and $W^{*,-}_{\nu}(p_-)$ and 
w.r.t. $\lambda_A \,, c^{+}(p_+)$ and $G^{*,-}(p_-)$:
\begin{eqnarray}
\label{new_eq.2} 
\lefteqn{\left. \frac{\delta^3 {\cal W}_{(\lambda)}(\gg)^{(1)}}
 {\delta \lambda_A(-p_+ - p_-) \delta c^+(p_+)  \delta W^{*,-}_\nu(p_-)}
\right|_{\phi=0} =}
\nonumber\\&&\mbox{}
  i \left(p_{+} + p_{-}\right)^\mu \gg^{(1)}_{\hat{A}_{\mu} c^{+} 
  W^{*,-}_{\nu}} (p_{+},p_{-}) - i e \left(  \gg^{(1)}_{c^{+} 
  W^{*,-}_{\nu}} (p_{-}) - \gg^{(1)}_{c^{+} W^{*,-}_{\nu}} (-p_{+}) 
\right)
\nonumber\\&=&0
\,,\\
\lefteqn{\left. \frac{\delta^3 {\cal W}_{(\lambda)}(\gg)^{(1)}}
{\delta \lambda_A(-p_+ - p_-) \delta c^+(p_+)  \delta G^{*,-} (p_-)}
\right|_{\phi=0} =}
\nonumber\\&&\mbox{}
  i \left(p_{+} + p_{-}\right)^\mu \gg^{(1)}_{\hat{A}_{\mu} c^{+} 
  G^{*,-}} (p_{+},p_{-}) - i e \left(  \gg^{(1)}_{c^{+} 
  G^{*,-}} (p_{-}) - \gg^{(1)}_{c^{+} G^{*,-}} (-p_{+}) 
\right)
\nonumber\\&=&0
\,.
\end{eqnarray} 
These WTIs do not involve any new Green functions since the two-point 
functions $ \gg^{(1)}_{c^{+} W^{*,-}_{\nu}}$ and
$\gg^{(1)}_{c^{+} G^{*,-}}$ 
are already fixed by means of the Faddeev-Popov equations~(\ref{e_2}) 
as discussed in Appendix~\ref{app:stiex}.
Notice that only the first equation of~(\ref{new_eq.2}) can be broken 
by the radiative corrections since the second one involves 
(by Lorentz invariance) only  finite quantities. 

To restore the identities of the sub-diagrams, we need a complete
set of counterterms. It is convenient to divide them
into three different sets 
\begin{equation}\label{org_1}
{\cal L}^{(1)}_{b\rightarrow s \gamma} = 
{\cal L}^{(1),WTI}_{b\rightarrow s \gamma} + 
{\cal L}^{(1),STI}_{b\rightarrow s \gamma} + 
{\cal L}^{(1),INV}_{b\rightarrow s \gamma} 
\,,
\end{equation}
organized in such a way that 
\begin{eqnarray}\label{org_cont}
  {\cal W}_{(\lambda)}\left(\int {\rm d}^4 x 
    {\cal L}^{(1),STI}_{b\rightarrow s \gamma}\right)
  \,\, = \,\,
  {\cal W}_{(\lambda)}\left(\int {\rm d}^4 x 
    {\cal L}^{(1),INV}_{b\rightarrow s \gamma}\right)
  &=& 0\,, 
  \nonumber \\
  {\cal S}_0 \left(\int {\rm d}^4 x 
    {\cal L}^{(1),INV}_{b\rightarrow s \gamma} \right) &=&0
  \,. 
\end{eqnarray}
This triangular organization of the counterterms
ensures that it is possible to restore 
the WTIs by only fixing the coefficients of
${\cal L}^{(1),WTI}_{b\rightarrow s \gamma}$. The counterterms 
${\cal L}^{(1),STI}_{b\rightarrow s \gamma}$ and 
${\cal L}^{(1),INV}_{b\rightarrow s \gamma}$ are invariant under the
background gauge transformations. 
${\cal L}^{(1),STI}_{b\rightarrow s \gamma}$ is
necessary to restore the STIs and with the help of
${\cal L}^{(1),INV}_{b\rightarrow s \gamma}$ the renormalization
conditions can be fulfilled.
In Section~\ref{sub:consi} and in Appendix~\ref{app:triangula} 
we will prove with the help of the consistency conditions that this procedure
is always possible provided no anomalies occur. 

The complete list of counterterms for three-point functions 
needed to restore the WTIs is given by 
\begin{eqnarray}\label{ta:cteq}   
  {\cal L}^{(1),WTI}_{b\rightarrow s \gamma} &=&   
  Z^{(1)}_{3} \hat{A}_{\lambda} \nabh^{\mu} W^{+,\lambda} W^{-}_{\mu} +   
  Z^{(1)}_{4} \hat{A}_{\lambda} \nabh^{\mu} W^{+}_{\mu} W^{-,\lambda} +   
  Z^{(1)}_{5} \hat{A}_{\lambda} W^{+,\lambda} \nabh^{\mu} W^{-}_{\mu}
  \nonumber \\   &&\mbox{} +
  Z^{(1)}_{6} \left(   
    \hat{A}_{\lambda} \nabh^{\lambda} W^{+}_{\mu} W^{-,\mu} +   
    \hat{A}_{\lambda} W^{+,\mu} \nabh^{\lambda} W^{-}_{\mu} \right) +   
  Z^{(1)}_{7} \hat{A}_{\lambda} W^{+}_{\mu} \nabh^{\mu} W^{-}_{\lambda}
  \nonumber \\   &&\mbox{} +
  Z^{(1)}_{8} \left( \epsilon_{\mu\nu\lambda\sigma}   
    \hat{A}^{\lambda} W^{+,\mu} \nabh^{\sigma} W^{-,\nu} 
    - \epsilon_{\mu\nu\lambda\sigma}   
    \hat{A}^{\lambda} W^{-,\mu} \nabh^{\sigma} W^{+,\nu} \right)
  \nonumber \\   &&\mbox{} +
  Z^{(1)}_{9} \hat{A}_{\lambda} G^{+} \nabh^{\lambda} G^{-} +   
  Z^{(1)}_{10} \hat{A}_{\lambda} G^{-} \nabh^{\lambda} G^{+} +   
  Z^{(1)}_{11} \hat{A}_{\lambda} G^{+} W^{-}_{\lambda}
  \nonumber \\   &&\mbox{} +
  Z^{(1)}_{12} \hat{A}_{\lambda} G^{-} W^{+}_{\lambda} +  
  \left( Z^{(1)}_{13} \hat{A}_{\lambda} \bar{s} P_L \gamma^{\lambda} b +
  Z^{(1)}_{14} \hat{A}_{\lambda} \bar{s} P_R \gamma^{\lambda} b + \right.
  \nonumber \\   &&\mbox{} +
  \left. Z^{(1)}_{15} \hat{A}_{\lambda} \bar{q}_{2}  P_L \gamma^{\lambda}
  q_{1} + 
  Z^{(1)}_{16} \hat{A}_{\lambda} \bar{q}_{2}  P_R \gamma^{\lambda} q_{1} +
  {\rm h.c.} \right)     
  \nonumber \\   &&\mbox{} +
  Z^{(1)}_{17} \hat{A}_{\lambda} c^{+} W^{*,-}_{\lambda} +   
  Z^{(1)}_{18} \hat{A}_{\lambda} c^{-} W^{*,+}_{\lambda} 
  \,,
\end{eqnarray}
where $\nabh$ is the covariant derivative w.r.t. $U_Q(1)$.
These counterterms correspond to the various Green functions occurring in 
the WTIs~(\ref{wti.1})--(\ref{wti.4}) and in the WTI~(\ref{new_eq.2}).
In Eq.~(\ref{ta:cteq}) they are partially written in
covariant form w.r.t. $U_Q(1)$ and also include counterterms to the WTIs
which, however, are not relevant for our specific
process under consideration. They would be needed, e.g., 
for the calculation of four-point functions like $\gg_{A_\mu A_\nu  W^+_\rho
  W^-_\sigma}$. 
We note that this is the preferable basis
for counterterms in order to analyze the
renormalization of the whole model~\cite{grassi}.

Besides the WTIs, we also have to take into account the following  
STIs. The general form for the two-point functions  
is also discussed in Appendix~\ref{app:stiex}.
  In Eqs.~(\ref{e_1}) and~(\ref{e_2}) they are
  given for a generic gauge field, $V_\mu$, and a generic scalar field
  $\Phi$. Note, that the Faddeev-Popov equation
  has to be used in order to obtain
  this simple form
\begin{eqnarray}
  \label{exap}
  \frac{\delta^{2} {\cal S}(\gg)}{\delta c^{+}(-p) \delta
    W^{-}_{\mu}(p)}\bigg|_{\phi=0} &=&  
  \gg_{c^{+} W^{*,-}_{\nu}}(p)  \gg_{W^{+}_{\nu}
    W^{-}_{\mu}}(p) +   \gg_{c^{+} G^{*,-}}(p)  \gg_{G^{+}
    W^{-}_{\mu}}(p) \,\,=\,\,0 
\,,
  \nonumber \\
  \frac{\delta^{2}  {\cal S}(\gg)}{\delta c^{+}(-p) \delta G^{-}(p)
    }\bigg|_{\phi=0}
  &=&  
  \gg_{c^{+} W^{*,-}_{\nu}}(p)  \gg_{W^{+}_{\nu}
    G^{-}}(p) +   \gg_{c^{+} G^{*,-}}(p)  \gg_{G^{+}
    G^{-}}(p) \,\,=\,\, 0
\,,
\end{eqnarray}
which at one-loop order become 
\begin{eqnarray}\label{exap_1}
\gg^{(1)}_{c^{+} W^{*,-}_{\nu}}(p) ( i \, M_W^2 \, g_{\nu\mu} ) + 
i \, p_\nu \gg^{(1)}_{W^{+}_{\nu} W^{-}_{\mu}}(p) +   
\gg^{(1)}_{c^{+} G^{*,-}}(p) (-i \, p_\mu \, M_W) +
i\, M_W \, \gg^{(1)}_{G^{+}  W^{-}_{\mu}}(p)
&=&0 
\,,
\nonumber \\
\gg^{(1)}_{c^{+} W^{*,-}_{\nu}}(p)   (-i \, p_\nu M_W)   +  
i \, p_\nu \,  \gg^{(1)}_{W^{+}_{\nu}G^{-}}(p) +    
\gg^{(1)}_{c^{+} G^{*,-}}(p) ( i \, p^2 ) + 
i\, M_W \, \gg^{(1)}_{G^{+} G^{-}}(p)
&=&0
\,.
\nonumber\\
\end{eqnarray}
Notice that the ghost fields do not couple directly to fermions. Hence,
at one-loop level these identities can be separated into two sets.
In fact,  by decomposing $\gg^{(1)} = \gg^{(1), ferm} + \gg^{(1), rest}$,
where $ \gg^{(1), ferm}$ contains only diagrams with virtual fermions and 
$\gg^{(1), rest}$ the remaining ones, the fermionic 
contributions satisfy the following simplified identities
\begin{eqnarray}\label{exap_2}
  i \, p_\nu \gg^{(1), ferm}_{W^{+}_{\nu} W^{-}_{\mu}}(p)    
  + i \,  M_W \, \gg^{(1), ferm}_{G^{+}  W^{-}_{\mu}}(p) &=&0 
  \,,
  \nonumber \\
  i \, p_\nu  \gg^{(1), ferm}_{W^{+}_{\nu}G^{-}}(p)     
  + i \, M_W \, \gg^{(1), ferm}_{G^{+} G^{-}}(p) &=& 0
  \,.
\end{eqnarray}
These identities have to be considered if the computation is done in the
't~Hooft-Veltman scheme.

In general, the two STIs in~(\ref{exap})
can be restored by counterterms which are invariant under the background gauge
transformations
(That this is possible is shown in Appendix~\ref{app:triangula}.).
The corresponding Lagrange density reads:
\begin{eqnarray}\label{ta:ctBRS} 
{\cal L}^{(1),STI}_{b\rightarrow s \gamma}&=& 
Z^{(1)}_{19} \nabh^{\mu} W^{+}_{\mu}  \nabh^{\nu} W^{-}_{\nu} +     
Z^{(1)}_{20} \nabh^{\mu} G^{+}  W^{-}_{\mu}
 \nonumber \\   &&\mbox{} +
Z^{(1)}_{21} \nabh^{\mu} G^{-}  W^{+}_{\mu} +    
Z^{(1)}_{22} \nabh^{\mu} G^{+}  \nabh_{\mu} G^{-} +   
Z^{(1)}_{23} G^{+} G^{-}
\nonumber\\     &&\mbox{} + 
Z^{(1)}_{24} \bar{b} P_L \not\! W^{+}_{\mu} q +
Z^{(1)}_{25} \bar{q}' P_L \not\! W^{+}_{\mu} s + {\rm h.c.}
\,.
\end{eqnarray}   
Note that the last two terms are background gauge invariant
because the $W$ boson transforms as an isovector of $SU(2)$. 

Also the following STI for the 
three-point functions have to be considered
\begin{eqnarray}
  \lefteqn{\frac{\delta^{3} {\cal S}(\gg)}{\delta c^{+}(-p_q-p_b) \delta
      \bar{q}(p_q) \delta b(p_b)}\Bigg|_{\phi=0} =}
  \nonumber \\&&\mbox{}
  \gg_{c^{+} W^{*,-}_{\nu}}(p_q+p_b)  \gg_{W^{+}_{\nu} \bar{q} b}(p_q,p_b) +   
  \gg_{c^{+} G^{*,-}}(p_q+p_b)  \gg_{G^{+} \bar{q} b}(p_q,p_b)
  \nonumber \\&&\mbox{}
  -\gg_{\bar{q} q'}(-p_q) \gg_{c^+ \bar{q}^{'*} b}( p_q,p_b) -
  \gg_{c^+ \bar{q} {q}^{'*}}( p_q, p_b) \gg_{\bar{q}' b}(p_b) 
  \nonumber\\
  &=&0 
  \,,
  \label{exap_4}
\end{eqnarray}
where the sum over the quark fields $q^\prime$ is understood.
Here $q$ and $q'$ are generic quark fields and $q^*$ and $q^{\prime*}$ are the
corresponding BRST sources.
An analogous equation where $c^+,q$ and $b$ is replaced by
$c^-,s$ and $q$ has to be taken into account.
We only need the one-loop expansion of these identities which reads
for~(\ref{exap_4})
\begin{eqnarray}
  \label{exap_5}
  0&=&
  \gg^{(1)}_{c^{+} W^{*,-}_{\nu}}(p_q + p_b)
  \left( i V_{q b} \gamma_\nu P_L \right)
  + i \left(p_q+p_b\right)_\nu \gg^{(1)}_{W^{+}_{\nu} \bar{q} b}(p_q,p_b)    
  \nonumber \\ &&\mbox{}
  +\gg^{(1)}_{c^{+} G^{*,-}}(p_q+p_b)
  \left(  i {V_{q b} \over M_W}
  \left[P_L m_q - P_R   m_b \right]\right) + 
  i\, M_W  \gg^{(1)}_{G^{+} \bar{q} b}(p_q,p_b) 
  \nonumber\\&&\mbox{}
  - i(- \not\!p_q - m_q) \gg^{(1)}_{c^+ \bar{q}^{*} b}( p_q,p_b) -
  \gg^{(1)}_{\bar{q} q'}(-p_q) \left(-i V_{q' b} P_L\right)
  \nonumber \\&&\mbox{} 
  - \left(i V_{q q'} P_R\right) \gg^{(1)}_{\bar{q}' b}(p_b) -
  \gg^{(1)}_{c^+ \bar{q} {b}^{*}}( p_q, p_b) i(\not\!p_b - m_b)
  \,.
\end{eqnarray}
Here $V_{q q'}$ are the CKM matrix elements. 
For convenience the prefactor $e/(s_W\sqrt{2})$ has been omitted.

As stated above, the study of the one-loop approximation
disentangles the different contributions coming from
fermionic and bosonic radiative corrections.
Unfortunately in the case of Eq.~(\ref{exap_5}) it is very hard to
disentangle the  
fermionic contributions because of the presence of external fermionic
fields. In addition we also have to note that the Green functions with
external ghost and anti-fields --- in contrast to the case of the
STIs for the $W$ boson two-point functions --- contain fermion loops 
already at one-loop order.
This is because there are vertices involving the anti-fields
$\bar{q}^{\prime*}$ and ${q}^{\prime*}$, ghosts and fermions
(see~\cite{STII,hollik_2,krau_ew}
and Appendix~\ref{app:sourceterms} for the Feynman rules). 

If we consider on-shell $b$ quarks
some Green functions vanish and the identity~(\ref{exap_4}) is simplified.
This can be heavily exploited 
in the algebraic one-loop analysis of \bsg~\cite{ferrari}. 
However, we have to remember that since these one-loop corrections appear as
sub-divergences for two-loop amplitudes these simplifications do not apply
here.

In Eq.~(\ref{exap_5}) Green functions with 
fermionic BRST sources are involved, like, e.g.,
$\gg^{(1)}_{c^+ \bar{q}^{'*} b}(p_q,p_b)$ and their hermitian
counterparts. Thus,  
according to rule~\ref{rule4} (cf. Section~\ref{sub:conventional})
one has to consider STIs for them. However,
in the case of the BFM, according to rule~$4^\prime$, 
we are left with the following linear WTIs:  
\begin{eqnarray}\label{WTI_anti}
\lefteqn{\left. \frac{\delta^4 {\cal W}_{(\lambda)}(\gg)}
{\delta \lambda_A(0) \delta c^+(-p_q -p_b)  
\delta \bar{q}^*_i(p_q)  \delta b(p_b)}
\right|_{\phi=0} =}
\nonumber \\&&\mbox{}
- i e \gg_{ c^+ \bar{q}^*_i b}(p_q,p_b)  + 
i e Q_q \gg_{ \bar{q}^*_i c^+ b}(-p_q-p_b,p_b) 
- i e Q_b \gg_{b c^+ \bar{q}^*_i}(-p_q-p_b,p_q)
\nonumber\\
&=&0
\,,
\nonumber \\
\lefteqn{\left. \frac{\delta^4 {\cal W}_{(\lambda)}(\gg)}
{\delta \lambda_A(0) \delta c^+(-p_q -p_b)  
\delta \bar{q}_i(p_q)  \delta b^*(p_b)}
\right|_{\phi=0} =} 
\nonumber\\&&\mbox{}
- i e \gg_{ c^+ \bar{q}_i b^*}(p_q,p_b)  + 
i e Q_q \gg_{ \bar{q}_i c^+ b^*}(-p_q-p_b,p_b)  
- i e Q_b \gg_{b^* c^+ \bar{q}_i  }(-p_q-p_b,p_q)
\nonumber\\
&=& 0
\,.
\end{eqnarray}
The corresponding identity where, e.g., $\lambda_A(0)$ is replaced by
$\lambda_Z(0)$ reads
\begin{eqnarray}\label{WTI_anti.1}
\lefteqn{\left. \frac{\delta^4 {\cal W}_{(\lambda)}(\gg)}
{\delta \lambda_Z(0) \delta c^+(-p_q -p_b)  
\delta \bar{q}^*_i(p_q)  \delta b(p_b)}
\right|_{\phi=0} =}
\nonumber\\&&\mbox{}
- M_Z \gg^{(1)}_{G_0 c^+ \bar{q}^*_i  b}(-p_q-p_b, p_q, p_b)
\nonumber\\&&\mbox{}
+ i e {c_W \over s_W } \gg_{ c^+ \bar{q}^*_i b}(p_q,p_b)  + 
i e Q^Z_q \gg_{ \bar{q}^*_i c^+ b}(-p_q-p_b,p_b) 
- i e Q^Z_b \gg_{b c^+ \bar{q}^*_i  }(-p_q-p_b,p_q) 
\nonumber\\
&=&0 
\,.
\end{eqnarray}
Here $Q^Z_{u/d} = ( Q_{u/d} s_W/c_W \mp 1/(2 s_W c_W) )$.
There are two other pairs
of equations 
where $\lambda_A(0)$ and $c^+(-p_q -p_b)$ are replaced by 
$\lambda_+(0)$ and $c_A(-p_q -p_b)$ or $\lambda_+(0)$ and $c_Z(-p_q
-p_b)$, respectively.
  
Finally we have to fix the invariant counterterms~\cite{STII,hollik_2,krau_ew}
${\cal L}^{(1),INV}_{b\rightarrow s \gamma}$
in order to fulfill the renormalization conditions. 
It contains $Z_W$, the wave function renormalization for the $W$
boson, its mass $M_W$, the wave function renormalization 
for the Goldstone boson, $Z_G$, and the corresponding mass
$M_G$ (which coincides with the product of the 
gauge parameter $\xi_W$ and $M_W^2$).
Furthermore
we have to fix the renormalization conditions for the fermions, namely 
the masses $m_{q_i}$ inside of loops\footnote{As already mentioned in
  Section~\ref{sub:BFM} rule~$4^\prime$, the 
  fermion wave function renormalization is fixed by WTI~(\ref{WTI_anti}). For
  further discussion see~\cite{grassi}.}
and $m_s$ and $m_b$ which are needed for the computations of the on-shell
amplitudes. Also the CKM elements and 
the couplings $\alpha_{QED}$ and $G_F$ have to be fixed.

For the CKM matrix there are 
two possible choices which can be adopted~\cite{gg}:
$(i)$ the use of the $\overline{\rm MS}$ scheme where only the poles are
subtracted~\cite{qmr,strumia} or $(ii)$
the definition given in~\cite{gg} which 
relies on subtractions at zero momentum.
For the general analysis of renormalization conditions in the background field
gauge we refer the reader to \cite{grassi}.
Our specific choices in the case of 
$b \rightarrow s \gamma$ will be discussed in Section~\ref{sec:bsg_coun}.
There we will see
that some of the identities are
automatically preserved by a conscious choice of
regularization. This will provide great simplifications.

%%%%%%%%%%%%%%%%%%%%%%%%%%%%%%%%%%%%%%%%%%%%%%%%%%%%%%%%%%%%

\setcounter{equation}{0}
\section{Renormalization of the identities}
\label{sec:renorma}

In Section~\ref{sub:QAP} we give a brief review to the Quantum Action
Principle (QAP) from a practical point of view 
and introduce some necessary notation. 
We present the principle algebraic procedure necessary 
to remove the breaking terms.
In Section~\ref{sub:consi} we briefly discuss the important practical
consequences of the consistency conditions.
Finally in the last subsection, we propose
our strategy which provides the possibility to remove the breaking terms
in an efficient way.

%%%%%%%%%%%%%%%%%%%%%%%%%%%%%%%%%%%%%%%%%%%%%%%%%%%%%%%%%%%%

\subsection{The Quantum Action Principle and the algebraic method} 
\label{sub:QAP}
The QAP  is the fundamental theorem of renormalization theory. It
guarantees the 
locality 
of the counterterms, and as a consequence, the polynomial character of the 
renormalization procedure. The QAP also implies that
all breaking terms of the STIs and WTIs are local and that they can be fully
characterized in terms of classical fields\footnote{Within 
this subsection, we will skip the prime on the effective action and on the 
functional identities since the following results are clearly not restricted 
to the BFM}.

Formally, the QAP states that within a specific 
renormalization framework derivatives of a 1PI generating functional $\gg$  
w.r.t. a parameter\footnote{ Here we mean all the
  parameters 
  of the renormalized theory: masses, couplings, vacuum expectation values,
  gauge fixing parameters, renormalization scales, infra-red (IR) regulators,
  \ldots 
  (see~\cite{zimm,QAP}).}
of the theory~\cite{QAP},
$\mu$, or w.r.t. a field~\cite{maiso} are local insertions $\Delta$ in the 
1PI Green functions
\begin{equation}
  \label{qap_1} 
  \frac{\partial \gg}{\partial \mu} = \Delta_\mu \cdot \gg 
  \,,
  \hspace{1cm} 
  \frac{\delta \gg}{\delta \phi(x)} = \Delta_\phi(x) \cdot \gg
  \,. 
\end{equation} 
The explicit meaning of the r.h.s. is the following: 
In analogy to~(\ref{gree.5})
we consider an (extended) action at lowest order
\begin{equation}  \label{gree.5new}
\g_0 = \int {\rm d}^4x \left( 
{\cal L}_{INV} + \sum_i \phi^*_i \, s \phi^i + {\cal L}_{GF} +{\cal
  L}_{\Phi\Pi} + \sum_r \rho_r \Delta^r  \right).   
\end{equation}
where the sum runs over all the possible local insertion $\Delta^r$. 
Then one has
\begin{eqnarray}  \label{qap_1.1}
\lefteqn{\Delta^r \cdot \gg  =}
\\&& \sum_{n=0}^{\infty} \frac{(-i)^n}{n!}\int 
\left( \prod_{j=0}^{n} {\rm d}^4p_j \right) \delta^4 \left( \sum_k p_k \right) 
\phi_1(p_1) \dots \phi_{n-1}(p_{n-1}) \rho_r(p_n) 
\gg^{(n)}_{\phi_1 \dots \phi_{n-1} \rho_r}(p_1, \dots, p_n) 
\,,
\nonumber
\end{eqnarray} 
with 
\begin{equation}  \label{gree.3new}
i^n 
\left. \frac{\delta^{n} \gg}{\delta 
\phi_1(p_1) \dots 
\delta \phi_{n-1}(p_{n-1})\rho_r(p_n)} \right|_{\phi=0} =  
\gg^{(n)}_{\phi_1 \dots \phi_{n-1} \rho_r}(p_1, \dots, p_n). 
\end{equation} 
Thus, $\Delta^r \cdot \gg$ generates the 1PI Green functions with an insertion 
of an integrated or local composite operator $\Delta^r$
\begin{equation}
\langle T \left( \phi_1 \dots \phi_n \Delta^r \right) \rangle^{1PI}.
\end{equation}
It can be decomposed into a basis of  
integrated monomials of fields and their derivative with
the same quantum numbers as the l.h.s. of (\ref{qap_1}).
Therefore the r.h.s. of~(\ref{qap_1})
can be  decomposed into the
classical insertion and  their radiative corrections:
\begin{equation}\label{qap_1.2} 
\Delta_\mu  \cdot \gg = \Delta_\mu +  {\cal O}(\hbar \Delta_\mu)
\,,
\hspace{1cm} 
\Delta_{\phi}(x) \cdot \gg =
\Delta_{\phi}(x) +  {\cal O}(\hbar \Delta_{\phi}(x))
\,. 
\end{equation} 

In the case of STIs, the QAP implies that the (subtracted) Green functions
$\g$, computed within a given scheme,
fulfill them up to local insertions $\Delta_{STI}$ in the 1PI Green
functions: 
\begin{equation}\label{qap_1a} 
{\cal S}( \g ) = \Delta_{STI} + {\cal O}(\hbar \Delta_{STI})
\,.
\end{equation} 
Here $\Delta_{STI}$ is an integrated, Lorentz invariant polynomial
(of the fields and their derivatives) 
with ultra-violet (UV) degree $\leq 4$ and
IR degree $\geq 3$ (assuming four space-time dimensions).

Although Eqs.~(\ref{qap_1}) and~(\ref{qap_1a})
apply to any renormalization scheme, 
the coefficients of the various $\Delta$s  depend on the
particular scheme adopted. In fact, the definitions of $\Delta_\mu,
\Delta_\phi(x)$ and $\Delta_{STI}$  rely on specific conventions for 
composite operators. Thus a renormalization description 
for the composite operators is necessary. Here one uses 
the concept of 
Normal Product Operators (NPO) introduced by Zimmermann~\cite{zimm}
or the conventional counterterm technique 
which is preferable from the practical point of view.

Once the breaking terms $\Delta_{STI}$ are given
we can discuss the main objective of the algebraic
method~\cite{brs,libro}. This essentially entails 
in a prescription to restore the identities by suitable local
non-invariant counterterms, $\g^{CT}$, such that at $n^{th}$ order
one has
\begin{eqnarray}\label{qap_new_1}
{\cal S}(\gg)^{(n)} \equiv {\cal S}_0(\g^{(n)}) + \sum^{n-1}_{j=1} (\gg^{(j)},
\gg^{(n-j)}) - \ {\cal S}_0(\g^{CT,(n)})  
=  {\cal O}(\hbar \Delta_{STI})
\,,
\end{eqnarray}
where the decomposition given in Eq.~(\ref{line_ST}) has been used.
Notice that the Green functions $\gg^{(j)}$ with $j < n$ are already fixed
and only $\g^{(n)}$ has to be adjusted by the local counterterms
$\g^{CT,(n)}$.
Thus, in practice the problem amounts to solve the algebraic equations 
\begin{eqnarray}\label{qap_new_2}
{\cal S}_0(\g^{CT,(n)}) =  \Delta_{STI}
\,,
\end{eqnarray}
where ${\cal S}_0$ is given in Eq.~(\ref{line_ST}).
This equation turns out to be solvable  in absence of
anomalies~\cite{brs,libro} where only the consistency conditions
have to be used
(cf. Section~\ref{sub:consi}).
Moreover, due to a non-trivial kernel of the 
operator ${\cal S}_0$ (i.e. the space of invariant counterterms), 
one is allowed to impose renormalization conditions tuning the 
free parameters of the model. 

This principal algebraic procedure does not
automatically lead to a practical advice for higher-loop
calculations.  
As already mentioned in the Introduction, regardless which regularization 
scheme one uses, the calculation of $\Delta_{STI}$
in~(\ref{qap_new_2})
is quite tedious and gets even more complicated at higher orders. 
In general, one has to
calculate all Green functions which occur in the complete 
set of STIs. Inserting them
in the STIs one fixes   $\Delta_{STI}$. 
The additional computations necessary in the conventional
algebraic method can be slightly reduced:  
instead of calculating all Green 
functions which occur in the full set of the STIs,
one can simplify  the problem by computing the 
Green functions in special points, namely for zero momentum $p=0$, for
on-shell momentum or for large external momenta.
As a consequence the breaking terms,
$\Delta_{STI}$,  are simply related 
to Green functions evaluated in these special points.
Clearly, if on-shell renormalization conditions are used 
in the calculation, the on-shell method is definitely superior to 
the zero-momentum subtraction. In the infinite-momentum scheme
one can take advantage of Weinberg's theorem~\cite{wein} (see also
Section~\ref{sub:remove}).

Despite of these simplifications, 
still all Green functions involved in the STI have to be taken
into account for the computation of the breaking terms in~(\ref{qap_new_2}).
In Section~\ref{sub:remove} we will present our strategy which drastically
reduces the additional work as will also be 
shown in the examples of Section~\ref{sec:hgg_coun} and~\ref{sec:bsg_coun}.

%%%%%%%%%%%%%%%%%%%%%%%%%%%%%%%%%%%%%%%%%%%%%%%%%%%%%%%%%%%%

\subsection{Consistency and renormalization conditions} 
\label{sub:consi}

One of the main tools of algebraic renormalization is 
provided by the algebraic relations between 
the functional operators ${\cal S'}_{\gg'}$ 
of the STIs (see Eq.(\ref{ST_bfm})  and the definition (\ref{line_ST})) , 
${\cal W}'_{(\lambda)}$ of the WTIs 
(given in Eq. (\ref{WTI}))
and of the other supplementary identities like the Faddeev-Popov equations
(see \cite{krau_ew,grassi}).
Beyond their relevance in the theoretical framework~\cite{libro},
the consistency conditions turn out to be important for practical
applications. 

The operators ${\cal S'}_{\gg'}$ and ${\cal W}'_{(\lambda)}$ form an algebra 
\begin{eqnarray}\label{cc.1} 
  {\cal S'}^2_{\gg'}= 0 \hspace{.4cm} {\rm if} \hspace{.4cm}  {\cal
    S'}(\gg')&=&0  
  \,,
  \nonumber \\ 
  {\cal S'}_{\gg'}({\cal W}'_{(\lambda)}(\gg')) 
  - {\cal W}'_{(\lambda)}({\cal S'}(\gg')) &=&0
  \,,  
  \nonumber\\ 
   {\cal W}'_{(\lambda)} ({\cal W}'_{(\beta)}(\gg')) - 
  {\cal W}'_{(\beta)} ({\cal W}'_{(\lambda)}(\gg')) &=& 
  {\cal W}'_{(\lambda \wedge \beta)}(\gg')
  \,,
\end{eqnarray} 
which, applied to the breaking terms 
$\Delta'_S$ and $\Delta'_W(\lambda)$ of the STI, respectively, WTI
\begin{eqnarray}\label{cc.2} 
  {\cal S}'(\g')&=& 
  \Delta'_S + {\cal O}(\hbar \Delta'_S)   
  \,,
  \nonumber\\ 
  {\cal W}'_{(\lambda)} (\g') &=& 
  \Delta'_W(\lambda) + {\cal O}(\hbar \Delta'_W)
  \,,
\end{eqnarray} 
leads to the so-called consistency conditions
\begin{eqnarray} 
{\cal S}'_0 (\Delta'_S) &=& 0   
\,,
\label{cc.3.1} \\ 
 {\cal S}'_0 (\Delta'_W(\lambda)) 
- {\cal W}'_{(\lambda)}(\Delta'_S) &=&0 \label{cc.3.2}
\,,
\\ 
{\cal W}'_{(\lambda)} (\Delta'_W(\beta)) - 
{\cal W}'_{(\beta)} (\Delta'_W(\lambda)) &=& 
\Delta'_W(\lambda \wedge \beta)
\,,
\label{cc.3.3} 
\end{eqnarray} 
where $(\lambda\wedge\beta)^a=f^{abc}\lambda^b\beta^c$.
These kind of equations are  called Wess-Zumino consistency condition.

The consistency conditions have very important practical consequences:
\begin{enumerate}
\item
In the definition of the possible breaking terms of the STIs and WTIs
one first admits any kind of local Lorentz invariant terms 
which have the proper quantum numbers. However, the consistency conditions 
of Eqs.~(\ref{cc.3.1})--(\ref{cc.3.3}) constrain those breaking terms further. 
Actually, they play  the key role in the algebraic
analysis of anomalies. They single out the possible candidates for
breaking terms which cannot be removed by suitable counterterms.
It is well known that this can be done  by means of
cohomological methods. For this important issue we refer 
the reader to the rich literature~\cite{coho}.
As an example, in the SM the consistency conditions single out 
the Adler-Bardeen-Jackiw anomaly. However, as is well known, 
the latter cancels out because of the 
specific choice of the fermion content in the SM. 

Once we know that the identity has no anomaly,
the practical constraints of the consistency conditions 
on the breaking terms (introduced by the non-invariant regularization)
can be worked out. This is achieved by
the condition that only those breaking terms 
are possible which can be produced by local counterterms.
We will explicitly  illustrate this  practical procedure 
for the process \bsg~in Section~\ref{sec:bsg_coun}.

\item
Another important consequence of the consistency conditions  
is the triangular structure of functional identities. This means that it is
possible  
to organize the set of functional identities
into a hierarchical structure in such a way that we can restore the
identities one after the other without spoiling those which are already
recovered --- as explicitly explained in Section~\ref{sub:bsg} in the 
example of $b \rightarrow s \gamma$. 

We have to solve the following problems
\begin{eqnarray}\label{eq:app_D_1} 
  {\cal S}'_0(\g'_{CT}) &=& \Delta'_S  
  \nonumber\\ 
  {\cal W}'_{(\lambda)} (\g'_{CT}) &=& \Delta'_W(\lambda)
  \,,  
\end{eqnarray} 
where $\g'_{CT}=\int {\rm d}^4x {\cal L}'^{CT}(x)$ with the 
conditions~(\ref{cc.3.1})--(\ref{cc.3.3}).

We know that, in absence of anomalies, the breaking of the
Wess-Zumino consistency conditions~(\ref{cc.3.3}) are solved by
the counterterms 
\begin{eqnarray}
{\cal W}'_{(\lambda)} (\g_{CT}^{' WTI}) &=& \Delta'_W(\lambda)  
\nonumber 
\end{eqnarray}   
Therefore, by introducing those counterterms in the Feynman amplitudes,
or equivalently in the action $\g'$: $\g'\rightarrow \g'-\g_{CT}^{' WTI}$,
we have the new system 
\begin{eqnarray}
  \label{eq:app_D_6} 
  {\cal S}'(\g' - \g_{CT}^{'WTI}) &=& \Delta'_S -   {\cal S}'_0(
  \g_{CT}^{' WTI}) \equiv \hat{\Delta'}_S 
  \,,
  \nonumber\\ 
  {\cal W}'_{(\lambda)} (\g' - \g_{CT}^{' WTI}) &=& \Delta'_W(\lambda)  - 
  {\cal  W}'_{(\lambda)} (\g_{CT}^{' WTI}) = 0
  \,,
  \nonumber 
\end{eqnarray} 
where the new breaking term $\hat{\Delta'}_S $ is explicitly background
gauge invariant because of~(\ref{cc.1}) and~(\ref{cc.3.2})
\begin{eqnarray}\label{eq:app_D_7} 
  {\cal W}'_{(\lambda)} ( \hat{\Delta'}_S) ={\cal W}'_{(\lambda)}( \Delta'_S)
  -   {\cal W}'_{(\lambda)} {\cal S}'_0(
  \g_{CT}^{' WTI}) = {\cal W}'_{(\lambda)}( \Delta'_S) -  {\cal S}'_0
  \left(  \Delta'_W(\lambda)\right) = 0. 
  \nonumber 
\end{eqnarray} 
This means that, in order to restore the STI, we need only background
gauge invariant counterterms. To that purpose, we add terms like 
$\g_{CT}^{'  WTI}$ so that
\begin{eqnarray}
\label{eq:app_D_8}
{\cal W}'_{(\lambda)} (\g_{CT}^{' STI}) = 0, ~~~~~ 
 {\cal S}'_0(\g_{CT}^{' STI})= \hat{\Delta'}_S. 
\end{eqnarray}
This proves that we can effectively disentangle the
two sets of identities. In Appendix~\ref{app:triangula} we specialize 
the argument to each subspace of counterterms. 
\end{enumerate}

In the following we analyze 
the  practical algebraic renormalization program 
within the BFM discussed in  Section~\ref{sub:BFM} in more detail. 

We consider the STI (\ref{ST}) and  separate the extended part from the 
conventional  one (namely the old STI (\ref{ST}))
by taking the derivative with respect to $\Omega$ and then 
setting them to zero. 
As mentioned in Section~\ref{sub:BFM},
we obtain a set of the following three different equations 
(in the following we regard only the example for $\Omega^a_\mu$, 
but it is understood that this is valid for every $\Omega$):
\begin{eqnarray}
  \label{eq:an_11}
&&  {\cal S} \left( \gg \right) = 0\,, ~~~~{\cal W}_{(\lambda)} (\gg) = 0\,, \\
&&   \left(  \frac{\delta\gg}{\delta \hat{G}^a_\mu} -
\frac{\delta\gg}{\delta G^a_\mu}\right)  =  
{\cal S}_{\gg} \left( \frac{\delta\gg'}{\delta {\Omega}^a_\mu}
\right)_{\Omega=0}  
\end{eqnarray}
where $\gg \equiv \gg'_{\Omega=0}$, ${\cal S}_{\gg}$ is the linear 
(conventional) STI operator (\ref{line_ST})  and 
${\cal W}_{(\lambda)}$ is the WTI operator defined in (2.24).

Notice that  reducing the equation for an arbitrary $\Omega$ to  $\Omega=0$ might hide some 
relevant Green functions with higher powers of $\Omega$. 
In particular in the Eqs.~(\ref{eq:an_11})  we consider only the 
first power of $\Omega$ and the functionals independent of $\Omega$. 
However, we would like to recall that $\Omega$ is a vector or a scalar
field from the Lorentz  
point of view, they have ghost number equal to +1 and dimension equal to 1. 
Therefore the higher powers of $\Omega$ cannot contribute to divergent Green functions or, 
equivalently, there are no counterterms with two powers of $\Omega$. 

At the quantum level they are broken by the breaking terms controlled 
by the QAP
\begin{eqnarray}
  \label{eq:an_2}
&&  {\cal S} \left( \g \right) = \Delta_S + {\cal O}(\hbar \Delta_S), ~~~~{\cal W}_{(\lambda)} (\g) =  \Delta_W(\lambda) +  {\cal O}(\hbar \Delta_W) \\
&&   \left(  \frac{\delta\g}{\delta \hat{G}^a_\mu} - \frac{\delta\g}{\delta G^a_\mu}\right)  - 
{\cal S}_{\g} \left( \frac{\delta\g'}{\delta {\Omega}^a_\mu} \right)_{\Omega=0} =  \Delta_{G^a_\mu} +  {\cal O}(\hbar \Delta_{G^a_\mu}) .
\end{eqnarray}
{Notice, furthermore, that there is no prime 
in the following equations  since we restrict our considerations 
to the functional $\g$ and to the breaking terms 
$\Delta_S$ and $\Delta_W$ in the case where $\Omega=0$.}
Finally we can use the consistency conditions
(cf. Eqs. (\ref{cc.3.1})-(\ref{cc.3.3})) and  
one can easily deduce the following set of constraints
\begin{eqnarray}
  \label{eq:an_3a}
&&{\cal S}_{0}  \left(\Delta_S\right) = 0 \\
  \label{eq:an_3b}
&&\left( \frac{\delta}{\delta \hat{G}^a_\mu} - \frac{\delta}{\delta
G^a_\mu} \right)  \left(\Delta_S\right) =  
{\cal S}_{0}  \left(\Delta_{G^a_\mu}\right) \\
  \label{eq:an_3c}
&&\left( \frac{\delta}{\delta \hat{G}^a_\mu} - \frac{\delta}{\delta
G^a_\mu} \right)  
\left(\Delta_{G^b_\nu}\right) - \left( \frac{\delta}{\delta
\hat{G}^b_\nu} - \frac{\delta}{\delta G^b_\nu} \right)  
\left(\Delta_{G^a_\mu}\right) = 0 \\
  \label{eq:an_3d}
&& {\cal S}_{0}  \left(\Delta_W(\lambda)\right) - 
{\cal W}_{(\lambda)}  \left(\Delta_S\right) = 0 \\
  \label{eq:an_3e}
&& {\cal W}_{(\lambda)}   \left(\Delta_{G^a_\mu}\right) - \left(
\frac{\delta}{\delta \hat{G}^a_\mu} - \frac{\delta}{\delta G^a_\mu}
\right) 
\left(\Delta_W(\lambda)\right)  
= f_{abc} \lambda^b \Delta_{G^c_\mu}. 
\end{eqnarray}
This list is not complete because 
we have also to add the consistency conditions 
between the breaking terms $\Delta_W(\lambda)$, 
but these are exactly the same as written in (3.10).

Using this equation we can study the two approaches singled out 
in Section~(\ref{sub:BFM}). 
In the first approach we restore the naive STIs, that means we compensate the 
breaking terms $\Delta_S $ with suitable counterterms. In addition we
consider the WTI and their breaking terms $\Delta_W(\lambda)$   
and we compensate them by using suitable counterterms. 
Those counterterms are related by the consistency conditions 
(\ref{eq:an_3d}) which we take into account in our explicit
examples in order to reduce the effort of our computation. 

By setting all the breaking terms $\Delta_S=\Delta_W(\lambda)=0$ to
zero and 
assuming that all the necessary counterterms are introduced, we have finally 
\begin{eqnarray}
  \label{eq:an_4a}
&& {\cal S}_{0}  \left(\Delta_{G^a_\mu}\right) = 0 \\
  \label{eq:an_4b}
&&\left( \frac{\delta}{\delta \hat{G}^a_\mu} - \frac{\delta}{\delta
G^a_\mu} \right)  
\left(\Delta_{G^b_\nu}\right) - \left( \frac{\delta}{\delta \hat{G}^b_\nu} -
\frac{\delta}{\delta G^b_\nu} \right)  \left(\Delta_{G^a_\mu}\right) = 0 \\ 
  \label{eq:an_4c}
&& {\cal W}_{(\lambda)}   \left(\Delta_{G^a_\mu}\right) = f_{abc} \lambda^b
\Delta_{G^c_\mu}.   
\end{eqnarray}
The first equation (\ref{eq:an_4a}) tells us that the breaking terms $\Delta_{G^c_\mu}$ of the 
remaining identities (which are needed in order to control 
the breaking of the relations between the background gauge fields and 
the quantum ones) are STI invariant breaking terms. Therefore 
we need only counterterms which satisfy the  STI. The second equation 
(\ref{eq:an_4b}) provides a consistency condition for the breaking terms 
$ \Delta_{G^c_\mu}$ themselves. 
It tells us that the breaking terms can only 
depend upon the linear combinations of the background gauge fields. 
The last equation, namely (\ref{eq:an_4c}), 
says that the breaking terms $\Delta_{G^c_\mu}$ should transform as vectors 
under transformations of the background gauge symmetry. 
It is  easy to see that the possible candidates for the breaking terms 
$\Delta_{G^c_\mu}$ are  very few {and they can be reabsorbed 
by adjusting counterterms with background fields.}

In the second approach 
we first restore all the WTI and all the identities like (\ref{eq:new_1}). 
That is we suppose that we have inserted the relevant counterterms which 
render $\Delta_{G^c_\mu} =  
\Delta_W(\lambda)= 0 $. From the consistency conditions above one easily
derives again the constraints on the breaking terms $\Delta_S$.

In order to select a practical criterium 
to use one approach instead of the other we have to study the process under consideration. 
In fact two situations can in general occur. If we consider only processes
at one- or two-loop, we have 
\begin{itemize}
\item the case where both, the quantum and the background version 
of a Green function, occur as sub-diagrams { or as a component of the connected Green functions 
involved in the process under consideration}. 
Here the second approach seems to be  advantageous.  
\item the case where the quantum sub-diagrams  are completely unrelated 
to the background Green functions involved in
process (for instance in our example in Section~\ref{sub:bsg}). 
Then identities (3.18) are not relevant and one is allowed to restrict
the algebraic analysis to the WTI and to the conventional part of the STI 
(3.17).  
\end{itemize}

%%%%%%%%%%%%%%%%%%%%%%%%%%%%%%%%%%%%%%%%%%%%%%%%%%%%%%%%%%%% 

\subsection{Removing the breaking terms}  
\label{sub:remove}

In the following a general procedure is presented which optimizes the
algebraic method. In~\cite{fg} this procedure was used to discuss the complete
renormalization of the Abelian Higgs model. In this paper we use the procedure
in the more general context of the SM.
The breaking terms to be computed are reduced to the 
evaluation of finite Green functions. Moreover, it is shown that 
this method is very powerful if in addition the BFM is adopted.

First a formal derivation of the various steps is presented. 
Afterwards the power of the method is demonstrated
in the case of the two-point function for the $W$ bosons 
which represents an important ingredient for the calculation of radiative
corrections to the  well-known $\rho$ parameter~\cite{rhopar} parameterizing 
the isospin breaking of a fermion doublet.  

Let us assume that the invariant vertex functions $\gg^{(m)}$
has already been constructed
up to order $m\le n-1$.
Thus, because of the QAP (see Section~\ref{sub:QAP}),
the subtracted functional $\g^{(n)}$ satisfies the broken STI\footnote{
In the present section we skip the prime on the effective action 
again.}
\begin{eqnarray}
{\cal S} (\Gamma)^{(n)} = {\cal S}_0 (\Gamma^{(n)})
+\sum_{j=1}^{n-1}
\left({{\rm I}\!\Gamma}^{(j)},{{\rm I}\!\Gamma}^{(n-j)}  \right)
=
\Delta^{(n)}
\,,
\label{invariance}
\end{eqnarray}
where the meaning of ${\cal S}_0$ and the compact notation for
$(\cdot, \cdot)$ is given in Eqs.~(\ref{line_ST}) and~(\ref{bracket}).
In the case of the absence of anomalies, we know from the general 
results of algebraic renormalization that one can add finite counterterms 
in the action in order
to restore the validity of the STIs. 
As we have seen in Section~\ref{sub:conventional}, the STIs
connect a large amount of Green function. In principle, all of them
have to be calculated in order to fix the breaking term $\Delta^{(n)}$.

The construction of an efficient and convenient method for the determination
of $\Delta^{(n)}$ consists of the following steps:
\begin{enumerate}
\item
We disentangle the various terms in $\Delta^{(n)}$
expanding it into an operator basis 
\begin{equation}
\Delta^{(n)} = \sum_i c_i \int {\rm d}^4 x {\cal M}_i(x)
\end{equation}
where ${\cal M}_i(x)$ are local Lorentz invariant monomials 
in the fields and their derivatives.
The basis is quite  restricted 
by additional symmetries preserved by the regularization procedure. 
The usual power counting poses an upper bound on their
mass dimension (which is independent of the loop order in the case
of power counting renormalizable theories). As we have seen in
Section~\ref{sub:consi}, 
$\Delta^{(n)}$ is also constrained by the Wess-Zumino consistency
condition~\cite{libro} given in~(\ref{cc.3.3}).
\item
One acts with the zero momentum subtraction operator ($1-T^\delta$) on both
sides of Eq.~(\ref{invariance}). Here $T^\delta$ denotes  the Taylor
operator in the external momenta up to a suitable degree $\delta$ (see
Appendix~\ref{app:sourceterms}, Eqs.~(\ref{definitionT2}) for its explicit
form). 
The locality of $\Delta^{(n)}$ ensures that it is possible to find a 
degree $\delta_n$ such that\footnote{In power counting renormalizable theories
  the degree $\delta$ is of course independent of the loop order $n$.} 
\begin{eqnarray} 
  \label{delta} 
  (1-T^{\delta_n}) {\cal S}(\Gamma)^{(n)} = ( 1-T^{\delta_n} ) \Delta^{(n)} =
  0
\,,
\end{eqnarray} 
and thus $\Delta^{(n)}$ is subtracted away.
At the moment we assume that the zero momentum subtraction 
is possible. This means that the vertex functions have to be
sufficiently regular at zero momenta. 
Below possible adjustments
for the case that IR problems occur
are discussed.

\item Clearly the l.h.s. of~(\ref{delta}) has not yet the correct 
form. Indeed we want to obtain a new STI for subtracted Green functions,
i.e. for $(1-T^{\delta_{pc}}) \g^{(n)}$, where $\delta_{pc}$ is the
naive power counting degree.
In general we have $\delta_n  \geq   \delta_{pc}$. To that purpose 
we commute the Taylor operation with the Slavnov-Taylor 
operator. It is convenient to adopt the decomposition~(\ref{invariance}) 
into a linearized operator plus bilinear terms.
The part involving the linearized operator leads to
\begin{eqnarray}
  (1-T^{\delta_n}) {\cal S}_0(\Gamma^{(n)})= {\cal S}_0
  \left( (1-T^{\delta_{pc}})\Gamma^{(n)}\right)+
  {\cal S}_0 \left( T^{\delta_{pc}} \Gamma^{(n)} \right) - T^{\delta_n}
  \left( {\cal S}_0(\Gamma^{(n)}) \right)
  \,,
  \label{oversubtraction}
\end{eqnarray}
which expresses the fact that ${\cal S}_0$ is in general not homogeneous 
in the fields. In particular this is the case for theories with spontaneous
symmetry breaking.
Notice that 
the Taylor operator is scale invariant, however, it does not commute
with spontaneous symmetry breaking. Furthermore
there might be IR problems in connection
with massless fields as will be discussed below. 
The difference between $\delta_n$ and
$\delta_{pc}$ leads to over-subtractions
in $\g^{(n)}$. Therefore
breaking terms generated by the last two terms on the r.h.s. of the
above equation have to be introduced.
Furthermore, the action of the Taylor operator can
be split into the naive contribution $\sum_{j=1}^{n-1}\left(
{{\rm I}\!\Gamma}^{(j)},{{\rm I}\!\Gamma}^{(n-j)}
\right)$ plus the local terms obtained by Taylor expansion. These local
terms also contribute to the new breaking terms. 

Finally, by applying the Taylor operator on~(\ref{invariance})
and by using~(\ref{delta}) and~(\ref{oversubtraction}) we obtain 
\begin{eqnarray}\label{rea}
 (1 - T^{\delta_n}) {\cal S}(\Gamma)^{(n)} &=& 
{\cal S}_0\left( (1-T^{\delta_{pc}}) \Gamma^{(n)}\right) +
\sum_{j=1}^{n-1}\left(
{{\rm I}\!\Gamma}^{(j)},{{\rm I}\!\Gamma}^{(n-j)}
\right) 
\\&&\mbox{}  
-\big[ T^{\delta_n} {\cal S}_0 - {\cal S}_0 T^{\delta_{pc}}\big]
\Gamma^{(n)} -
T^{\delta_n}\sum_{j=1}^{n-1}\left(
{{\rm I}\!\Gamma}^{(j)},{{\rm I}\!\Gamma}^{(n-j)}
\right) =  0
\nonumber
\,.
\end{eqnarray}
The terms in the second line of~(\ref{rea}) represent 
the new local breaking terms which correspond to the
subtracted function at the order $n$, 
$(1-T^{\delta_{pc}}) \Gamma^{(n)}$.
Thus, it is convenient to define the new breaking terms as
\begin{equation}
  \Psi^{(n)} =\big[ T^{\delta_n} {\cal S}_0 - {\cal S}_0 T^{\delta_{\rm
      pc}}\big] 
  \Gamma^{(n)} + 
  T^{\delta_n}\sum_{j=1}^{n-1}\left(
    {{\rm I}\!\Gamma}^{(j)},{{\rm I}\!\Gamma}^{(n-j)}
  \right)
  \,.
  \label{newterms}
\end{equation}
We emphasize that they
are universal in the sense that they do not depend  
on the specific regularization used in the calculation --- in contrast to
$\Delta^{(n)}$ in Eq.~(\ref{invariance}).  

\item
Now the principal construction of an invariant Green function
$\gg^{(n)}$ is clear:
First, one has to calculate the universal terms $\Psi^{(n)}$. 
This step  consists of the evaluation of a set
of finite amplitudes and their derivatives at zero momenta
as one can read off Eq.~(\ref{newterms}). 
There, the functions ${{\rm I}\!\Gamma}^{(m)}$
are computed at the lower orders in the perturbative expansion.
They are supposed to satisfy the STI at every order $m<n$.
One strategy to further simplify the calculation of the universal
breaking terms, $\Psi^{(n)}$, is to reduce the number of bilinear 
contributions by a suitable choice of the renormalization 
conditions. Second, one has to find the
counterterms $\Xi^{(n)}$ which satisfy
\begin{eqnarray}
  S_0(\Xi^{(n)}) = - \Psi^{(n)} 
  \label{counterterms}
\,.
\end{eqnarray}
Finally, the correct vertex function reads
\begin{eqnarray}
  {{\rm I}\!\Gamma}^{(n)} = (1-T^{\delta_{\rm pc}})\Gamma^{(n)} + \Xi^{(n)}
  \label{final}
  \,.
\end{eqnarray}
We stress that only the universal breaking terms have to be computed
which
drastically simplifies the determination of the non-invariant counterterms. 

\item
As mentioned above, in the presence of massless particles 
zero-momentum subtractions
of the regularized function $\Gamma^{(n)}$ might lead to
IR divergences. In principle one can circumvent this problem 
by using the Lowenstein scheme~\cite{IR}. In this case one has to
introduce a generalized Taylor expansion which also takes care of
the soft mass parameter $s$ of the Lowenstein scheme. 
Consequently one has to analyze the new breaking terms arising from
the commutator between this new operator and the STI, respectively,
the WTI operator. Moreover, one can take advantage of  the fact that 
the breaking terms are IR safe for principal reasons 
provided there are no IR anomalies in the model.
In the phenomenological examples we discuss in the following 
sections no IR problems occur. 

There is also an alternative path for the case where IR
problems are induced by zero-momentum subtractions.
The whole procedure proposed in this section, that is 
the translation of the ``conventional'' breaking terms into
``universal'' ones, also leads to drastic simplifications if  
the subtractions are performed for infinite instead of zero momenta.
The former procedure is even preferable if four- and five-point functions
occur in the STIs.
It obviously circumvents all IR problems mentioned above.

\item 
We emphasize that there is the free choice of the renormalization conditions 
which corresponds to a specific choice of the invariant counterterms.
A change in the renormalization conditions leads to a
change of the basis of the breaking terms.
In practice, this means that one is allowed to shift 
the breaking terms of some WTIs and some STIs to
others which are not relevant for the specific process under the
consideration.

\item
There are also cases where the zero
momentum subtraction is not very practical. 
Let us consider a STI (or a
WTI) involving Green functions with four external legs as a minimum
(e.g. $\g_{W_\mu W_\nu A_\rho A_\sigma}$) or with high-dimensional
fields like the BRST sources (e.g. $\g_{c^+ \bar{q}^*_i b}$ where
$\bar{q}^*_i$ is the BRST source for the quark field $q_i$). As is easy to
show they are fixed by STIs involving convergent Green functions (see
Sections~\ref{sub:bsg} and~\ref{sub:restSTI}). Therefore
zero-momentum subtraction introduces breaking terms which
are very tedious to compute as they in general involve Green functions with
five or six external legs.
Very often quite a lot of diagrams contribute.
Thus sometimes it is more convenient to choose
another subtraction technique.
One possibility would be to consider large external momenta.
Using Weinberg's theorem~\cite{wein},
the coefficients of the breaking terms are computed directly
by Green functions with only a small number of external legs,
like $\g_{c^+ \bar{q}^*_i b}$.

\item
The subtraction technique for STIs and WTIs 
correlated to Green functions like $\g_{c^+ \bar{q}^*_i b}$ or similar ones
are more involved as can be seen from Eq.~(\ref{WTI_anti.1}).
Thus it is convenient to decompose the 
STI~(\ref{invariance}) into different terms:
\begin{eqnarray}
{\cal S}_0 (\Gamma^{(n)})
+\sum_{j=1}^{n-1}
\left({{\rm I}\!\Gamma}^{(j)},{{\rm I}\!\Gamma}^{(n-j)}  \right)
&\equiv& 
\sum_{\alpha \in {\cal U}} c_{\alp} \Gamma^{(n)}_\alpha + 
\sum_{\alp \in {\cal U}'} d_{\alp} \gg^{(n)}_\alpha +\sum_{j=1}^{n-1}
\left({{\rm I}\!\Gamma}^{(j)},{{\rm I}\!\Gamma}^{(n-j)}  \right)
\nonumber\\&=&\Delta^{(n)}
\,,
\label{invariance_2}
\end{eqnarray}
where ${\cal U}$, respectively, ${\cal U}'$ are index sets
corresponding to the $n$-loop Green functions
$\Gamma^{(n)}_\alpha$ and $\gg^{(n)}_\alpha$.
Remember that the finite parts of $\Gamma^{(n)}_\alpha$ still have to be fixed
whereas this is already done for $\gg^{(n)}_\alpha$.
The functions $\gg^{(n)}_\alpha$ are conveniently computed using other
techniques, e.g. the one described in 7.
The coefficients $c_{\alpha}$ and $d_{\alpha}$ 
are functions of the external momenta. 

Now the Taylor operator $(1-T^\delta)$ can again be applied
leading to the following result
\begin{eqnarray}
&& \sum_{\beta \in {\cal U} } c_{ \alp}  
\left( (1-T^{\delta_{pc}}) \Gamma^{(n)}_\alpha \right) +
\sum_{\alp \in {\cal U}'} d_{\alp} \gg^{(n)}_\alpha +
\sum_{j=1}^{n-1}\left(
{{\rm I}\!\Gamma}^{(j)},{{\rm I}\!\Gamma}^{(n-j)}
\right) + 
\\
&& - \sum_{\alp \in {\cal U} } 
\big[ T^{\delta_n}  c_{\alp}  -  c_{\alp} T^{\delta_{pc}}\big]
\Gamma^{(n)}_\alpha - 
T^{\delta_n}\left[  \sum_{\alp \in {\cal U}'} d_{\alp} \gg^{(n)}_\alpha
  + \sum_{j=1}^{n-1}\left( 
{{\rm I}\!\Gamma}^{(j)},{{\rm I}\!\Gamma}^{(n-j)}
\right) \right]=  0
\nonumber
\,,
\label{rearrangement_2}
\end{eqnarray}
where the new contribution $T^{\delta_n} \sum_{\alp \in {\cal U}'} d_{\alpha} 
\gg^{(n)}_\alpha$ appear in the breaking terms $\Psi^{(n)}$. It
depends on 
the Green functions $\gg^{(n)}_\alpha $. This modification of our formalism
will be used in Section~\ref{sec:bsg_coun} for the example $b\rightarrow
  s\gamma$.
\end{enumerate}

In order to illustrate the method elaborated above we
consider the two-point functions for the $W$ boson.
All the ingredients to consider the STIs involved in the
calculation of the
two-point functions in a non-symmetric regularization
scheme are now introduced.
In Appendix~\ref{app:stiex} it is shown
that the two-point functions and their 
STIs (plus the ghost equation) form a closed set of relations.
It is also known that the STIs~(\ref{exap}) 
are broken by local terms only. According to the QAP we have:
\begin{eqnarray}\label{exa_1}
\g_{c^{+} W^{*,-}_{\nu}}(p)  \g_{W^{+}_{\nu}
  W^{-}_{\mu}}(p) +   \g_{c^{+} G^{*,-}}(p)  \g_{G^{+}
  W^{-}_{\mu}}(p)  
&=&  
\Delta_{c^{+} W^{-}_{\mu}}(p) + {\cal O}(\hbar \Delta)
\,,
\nonumber \\
\g_{c^{+} W^{*,-}_{\nu}}(p)  \g_{W^{+}_{\nu}
  G^{-}}(p) +   \g_{c^{+} G^{*,-}}(p)  \g_{G^{+}
  G^{-}}(p) 
&=&   
\Delta_{c^{+} G^{-}}(p) + {\cal O}(\hbar \Delta)
\,.
\end{eqnarray}
The breaking terms
$\Delta_{c^{+} W^{-}_{\mu}}(p)$ and $\Delta_{c^{+} G^{-}}(p)$ 
are finite and exhibit the
following decomposition in terms of independent monomials
\begin{eqnarray}\label{exa_2}
  \Delta_{c^{+} W^{-}_{\mu}}(p) &=& \alpha_1 p^2 p_{\mu} +
  \alpha_2 p_{\mu}  
   \,,
  \nonumber \\  
  \Delta_{c^{+} G^{-}}(p) &=& \alpha_3 p^2 +  \alpha_4 
  \,.
\end{eqnarray}

According to our procedure we apply the Taylor operator to the 
various Green functions involved in the STIs
\begin{eqnarray}
  \label{ep_1}
  \gh_{W^{+}_{\nu}
    W^{-}_{\mu}}(p) &=& ( 1 - T^2_p)  \g_{W^{+}_{\nu}
    W^{-}_{\mu}}(p)
  \,,
  \nonumber \\
  \gh_{G^{+} W^{-}_{\mu}}(p) &=& ( 1 - T^2_p)  \g_{G^{+}
    W^{-}_{\mu}}(p) \,\,=\,\, -i p_\mu ( 1 - T^0_p)  \g_{G^{+}
    W^{-}}(p^2)
  \,,
  \nonumber \\
  \gh_{W^{+}_{\mu} G^{-}}(p) &=& ( 1 - T^2_p)  \g_{W^{+}_{\mu}
    G^{-}}(p) \,\,=\,\, -i p_\mu ( 1 - T^0_p)  \g_{W^{+}
    G^{-}}(p^2)
  \,,
  \nonumber \\
  \gh_{G^{+} G^{-}}(p) &=& ( 1 - T^2_p)  \g_{G^{+} G^{-}}(p) 
  \,,
  \nonumber \\
  \gh_{c^{+} W^{*,-}_{\nu}}(p) &=& ( 1 - T^1_p)  \g_{c^{+}
    W^{*,-}_{\nu}}(p) \,\,=\,\, 
  -i p_\mu ( 1 - T^0_p)  J(p^2)
  \,\,=\,\, -i p_\mu \hat{J}(p^2)
  \,,
  \nonumber \\
  \gh_{c^{+} G^{*,-}}(p) &=& ( 1 - T^1_p)  \g_{c^{+} G^{*,-}}(p) \,\,=\,\, 
  ( 1 - T^0_p) I(p^2)
  \,\,=\,\, \hat{I}(p^2)
  \,,
\end{eqnarray}
with 
$\g_{c^{+} W^{*,-}_{\nu}}(p)=-i p_\nu J(p^2)$ and
$\g_{c^{+} G^{*,-}}(p)=I(p^2)$.
The UV index is set equal to the superficial UV degree. 

In a next step we can compute 
the commutator  between the Slavnov-Taylor operator and the Taylor
subtraction. This leads us to the universal breaking terms
(cf. Eqs.~(\ref{rea}) and~(\ref{newterms}))
\begin{eqnarray}\label{ep_2}
  \Psi^{(1)}_{c^+W^-_\mu} &=& i p_\mu \left[ 
    \gh^{L,(1)}_{W^+W^-}(p^2) - i M_W \gh^{(1)}_{G^+W^-}(p^2) - 
  i M_W^2 \hat{J}^{(1)}(p^2) - M_W \hat{I}^{(1)}(p^2) \right]
  \nonumber\\
  &=& ip_\mu p^2 \left[ 
  - i M_W  \frac{\partial}{\partial p^2} \g^{(1)}_{G^+W^-}(0) 
  - M_W I^{(1)\prime}(0) 
  - i M^2_W J^{(1)\prime}(0)  
  \right] 
  \,,
  \\
  \Psi^{(1)}_{c^+G^-} &=&
  p^2 \gh^{(1)}_{W^+G^-}(p^2) + i M_W \gh^{(1)}_{G^+G^-}(p^2) - p^2 M_W
  \hat{J}^{(1)}(p^2) + i\, p^2 \hat{I}^{(1)}(p^2)  \,\,=\,\, 0
  \nonumber
  \,.   
\end{eqnarray}
where $\frac{\partial}{\partial p^2} \g^{(1)}_{G^+W^-}(0) \equiv 
\left. \frac{\partial}{\partial p^2} \g^{(1)}_{G^+W^-}(p^2)\right|_{p=0}$.  
$\Gamma^{L,(1)}_{W^+W^-}$ denotes the longitudinal part
of the $W$ boson two-point function defined through
\begin{eqnarray}
  \Gamma_{W^+_\mu W^-_\nu}(p) &=&
  \left(g_{\mu\nu}-\frac{p_\mu p_\nu}{p^2}\right) \Gamma^T_{W^+W^-}(p^2)
  + \frac{p_\mu p_\nu}{p^2} \Gamma^L_{W^+W^-}(p^2)
  \,.
  \nonumber
\end{eqnarray}
In the second equation of~(\ref{ep_2}) there is no breaking term at all
while in the first one only the term proportional to $p^2 p_\mu$
survives the Taylor operation. 
It can be re-absorbed by introducing the counterterm 
\begin{eqnarray}\label{ep_3}
  \Xi^{(1)} &=&
  {M_W}{\left( - \frac{\partial}{\partial p^2} \g^{(1)}_{G^+W^-}(0) 
      + i I^{(1)\prime}(0)
      - M_W J^{(1)\prime}(0)\right)}
  \int {\rm d}^4x \partial^\mu W^{+}_{\mu} \partial^\nu W^{-}_{\nu}
  \,,
\end{eqnarray}
which is manifestly universal because it is computed in terms of finite Green
functions which do not require any regularization procedure.
This is the major difference between
the direct computation of the breaking terms 
and the one discussed here.
In the former case the complete calculation of the
coefficients $\alpha_1$, $\alpha_2$, $\alpha_3$ and $\alpha_4$
have to be performed.
On the contrary, in our approach this is in part automatically
taken into account by the subtraction procedure. In the present example,
we only have to compute the residual coefficient $\alpha_1$
which arises in Green functions involving ghost
fields and $W$--$G$ mixing.
The other breaking terms of Eq.~(\ref{exa_2}) are automatically removed by
the specific choice of renormalization conditions given by the Taylor
subtraction in~(\ref{ep_1}). In other words  
it is possible to remove the breaking terms
$\alp_{i}$ ($i=2,3,4$) by invariant counterterms.

Note that in Eq.~(\ref{ep_3}) only the first term in the
bracket gives a contribution
to the fermionic part since at one-loop order
the ghost two-point functions 
do not receive any contribution from fermionic fields. 

Before we discuss the additional simplifications of our 
method in connection with the BFM, we mention that our specific procedure 
allows us to work out the corresponding formulae at the $n$-loop level 
explicitly:
\begin{eqnarray}\label{ep_4.0}
\Psi^{(n)}_{c^+W_\mu^-}  &=& ip_\mu\Bigg\{
  \gh^{L,(n)}_{W^+W^-}(p^2) - iM_W \gh^{(n)}_{G^+W^-}(p^2) - 
  i M_W^2 \hat{J}^{(n)}(p^2) - M_W \hat{I}^{(n)}(p^2) 
  \nonumber\\
  &&
  - \sum^{n-1}_{m=1} \left[ {\cal J}^{(m)}(p^2) \gg^{L,(n-m)}_{W^+W^-}(p^2) 
    + {\cal I}^{(m)}(p^2)  \gg^{(n-m)}_{G^+W^-}(p^2) \right]
  \Bigg\}
  \nonumber \\
  &=& ip_\mu\Bigg\{
  p^2 \left( -iM_W\frac{\partial}{\partial p^2} \g^{(n)}_{G^+W^-}(0) 
    - M_W I^{(n)\prime}(0) -
    i M_W^2 J^{(n)\prime}(0) 
  \right) 
  \nonumber \\
  && \mbox{} 
  -  p^2  
  \sum^{n-1}_{m=1} 
  \left[ 
      {\cal J}^{(m)\prime}(0) \gg^{L,(n-m)}_{W^+W^-}(0) + 
      {\cal J}^{(m)}(0) \frac{\partial}{\partial p^2} 
      \gg^{L,(n-m)}_{W^+W^-}(0) 
    \right. 
  \nonumber \\
  &&
  \left. \mbox{} \qquad
      +{\cal I}^{(m)\prime}(0)  \gg^{(n-m)}_{G^+W^-}(0) + 
      {\cal I}^{(m)}(0) \frac{\partial}{\partial p^2} \gg^{(n-m)}_{G^+W^-}(0)
  \right]
  \nonumber \\
  && \mbox{} 
  - \sum^{n-1}_{m=1} 
  \left[ 
    {\cal J}^{(m)}(0) \gg^{L,(n-m)}_{W^+W^-}(0) + 
    {\cal I}^{(m)}(0)  \gg^{(n-m)}_{G^+W^-}(0)  
  \right]
  \Bigg\}
  \,,
  \nonumber \\
  \Psi^{(n)}_{c^+G^-}
  &=&
  p^2 \gh^{(n)}_{W^+G^-}(p^2) +i M_W \gh^{(n)}_{G^+G^-}(p^2) - p^2 M_W 
  \hat{J}^{(n)}(p^2) +i p^2 \hat{I}^{(n)}(p^2) 
  \nonumber \\
  &&
  + \sum^{n-1}_{m=1} \left[ 
    -p^2{\cal J}^{(m)}(p^2) \gg^{(n-m)}_{W^+G^-}(p^2) + 
    {\cal I}^{(m)}(p^2)  \gg^{(n-m)}_{G^+G^-}(p)
  \right]
  \nonumber \\
  &=&
  \sum^{n-1}_{m=1} 
  \left[ 
      -p^2{\cal J}^{(m)}(0) \gg^{(n-m)}_{W^+G^-}(0) 
      + {\cal I}^{(m)}(0)  \gg^{(n-m)}_{G^+G^-}(0) 
    \right. 
  \nonumber \\&&
  \left. \mbox{} \qquad
      + p^2{\cal I}^{(m)\prime}(0)  \gg^{(n-m)}_{G^+G^-}(0)
      + p^2{\cal I}^{(m)}(0) \frac{\partial}{\partial p^2}
      \gg^{(n-m)}_{G^+G^-}(0) 
  \right]
  \,,
\end{eqnarray}
where 
$- ip_\mu {\cal J}^{(k)}(p^2) = \gg^{(k)}_{c^+ W^{*,-}_\mu}(p)$
and 
${\cal I}^{(k)}(p^2) = \gg^{(k)}_{c^+ G^{*,-}}(p)$
($k=1,\ldots, n-1$)
represent the renormalized symmetric counterparts of
$\g^{(k)}_{c^{+} W^{*,-}_{\nu}}(p)=-i p_\nu J^{(k)}(p^2)$ and
$\g^{(k)}_{c^{+} G^{*,-}}(p)=I^{(k)}(p^2)$.
Already at two-loop order
the complete set of counterterms is needed.
They are, however, expressed in terms of one-loop 
Green functions and only for $\alpha_1^{(2)}$ a real two-loop 
calculation is needed.
The closed formulae for the coefficients of the
breaking terms, $\alpha_1^{(n)}$, $\alpha_2^{(n)}$, $\alpha_3^{(n)}$ and
$\alpha_4^{(n)}$, where the superscript ``$(n)$''
reminds on the number of loops,
allow us to adopt special renormalization conditions which
simplify the expressions.
The choice
\begin{equation}
  \label{ep_5}
  {\cal J}^{(0)}(0) = 1,\quad {\cal I}^{(0)}(0) = M_W,\quad
  {\cal J}^{(k)}(0) = 0,\quad {\cal I}^{(k)}(0) = 0,\quad
  k\ge1
  \,,
\end{equation}
in combination with the adjustment of the wave function
renormalization of the ghost and  Goldstone field,
$Z_c^{\pm}$ and $Z_G^{\pm}$, reduces the breaking terms to
\begin{eqnarray}\label{ep_4}
  \Psi^{(n)}_{c^+W_\mu^-}
  &=& ip_\mu\Bigg\{
  p^2 
  \left( -iM_W\frac{\partial}{\partial p^2} \g^{(n)}_{G^+W^-}(0) 
    - M_W I^{(n)\prime}(0) 
    - i M_W^2 J^{(n)\prime}(0)\right)  
  \nonumber \\
  &&\mbox{}
  -  p^2  
  \sum^{n-1}_{m=1} 
  \left[ 
      {\cal J}^{(m)'}(0) \gg^{L,(n-m)}_{W^+W^-}(0)
      +  {\cal I}^{(m)'}(0)  \gg^{(n-m)}_{G^+W^-}(0)  
  \right]
  \Bigg\}
  \,,
  \nonumber \\
  \Psi_{c^+G^-}^{(n)}
  &=&\mbox{}
  p^2  
  \sum^{n-1}_{m=1} 
      {\cal I}^{(m)'}(0)  \gg^{(n-m)}_{G^+G^-}(0)
  \,,
\end{eqnarray}
where the coefficient $\alpha_2^{(n)}$ and $\alpha_4^{(n)}$ are zero to all
orders. Therefore, only the counterterms
\begin{equation}
  \label{ep_7}
  \Xi^{(n)}
  = - \int {\rm d}^4x\left( {\alpha_1^{(n)}}  \partial^\mu W^{+}_{\mu}
   \partial^\nu W^{-}_{\nu} +  {\alpha_3^{(n)}}  \partial_\mu G^{+}
  \partial^\mu G^{-} \right)
  \,,
\end{equation}
are needed to restore the symmetry.
The appearance of the lower order terms is a special feature of the
STI. In fact this is strictly related to the renormalization of the
BRST transformations at the quantum level.

Finally, we show that the BFM in combination 
with our subtraction procedure provides a further simplification in the 
S-matrix calculation.
Instead of using the conventional approach, the S-matrix
elements can also be computed by the BFM.
In that case the above identities are replaced by 
\begin{eqnarray}\label{ep_8}
  \hat{\Psi}^{(1)}_{\lambda_+W_\mu^-} 
  &=&ip_\mu\left(  \gh^{L,(1)}_{\hat{W}^+\hat{W}^-}(p) 
    - i M_W \gh^{(1)}_{\hat{G}^+ \hat{W}^-}(p)\right) \,\,=\,\,
  ip_\mu\left(
    - i M_W p^2 \frac{\partial}{\partial p^2} \g^{(1)}_{\hat{G}^+\hat{W}^-} 
  \right)
  \,,
  \nonumber \\
  \hat{\Psi}^{(1)}_{\lambda_+G^-} &=&
  p^2 \gh^{(1)}_{\hat{W}^+\hat{G}^-}(p) 
  + i M_W \gh^{(1)}_{\hat{G}^+\hat{G}^-}(p) \,\,=\,\,0
  \,,
\end{eqnarray}
where $\hat{\Psi}^{(1)}_{\lambda_+W_\mu^-}$ and
$\hat{\Psi}^{(1)}_{\lambda_+G^-}$
are the universal breaking terms of the WTIs~(\ref{bfm_2})
and~(\ref{bfm_3.0}), respectively.
We stress that the same WTIs hold to all orders without any changes.
This is due to the linearity of the identities. No ghost fields occur in these 
equations.
As a consequence,  only the counterterm
\begin{equation}\label{ep_9}
  \Xi^{(1)} =
  { M_W\left( - \frac{\partial}{\partial p^2} \g^{(1)}_{\hat{G}^+\hat{W}^-}(0)
  \right)} 
  \int {\rm d}^4x  \partial^\mu \hat{W}^{+}_{\mu} \partial^\nu
  \hat{W}^{-}_{\nu}  
  \,,
\end{equation}
has to be introduced in order
to restore the background gauge symmetry. We mention that the corresponding
counterterm at the order $n$ is given by the same formula. 

At this point we again want to underline the
important difference between the BFM approach and the conventional
one. In the former case, the computation of the amplitude can be
performed by using Green functions with external background fields
instead of gauge bosons and scalars. For these Green
functions only WTIs are needed to define their finite parts
correctly. However, at higher orders also quantum two-point functions
appear as sub-diagrams. This implies that
STIs have to be employed at lower orders.
By using the method
exposed here we can immediately see that for one-loop STIs already 
the zero momentum 
subtraction partially restores the identities while
only few additional counterterms are needed.
Furthermore the restoration of the WTIs is simple anyway.
The method of subtractions
provides a very powerful technique to renormalize a model in a
non-invariant regularization scheme, especially if in addition the
BFM is adopted.

In the forthcoming sections, we will see that in the case of
an external photon field (as in~\bsg~and~\hgg) there are further  
simplifications due to $U(1)$ group of the SM.

%%%%%%%%%%%%%%%%%%%%%%%%%%%%%%%%%%%%%%%%%%%%%%%%%%%%%%%%%%%%

\setcounter{equation}{0}
\section{Two-loop QCD corrections to \hgg} 
\label{sec:hgg_coun}

In the minimal version of the SM the only not yet discovered particle
is the Higgs boson. Up to now only limits on its mass exist where the
lower one is provided by  the direct search at LEP.
The upper limit is obtained from the combination of precision
measurements and quantum corrections.
A Higgs boson in the intermediate mass range,
i.e. $M_W<M_H<2M_W$, is very promising. 
Then the loop-induced decay into two photons is one possibility for
its detection. Next to the diagram involving gauge bosons also the
one where the Higgs boson couples to top quarks accompanied with
additional QCD corrections is important.
In the latter case an expansion for a heavy top quark mass 
is sufficient in order to obtain a sensible approximations to the exact
result. In the following we will exemplify our approach to the
algebraic renormalization developed in the previous sections
in this limit. The calculations are performed in the framework of
Analytical Regularization which will be introduced in the first subsection.

\subsection{Analytical Regularization}
\label{sub:ana_reg}

Although quite a lot of different regularization prescriptions have
been developed since the invention of quantum field theory
by far most practical applications have been performed in the framework of
Dimensional Regularization~\cite{Hoo,dim_reg}.
Meanwhile quite some technology is available which allows the computations 
of rather complicated Feynman diagrams sometimes even at the four-loop level.
One of the main reasons for this development is that most of the
symmetries are preserved. Furthermore, very often the occurring
$d$ dimensional integrals are even simpler to evaluate than the 
four dimensional ones.

A regularization method which was invented even before the advent of the
dimensional one is called Analytical Regularization
(for the practical details we refer to~\cite{Spe}). In the
dimensional method the space-time over which the integration is
performed is changed from four to a complex number, $d$, and the limit
$d\to4$ is taken at the very end. The divergences then appear as poles
in $1/(d-4)$.
On the contrary in the case of Analytical Regularization the
space-time is kept fixed and only the exponents of the denominators
are modified.
One of the big advantages of this approach is that no problems in
connection with $\gamma_5$ (or, more generally, the $\epsilon$ tensor)
appears.
The integrals which have to be solved can be  slightly more difficult.
It is, e.g., not possible to perform a partial fractioning as
the exponents are no integer numbers any longer. For some classes of
diagrams it is nevertheless possible to take over the
results obtained in Dimensional Regularization. As an example we
consider the massless scalar one-loop integral. For $d\not=4$ and
arbitrary exponents the result reads:
\begin{eqnarray}
  \lefteqn{\frac{1}{i}\int {{\rm d}^dq\over (2\pi)^d} 
  {1\over  ((q-p)^2 + i \eta )^a 
    (q^2 + i \eta )^b} \,\,=\,\,}
  \nonumber\\&&\mbox{}
  { (-p^2)^{d/2-a-b}\over (4\pi)^{d/2} (-1)^{-a-b}}
  \frac{\Gamma(a+b-d/2)\Gamma(d/2-a)\Gamma(d/2-b)}
  {\Gamma(d-a-b)\Gamma(a)\Gamma(b)}
  \,.
  \label{eq:ml1loop}
\end{eqnarray}
This result can immediately be interpreted
in the framework of Analytical Regularization
after choosing $d=4$
and $a=a_1+\lambda$ and $b=b_1+\mu$ where $a_1$ and $b_1$ are
integers and $\lambda$ and $\mu$ adopt the role of the regulator.
Possible divergences appear in the form $1/\lambda$, $1/\mu$ and
$1/(\lambda+\mu)$ after
the r.h.s. of Eq.~(\ref{eq:ml1loop}) is expanded in a Laurent series. This
also leads to logarithms with dimensional arguments. Therefore, one has to
introduce an arbitrary mass scale, $\mu$, such that this is corrected for.

Dealing with different regulators for different denominators
would be  in general quite tedious. In our case the
structure of the  divergences of the diagrams allows us to choose non-integer
exponents only for the gluon line --- the fermion lines have not to be 
regulated at all~\cite{breite}.

%%%%%%%%%%%%%%%%%%%%%%%%%%%%%%%%%%%%%%%%%%%%%%%%%%%%%%%%%%%%

\subsection{Breaking terms for \hgg}
\label{sub:break_hgg}

According to Section~\ref{sub:hgg} we have to discuss the structure of
the breaking terms of the STIs for the amplitude~(\ref{ide.hgg.6}) and the
sub-divergences~(\ref{wti.hgg.8}). Recall that we are only  
interested in the two-loop QCD corrections to the fermionic
contributions $\gg^{(2), QCD}_{H A_\mu A_\nu}$.    

The most general Lorentz invariant breaking terms corresponding
to~(\ref{wti.hgg.8}) and~(\ref{ide.hgg.4}) are given by
\begin{eqnarray}
  \label{ide.hgg.4.1} 
  \Delta^{(1)}_{c_A \bar{q}_i q_i} (\bar{p},p) &=&  
  \delta^{(1)}_3 \left( \not\!\bar{p} - \not\!{p} \right) +  \delta^{(1)}_4
  \,,   
  \nonumber\\
  \Delta_{c_AA_\mu H} (q_2,p) &=& 
    \left(\delta_1^{(1)}+\delta_1^{(2)}\right)q_2^\mu
    +\left(\delta_2^{(1)}+\delta_2^{(2)}\right)p^\mu
  \,.
\end{eqnarray} 
where $\delta^{(i)}_{1}$ and $\delta^{(i)}_{2}$ are the  
one- and two-loop coefficients needed for the vertex contribution 
$\gg^{(i)}_{A_\mu A_\nu H}$ $(i=1,2)$.  
$\delta^{(1)}_{3}$ and $\delta^{(1)}_{4}$ are the sub-leading breaking 
terms for the sub-graph $\gg^{(1),QCD}_{A_\mu \bar{q} q}(\bar{p},p)$. 

The breaking terms are removed by 
expanding the Green function
$\g^{(i)}_{H A_\rho A_\nu}(q_1,q_2)$ w.r.t. its external  
momenta and discarding the first non trivial contribution as  the
operators $(1-T^{1}_{q_1,q_2})$ have to be applied.
Here $T^{1}_{q_1,q_2}$ denotes the Taylor operator up to first order
in $q_1$ and $q_2$.

For dimensional reasons the constant term $\delta^{(1)}_4$ has to be zero.
The computation of $\delta^{(1)}_3$ is discussed in the 
forthcoming subsection.
Again it is enough to apply the operators
$(1-T^0_{\bar{p}p})$ and $(1-T^1_{p})$ to
the corresponding Green functions,  
$\g^{(1),QCD}_{A_\mu \bar{q}_i q_i}(\bar{p},p)$  and  
$\g^{(1),QCD}_{\bar{q}_i q_i}(p)$, respectively.
This corresponds to adjust the free 
parameter $Z_1$ in Eq.~(\ref{lag.hgg.3}).

%%%%%%%%%%%%%%%%%%%%%%%%%%%%%%%%%%%%%%%%%%%%%%%%%%%%%%%%%%%%

\subsection{Results}
\label{sub:results}

Since we restrict ourselves to two-loop QCD corrections, 
only diagrams involving the top quark have to be taken into account.
Some sample diagrams are shown in Fig.~\ref{fig:hgg}. 
For our purpose it is enough to consider the limit where the top quark
mass is much larger than the Higgs boson mass. This allows us to perform an
expansion of the vertex diagrams in the external momenta and thus reduce the
integrals to be solved to vacuum graphs.
We note that the result is known in Dimensional
Regularization~\cite{Shifman:1979eb,Djouadi:1991tk}
which allows us for a comparison at the end.

It is convenient to split the contributions to the $H\gamma\gamma$ vertex
according to the loop order in the following way 
\begin{eqnarray}
\Gamma_{H A^{\mu} A^{\nu}}(q_1,q_2)
&=&
\Gamma_{H A^{\mu} A^{\nu}}^{(1)}(q_1,q_2)
+ \frac{\alpha_s}{\pi}C_F\Gamma_{H A^{\mu} A^{\nu}}^{(2)}(q_1,q_2)
+\ldots
\,,
\end{eqnarray}
where the momentum and polarization vector of the photons
is given by $(q_1,\epsilon^\mu)$ and $(q_2,\epsilon^\nu)$, respectively.
The general Lorentz decomposition of the vertex diagrams 
is given by\footnote{In the limit of on-shell photons there is no contribution
  from $C$ to the decay rate.}:
\begin{eqnarray}
\Gamma_{H A^{\mu} A^{\nu}}^{(i)}(q_1,q_2)
&=&
g^{\mu\nu}\,q_1.q_2\,A^{(i)}(q_1,q_2)
+q_1^\nu q_2^\mu   B^{(i)}(q_1,q_2)
+q_1^\mu q_2^\nu   C^{(i)}(q_1,q_2)
\,.
\end{eqnarray}
Note that in regularization schemes where gauge invariance is preserved ---
like Dimensional Regularization in the absence of chiral fermions --- one
has according to~(\ref{ide.hgg.6})
$A=-B$.
In the case of Analytical Regularization this is not true
and some breaking terms occur. 
However, it turns out that these breaking terms can be
removed by Taylor subtraction. No new universal breaking terms 
have to be introduced as in the case of over-subtraction. 
Therefore the subtracted Green function (using Analytic Regularization) 
fulfills Eq.~(\ref{ide.hgg.6}) at the one- and two-loop level
\begin{eqnarray}
- i q_1^{\mu} 
\left(1-T_{q_1,q_2}^0\right)\Gamma^{(i)}_{H A_\mu A_\nu}(q_1,q_2) 
&=& 0,~~~~i=1,2
\,
\label{eq:hggsti}
\end{eqnarray}
where $q_1$ and $q_2$ are the momenta of the photons.
This relation  can also be rewritten in terms of the functions $A$, $B$ and
$C$ with accordingly adjusted Taylor operators.

Let us now consider the one-loop calculation, which constitutes the Born
result, in more detail. 
The evaluation of first three terms in the expansion for large top quark mass
leads to: 
\begin{eqnarray}
A^{(1)} &=& - B^{(1)} + \hat{A}\frac{3}{4\bar\tau}
\nonumber\\
&=& 
\hat{A}\left(
\frac{3}{4\bar\tau}
+1
+\frac{7}{30}\bar\tau
+\frac{2}{21}\bar\tau^2
+ {\cal O}(\bar\tau^3)
\right)
\,,
\label{eq:1loop}
\end{eqnarray}
with $\hat{A}=N_cQ_t^2 2\alpha/(3\pi v)$ and $\tau=2q_1q_2/(4m_t^2)$.
$m_t$ is the top quark mass in the $\overline{\rm MS}$ scheme.
As Eq.~(\ref{eq:1loop})
constitutes the lowest order, $m_t$ may as well be replaced by the on-shell
mass, $M_t$.
Clearly the application of the operator $(1-T_{q_1,q_2}^{1})$ to
$q_1.q_2A^{(1)}$  
removes the breaking term proportional to $1/\bar\tau$
and leads to the invariant Green function $\gg^{(1)}_{H A_\mu A_\nu}$.

At two-loop order the situation is different.
Here, the divergent sub-diagrams have to be carefully
investigated in a first step 
in order to fix the renormalization constants $Z_1$, $Z_2$, $Z_m$
and $Z_Y$ defined in Eq.~(\ref{lag.hgg.3}).
As described in Section~\ref{sub:hgg}, 
it is possible to fix  $Z_2$ and $Z_m$ in the 
${\overline{\rm MS}}$ scheme and to choose
$Z_Y=Z_m$.
From the examination of the fermion propagator we get
\begin{eqnarray}
Z_2 &=& 1 + \frac{\alpha_s}{\pi}C_F\frac{-2+\xi}{8\lambda}
\,,
\nonumber\\
Z_m &=& Z_Y \,\,=\,\, 1 +
\frac{\alpha_s}{\pi}C_F\left(-\frac{3}{4\lambda}\right) 
\,.
\end{eqnarray}
Here, $\xi$ is the gauge parameter appearing in the gluon propagator defined
in the conventional way  $i(-g^{\mu\nu}+\xi p^\mu p^\nu/p^2)/(p^2+i\eta)$.
Thus we are essentially left with $Z_1$.
Its divergent and also finite part is fixed by the QED-WTI given
in Eq.~(\ref{wti.hgg.8}).
We again use Taylor expansion to get
\begin{eqnarray}
  i (\bar{p} + p)^{\mu}\left(1-T_{\bar{p}, p}^0\right)
  \Gamma^{(1),QCD}_{A_\mu \bar{q}q}(\bar{p},p)   
  + i e Q_q 
  \left(1-T_{\bar{p}, p}^1\right) \left[ 
    \Gamma^{(1),QCD}_{\bar{q}q}(p) 
    - \Gamma^{(1),QCD}_{\bar{q}q}(-\bar{p}) 
  \right]
\nonumber\\
  \,\,=\,\, 0
\,,
\end{eqnarray}
where $p$ and $\bar{p}$ in the three-point function denote the momenta of the
quark and anti-quark, respectively.
Note that also here no over-subtraction is introduced.
The Taylor coefficient $T_{\bar{p},p}^0
  \Gamma^{(1),QCD}_{A_\mu \bar{q}q}(\bar{p},p)$ fixes $Z_1$.
It turns out that the finite part of $Z_1$ is zero and the divergent part
fulfills the WTI. This means
that the Green functions obtained in Analytic Regularization
fulfill the WTI even without Taylor subtraction.
Thus we have  
\begin{eqnarray}
Z_1 &=& 1 + \frac{\alpha_s}{\pi}C_F\frac{-2+\xi}{8\lambda}
\,.
\end{eqnarray}

At this point it is convenient to introduce the notation:
\begin{eqnarray}
q_1.q_2 A^{(i),t} &=& \left(1-T_{q_1,q_2}^0\right) q_1.q_2A^{(i)}
\,,
\end{eqnarray}
and in analogy for the functions $B$ and $C$.
Then the renormalized function $A$ up to
${\cal O}(\alpha\alpha_s)$ can be written in the form 
\begin{eqnarray}
A(q_1,q_2) 
&=& A^{(1),t}\bigg|_{m_t^0=Z_m m_t} \left(1+\delta Z_2
  +2\delta Z_1-3\delta Z_2\right)
+ \frac{\alpha_s}{\pi}C_F A^{(2),t}
\nonumber\\
&=& A^{(1),t}\bigg|_{m_t^0=Z_m m_t}
+ \frac{\alpha_s}{\pi}C_F A^{(2),t}
\,,
\end{eqnarray}
where $m_t^0$ is the bare top mass and the condition $Z_Y=Z_m$ has been
used. In the first line the first
occurrence of $\delta Z_2$ origins from the Higgs-quark vertex. Furthermore,
all 
three quark propagators deliver a $\delta Z_2$ and each photon-quark vertex
leads to a $\delta Z_1$. After the second equality sign $\delta Z_1=\delta
Z_2$ has been 
used.

In Analytical Regularization the two-loop result reads in the
$\overline{\rm MS}$ scheme:
\begin{eqnarray}
A^{(2),t} &=& \hat{A}\left[-\frac{3}{4}
+\left( \frac{181}{360} -\frac{7}{20}\ln\frac{\mu^2}{m_t^2}
 \right)\bar{\tau}
+\left( \frac{1541}{4725} -\frac{2}{7}\ln\frac{\mu^2}{m_t^2}
 \right)\bar{\tau}^2
+ {\cal O}(\bar\tau^3)
\right]
\,.
\label{eq:2loopms}
\end{eqnarray}

In order to compare with the results known from Dimensional Regularization
physical conditions have to be imposed. We choose to express the result in
terms of the on-shell mass, $M_t$. It is defined through the zero of the
inverse 
propagator where the external momentum is on the mass shell.
In Analytical Regularization the transformation from the $\overline{\rm MS}$
mass to the on-shell one reads:
\begin{eqnarray}
m_t &=& M_t \left[1+\frac{\alpha_s}{\pi} C_F \left( -\frac{3}{8} -
    \frac{3}{4}\ln\frac{\mu^2}{M_t^2} \right)\right]
\,.
\end{eqnarray}
The substitution of this equation in~(\ref{eq:1loop}) leads to
\begin{eqnarray}
A^{(2),t} &=& \hat{A}\left[-\frac{3}{4}
+ \frac{61}{90}     \tau
+ \frac{2216}{4725} \tau^2
+ {\cal O}(\tau^3)
\right]
\label{eq:2loopos}
\,.
\end{eqnarray}
Finally,  the decay rate is given by
\begin{eqnarray}
\Gamma(H\to\gamma\gamma) &=&
\left| A^{(1),t}+\frac{\alpha_s}{\pi}C_FA^{(2),t} \right|^2 
\frac{M_H^3}{64\pi}
\,,
\end{eqnarray}
where the functions $A^{(i)}$ have to be evaluated for $q_1.q_2=M_H^2/2$.
The decay rate $\Gamma(H\to\gamma\gamma)$ coincides with the result obtained
in Dimensional Regularization~\cite{Shifman:1979eb,Djouadi:1991tk}.

Although the result presented in this section is quite simple, the main  
steps of the Algebraic Renormalization have been touched. We have also seen 
that in this specific example our algebraic procedure is as efficient as 
Dimensional Regularization due to the linearity of the STI involved.

%%%%%%%%%%%%%%%%%%%%%%%%%%%%%%%%%%%%%%%%%%%%%%%%%%%%%%%%%%%%

\setcounter{equation}{0}
\section{Two-loop electroweak corrections to  \bsg} 
\label{sec:bsg_coun}

In this section, the electroweak two-loop corrections 
to $b \rightarrow s \gamma$ are discussed.
The inclusive mode of this rare decay is already measured and plays 
an important role in the search for physics beyond the SM
(see e.g.~\cite{Greub}). In the literature a partial
calculation of the two-loop electroweak corrections has been
performed~\cite{Czarnecki:1998tn}.
In the limit of a heavy top quark 
and/or a heavy  Higgs boson also a complete calculation
exists~\cite{strumia}. In all cases the naive dimensional 
scheme for $\gamma^5$ has been used.
 
In our analysis, we use the BFM in combination with our practical 
algebraic framework. All relevant WTIs and STIs are explicitly
derived in Section~\ref{sub:bsg}. 
Special care is taken in order to explicitly illustrate every step of 
our subtraction method presented in Section~\ref{sub:remove}.
A comparison with the conventional algebraic method is made.

In the following $\Delta^{{\cal W},(n)}$ and $\Delta^{{\cal S},(n)}$  
denote the $n$-loop breaking terms to the WTI and the
STI, respectively.
For the differentiations w.r.t. fields it is convenient to introduce the
notation:
\begin{eqnarray}
  \frac{\delta^{n} \Delta^{{\cal W/S},(n)}}{ \delta  \phi_1(p_1)\dots\delta
    \phi_n(p_n)}\bigg|_{\phi=0}
  &=& \Delta^{{\cal W/S},(n)}_{\phi_1\dots 
    \phi_n}(p_1, \dots, p_n)
\,.
\end{eqnarray}

%%%%%%%%%%%%%%%%%%%%%%%%%%%%%%%%%%%%%%%%%%%%%%%%%%%%%%%%%%%%

\subsection{Restoring the WTIs}
\label{sub:restWTI}

Let us in a first step consider the WTI of Eq.~(\ref{bsg1})
\begin{eqnarray}\label{bsg1.1}
  \lefteqn{ {{ \delta^3 {\cal W}_{(\lambda)}(\g)^{(2)} } \over \delta
       \lambda_{A}(-p_s-p_b) 
       \delta \bar{s}(p_s)  \delta b(p_b)}\bigg|_{\phi=0} 
    =}
  \nonumber \\ && \hspace{-2cm}
  i (p_{s} + p_{b})^{\mu}    
  \g^{(2)}_{\hat{A}_{\mu} \bar{s} b}(p_{s},p_{b}) + i e Q_{d} \left[    
    \g^{(2)}_{\bar{s} b}(p_{b})   
    - \g^{(2)}_{\bar{s} b}(-p_{s})  \right] \,\,=\,\,
  \Delta^{{\cal W},(2)}_{\lambda_A
    \bar{s} b}(p_s, p_b)
  \,.
\end{eqnarray}
The corresponding counterterms~(\ref{twocnt}) have already been 
presented in Section~\ref{sub:bsg}.
They are fixed by hermiticity\footnote{ 
  One can eventually choose non-hermitian currents. Then, however, the number
  of counterterms increases~\cite{ferrari}.}
and Lorentz invariance. Furthermore they are constrained by the
consistency conditions given in the Section~\ref{sub:consi}.
In order to impose them we follow the suggestion presented there.
Since we know that in the SM all breaking terms
can be removed by non-invariant counterterms or, equivalently, no anomalies
are present, 
we insert our general ansatz of Eq.~(\ref{twocnt}) into (\ref{bsg1.1}) 
and deduce the most general breaking terms compatible with the consistency
conditions.
At the two-loop level the most general form reads:
\begin{eqnarray}\label{break_WTI_2}
  \Delta^{{\cal W},(2)}_{\lambda_A \bar{s} b}(p_s, p_b) &=&  
  \alpha^{1,L}_{sb} P_L \not\!p_s
  + \alpha^{\prime,1,L}_{sb} P_L \not\!p_b
  + \alpha^{1,R}_{sb} P_R \not\!p_s 
  + \alpha^{\prime,1,R}_{sb} P_R \not\!p_b
  \nonumber\\&&\mbox{}
  + \beta^{1,L}_{sb} P_L + \beta^{1,R}_{sb} P_R 
  \,.
\end{eqnarray}
Actually the coefficients $\beta^{1,L/R}_{sb}$ are zero
and $\alpha^{\prime,1,L/R}_{sb}=-\alpha^{1,L/R}_{sb}$
as can be seen by using the consistency conditions in combination with
the invariance under charge conjugation.
$\alpha^{1,L}_{sb}$ and $\alpha^{1,R}_{sb}$ are fixed by~(\ref{bsg1.1}):
\begin{eqnarray}\label{2twoco}   
  \alpha^{1,L/R}_{sb} P_{R/L} &=& \frac{1}{4}    
  P_{R/L} \gamma^{\alp} \left( \frac{\partial}{\partial p^{\alp}_s}   
    \left. { \delta^3 {\cal W}_{(\lambda)}(\g)^{(2)} \over \delta
  \lambda_{A}(-p_s-p_b)    
         \delta \bar{s}(p_s) \delta b(p_b)}\right|_{\phi=0} \right)\Bigg|_{p_s
  = p_b =0} 
  \nonumber \\ 
  &=& i
  P_{R/L} \gamma^{\alp} \left( \g^{(2)}_{\hat{A}_{\alp} \bar{s} b}(0,0) - 
    e Q_{d}  \left. \frac{\partial}{\partial p^{\alp}_s}  \g^{(2)}_{
        \bar{s} b}(-p_{s})\right|_{p_s=0}  \right) 
  \,.
\end{eqnarray}   
This equation shows that at the two-loop level the corresponding finite
counterterms are identical to the first terms of the Taylor expansion of the
Green function.
Therefore, by using our subtraction method we  automatically restore ---  up to
sub-divergences --- the WTI at two loops.

Let us move to the discussion of sub-divergences.
In the conventional algebraic method one starts again with the 
analysis of the breaking terms to the various WTIs (\ref{bsg1}),
(\ref{wti.1})--(\ref{wti.4})  
in the same manner as above.
For example considering  Eq.~(\ref{wti.3}) one gets the following
general breaking term:
\begin{eqnarray}\label{break_WTI_0}
  \!\!\!\!\!
  \Delta^{{\cal W},(1)}_{\lambda_A W^+_\mu W^-_\nu}(p^+,p^-)&=& 
  \delta_1 \,  p^+_\mu p^-_\nu + 
  \delta_2 \,  p^-_\mu p^+_\nu + 
  \delta_3 \,  p^+_\mu p^+_\nu + 
  \delta_4 \,  p^-_\mu p^-_\nu + 
  \delta_5 \,  g_{\mu \nu} p^+ \cdot p^- +
  \nonumber \\ &&\mbox{} +
  \delta_6 \,  g_{\mu \nu} p^+ \cdot p^+ +
  \delta_7 \,  g_{\mu \nu} p^- \cdot p^- +
  \delta_8 \,  g_{\mu \nu} + 
  \delta_9 \,  \epsilon_{\mu\nu\rho\sigma}  p^{+,\rho} p^{-,\sigma}
  \,,
\end{eqnarray}
where $p^\pm$ are the momenta of the $W^\pm$ bosons.
From hermiticity one obtains $\delta_3=\delta_4$ and 
$\delta_6=\delta_7$.
Using the consistency conditions, i.e. inserting  
(\ref{ta:cteq}) into Eq.~(\ref{wti.3}),
leads to  $\delta_8=0$, $\delta_5+2 \delta_6=0$ and 
$\delta_1+\delta_2 + 2 \delta_3=0$. Therefore, 
the r.h.s. of Eq.~(\ref{break_WTI_0}) 
reduces to the following four independent structures 
\begin{eqnarray}\label{break_WTI_0.2}
  \Delta^{{\cal W},(1)}_{\lambda_A W^+_\mu W^-_\nu}(p^+,p^-)&=& 
  \alp_2 \left( p^+_\mu p^+_\nu - 2 p^+_\mu p^-_\nu + p^-_\mu p^-_\nu \right)
  +  
  \alp_3 \left( p^+_\mu p^+_\nu - 2 p^-_\mu p^+_\nu + p^-_\mu p^-_\nu \right)
  +  
  \nonumber \\ &&\mbox{}
  +
  \alp_4 g_{\mu\nu} \left( p^+_\rho  p^{+,\rho} - 2 p^+_\rho  p^{-,\rho} +
    p^-_\rho  p^{-,\rho} \right) +
  \alp_5 \epsilon_{\mu\nu\rho\sigma} p^{+,\rho} p^{-,\sigma} 
  \,.
\end{eqnarray}
The other breaking terms could be treated in analogy leading to the 
results
\begin{eqnarray}\label{break_WTI_1}
  \Delta^{{\cal W},(1)}_{\lambda_A W^+_\mu G^- }(p^+,p^-) &=& 
  \alp_6 \left( p^+_\mu - p^-_\mu\right)
  \,,
  \nonumber \\
  \Delta^{{\cal W},(1)}_{\lambda_A G^+ W^-_\mu}(p^+,p^-) &=& 
  \alp_7 \left( p^+_\mu  - p^-_\mu \right)
  \,,
  \nonumber\\
  \Delta^{{\cal W},(1)}_{\lambda_A G^+ G^-}(p^+,p^-) &=& 
  \alp_8   \left( p^+_\rho  p^{+,\rho} - 2  p^+_\rho  p^{-,\rho} + p^-_\rho
    p^{-,\rho} \right) 
  \,,
  \nonumber \\
  \Delta^{{\cal W},(1)}_{\lambda_A \bar{u}_i u_j}(p_i, p_j) &=& 
  \alpha^{L}_{9,ij} P_L \left( \not\!p_i - \not\!p_j \right) + 
  \alpha^{R}_{9,ij} P_R \left( \not\!p_i - \not\!p_j \right) 
  \,,
  \nonumber \\
  \Delta^{{\cal W},(1)}_{\lambda_A \bar{d}_i d_j}(p_i, p_j) &=&  
  \alpha^{L}_{10,ij} P_L \left( \not\!p_i - \not\!p_j \right) + 
  \alpha^{R}_{10,ij} P_R \left( \not\!p_i - \not\!p_j \right) 
  \,,
\end{eqnarray}
where $\alpha^{9,L}_{ij},\dots, \alpha^{10,R}_{ij}$ are matrices in
the flavour space for $u$-, respectively, $d$-type quarks.

Clearly, the calculation of all the coefficients  in  
Eqs.~(\ref{break_WTI_0.2}) and~(\ref{break_WTI_1})
involves all
Green functions of the various WTIs and is quite tedious.
However, a closer look to  these equations shows that
already a simple Taylor expansion of the Green functions
can remove all possible breaking terms. 
Therefore, the Taylor subtracted Green functions (up to the order of the 
superficial degree) 
\begin{eqnarray}\label{eq:sub_1}
\gh^{(1)}_{\hat{A}_{\mu} b \bar{s}}(p_{b},p_{s}) &=& 
(1 - T^0_{p_b,p_s}) \g^{(1)}_{\hat{A}_{\mu} b \bar{s}}(p_{b},p_{s}) 
\,,
 \nonumber \\  
\gh^{(1)}_{\hat{A}_{\mu} W^{+}_{\nu} W^{-}_{\rho}} (p^+,p^-) &=&  
(1 - T^1_{p^+,p^-}) \g^{(1)}_{\hat{A}_{\mu} W^{+}_{\nu} W^{-}_{\rho}}
(p^+,p^-) 
\,,
\end{eqnarray}
and analogously for the functions
$\gh^{(1)}_{\hat{A}_{\mu} G^{+} G^{-}} (p^+,p^-)$,
$\gh^{(1)}_{\hat{A}_{\mu} G^{+} W^{-}_\rho} (p^+,p^-)$,
$\gh^{(1)}_{\hat{A}_{\mu} W^{+}_{\rho} G^{-}} (p^+,p^-)$,
$\gh^{(1)}_{\hat{A}_{\mu} c^{+} W^{*,-}_\rho} (p^+,p^-)$,
$\gh^{(1)}_{b \bar{s}}(p_{b},p_{s})$,
$\gh^{(1)}_{u_{i} \bar{u}_{j}}(p_{i},p_{j})$
and
$\gh^{(1)}_{d_{i} \bar{d}_{j}}(p_{i},p_{j})$
automatically satisfy the WTIs.
Thus, they coincide with the
symmetric Green functions denoted by $\gg^{(1)}_{\hat{A}_{\mu} 
  b \bar{s}}(p_{b},p_{s}),
\dots$, $\gg^{(1)}_{d_{i} \bar{d}_{j}}(p_{i},p_{j})$. 

At this point we mention again the subtleties involved in this procedure:  
if the degree of the Taylor expansion (necessary to remove the breaking terms) 
is larger than the  superficial divergence of the corresponding Green
function, local over-subtraction will produce new breaking terms to the WTI. 
Due to the fact that the WTIs~(\ref{bsg1}), (\ref{wti.1})--(\ref{wti.4}) 
do not explicitly depend on mass terms, no over-subtraction is introduced in
this specific case. However,  we will encounter this
problem in the next subsection.

The final result of the above analysis is 
that our prescription in combination with the BFM is a very powerful technique
to get rid of the breaking terms. In
particular no breaking terms for the WTIs must be computed and only an
overall subtraction has to be performed.
Regarding the WTIs with an external photon our procedure is 
as efficient as an invariant scheme since in this case simple
identities (cf. Eqs.~(\ref{bsg1.1}) and~(\ref{ide.hgg.4.1})),
like in QED, are obtained.
However, we stress that the finite counterterms corresponding to the first
terms of the Taylor expansion have to be taken into account
at higher orders, i.e. in our case the finite one-loop counterterms
enter the two-loop calculation.

%%%%%%%%%%%%%%%%%%%%%%%%%%%%%%%%%%%%%%%%%%%%%%%%%%%%%%%%%%%%

\subsection{Restoring the STIs}
\label{sub:restSTI}

Note that the STIs for the two-point functions (cf. Eq.~(\ref{exap})) have
already been discussed in Section~\ref{sub:remove} within our technique 
(see below Eq.~(\ref{ep_1})).
The universal counterterms in the latter approach are given in
Eq.~(\ref{ep_3}). 
Thus, in this section we only have to consider
the remaining identity of Eqs.~(\ref{exap_4}).

In the conventional algebraic method, one would analyze 
the most general breaking terms,
$\Delta^{{\cal S},(1)}_{c^{+}\bar{q}_i b}(p_q,p_b)$ and
$\Delta^{{\cal S},(1)}_{c^{-} \bar{s} q_i}(p_s,p_q)$, 
of these
STI (\ref{exap_4}) and the corresponding one obtained by the obvious
replacements.
Then one has to demonstrate how they  are restricted 
on the basis of the consistency conditions 
by inserting the most general counterterms into
the STIs. One could calculate these breaking terms 
in terms of the Green functions appearing in~(\ref{exap_4})
by projecting out the different
momentum structures of the breaking terms.

However, the latter information is not needed if again our
Taylor subtraction method is applied.
Let us in the following explicitly consider 
$\Delta^{{\cal S},(1)}_{c^{+}\bar{q}_i b}(p_q,p_b)$.
Similar equations hold for
$\Delta^{{\cal S},(1)}_{c^{-} \bar{s} q_i}(p_s,p_q)$.
The family index  is suppressed whenever there is no
source of misunderstanding.
According to the
formula~(\ref{rea}) we apply the Taylor
operator $(1-T^1_{p_b,p_q})$ to~(\ref{exap_4}) respectively
to~(\ref{exap_5}) 
\begin{eqnarray}
  \!\!\!\!
  ( 1 - T^1_{p_b,p_q}) 
  \left( \frac{\delta^{3} {\cal S}(\g)^{(1)}}{\delta c^{+}(-p_q-p_b) \delta
      \bar{q}(p_q) 
      \delta b(p_b)}\bigg|_{\phi=0} \right) &=& ( 1 - T^1_{p_b,p_q})
  \Delta^{{\cal S},(1)}_{c^{+} \bar{q} b}(p_q,p_b) \,\,=\,\,0
  \,,
  \label{eq:1mTdel}
\end{eqnarray}
where the degree of the Taylor operator is chosen to be the
lowest one which cancels the breaking terms
as is explained in Section~\ref{sub:remove}.
Let us in the following explicitly write down the individual terms
and subsequently discuss how they are treated.
Eq.~(\ref{eq:1mTdel}) in combination with~(\ref{exap_5}) leads to
\begin{eqnarray}
  0&=&
  ( 1 - T^1_{p_b,p_q}) \Gamma^{(1)}_{c^{+} W^{*,-}_{\nu}}(p_q + p_b)
  \left( i V_{qb} \gamma_\nu P_L \right)
  + i \left(p_q+p_b\right)_\nu
  ( 1 - T^0_{p_b,p_q}) \Gamma^{(1)}_{W^{+}_{\nu} \bar{q} b}(p_q,p_b)
  \nonumber \\&&\mbox{}+  
  ( 1 - T^0_{p_b,p_q}) \Gamma^{(1)}_{c^{+} G^{*,-}}(p_q+p_b) 
  \left(i {V_{q b} \over M_W} 
    \left[P_L m_q - P_R   m_b \right] \right) 
  \nonumber \\&&\mbox{}+  
  i\, M_W   ( 1 - T^1_{p_b,p_q}) \Gamma^{(1)}_{G^{+} \bar{q} b}(p_q,p_b) 
  \nonumber\\&&\mbox{}- 
  ( 1 - T^1_{p_b, p_q}) \left[ i(- \not\!p_q - m_q) \Gamma^{(1)}_{c^+
      \bar{q}^{*} 
      b}( p_q,p_b) \right]  
  - ( 1 - T^1_{p_q}) \Gamma^{(1)}_{\bar{q} q'}(-p_q) 
  \left(-i V_{q' b} P_L\right)
  \nonumber \\&&\mbox{}- 
  \left(i V_{q q'} P_R\right) \left( 1 - T^1_{p_b}\right)
  \Gamma^{(1)}_{\bar{q}' b}(p_b) 
  - ( 1 - T^1_{p_b, p_q}) \left[ \Gamma^{(1)}_{c^+ \bar{q} {b}^{*}}( p_q, p_b)
    i(\not\!p_b - m_b) \right] 
  \,.
  \label{eq:re_sti_1}
\end{eqnarray}
where in the second line we have exploited that the first derivative of
the scalar function 
$\g^{(1)}_{c^{+} G^{*,-}}$
vanishes.
In the second term of the first line the operator $T^1_{p_b,p_q}$ has already
been commuted with the momenta.
Eq.~(\ref{eq:re_sti_1}) tells us that in most cases the superficial UV
degree coincides with the one necessary to remove the breaking terms.
Let us in the following consider those terms in more detail
for which the Taylor expansion
provides an over-subtraction.
$\g^{(1)}_{G^{+} \bar{q} b}(p_q,p_b)$ has a logarithmic superficial divergence
and thus we have
\begin{eqnarray}\label{eq:re_2}
  ( 1 - T^1_{p_b,p_q}) \g^{(1)}_{G^{+} \bar{q} b}(p_q,p_b) = 
  ( 1 - T^0_{p_b,p_q}) \g^{(1)}_{G^{+} \bar{q} b}(p_q,p_b) 
  + ( T^0_{p_b,p_q} - T^1_{p_b,p_q})  
  \g^{(1)}_{G^{+} \bar{q} b}(p_q,p_b)
  \,,
\end{eqnarray} 
where the last term corresponds to an over-subtraction.
Also the terms proportional to the equation of motion for fermions lead 
to over-subtractions 
\begin{eqnarray}\label{eq:re_3}
  \lefteqn{
 ( 1 - T^1_{p_b,p_q})  \left[ \g^{(1)}_{c^+ \bar{q} {b}^{*}}( p_q,
      p_b) (m_b - \not\!p_b) \right]  =  
}
  \nonumber\\&&
  m_b ( 1 - T^1_{p_b,p_q}) \g^{(1)}_{c^+ \bar{q} {b}^{*}}( p_q, p_b) - 
  \not\!p_b   ( 1 - T^0_{p_b,p_q}) \g^{(1)}_{c^+ \bar{q} {b}^{*}}( p_q, p_b)
  \,,
\end{eqnarray}
as $\g^{(1)}_{c^+ \bar{q} {b}^{*}}( p_q, p_b)$ is only logarithmically
divergent.
These over-subtractions lead to the new universal (regularization
independent) breaking terms 
\begin{eqnarray}
  \Psi^{{\cal S},(1)}_{c^{+} \bar{q} b}(p_q,p_b) &=& 
  i \left( T^1_{p_b,p_q} -  T^0_{p_b,p_q} \right)
  \left(
    M_W  \g^{(1)}_{G^+ \bar{q} b}(p_q,p_b)  
  \right.\nonumber\\&&\left.\mbox{}\qquad\qquad
    + m_{q} \g^{(1)}_{c^+ \bar{q}^* b}(p_q,p_b)   
    + m_{b} \g^{(1)}_{c^+\bar{q} b^*}(p_q,p_b)  
  \right)
  \,,
  \nonumber\\
  \Psi^{{\cal S},(1)}_{c^{-} \bar{s} q}(p_s,p_q) &=&   
  i \left( T^1_{p_s,p_q} -  T^0_{p_s,p_q} \right)
  \left( 
    M_W  \g^{(1)}_{G^- \bar{s} q}(p_s,p_q)  
  \right.\nonumber\\&&\left.\mbox{}\qquad\qquad
    + m_{s} \g^{(1)}_{c^- \bar{s} q^* }(p_s,p_q)
    + m_{q} \g^{(1)}_{c^-\bar{s}^*   q }(p_s,p_q)
  \right)
  \,,
  \label{re_sti_2.0}
\end{eqnarray}  
where the Green functions 
$\g^{(1)}_{c^+ \bar{q}^*_i b}$,
$\g^{(1)}_{c^+ \bar{q}_i b^*}$,
$\g^{(1)}_{c^- \bar{s} q^*_i}$ and
$\g^{(1)}_{c^- \bar{s}^* q_i}$
will be discussed below.
Inspection of the most general counterterms available in the STI
(\ref{exap_4}), 
namely those provided in~ Eq.~(\ref{ta:ctBRS})
\begin{eqnarray}\label{re_sti_3}
\Xi^{(1)}_{W^{+}_{\mu} \bar{q}_i b} &=& Z_{i, 1}  P_L \gamma^\mu + Z_{i,
  2}  P_R \gamma^\mu  
\,,
\nonumber \\
\Xi^{(1)}_{G^{+} \bar{q}_i b} &=& Z_{i, 3}  P_L  + Z_{i, 4}  P_R 
\,,
\nonumber \\
\Xi^{(1)}_{\bar{q}_i b} &=& Z_{i, 5}  P_L \not\!p_b + Z_{i, 6}  P_R
\not\!p_b + Z_{i, 7}  P_L  + Z_{i, 8}  P_R 
\,,
\nonumber \\
\Xi^{(1)}_{\bar{q}_i q_j} &=& Z_{ij, 9}  P_L \not\!p_q + Z_{ij, 10}  P_R
\not\!p_q + Z_{ij, 11}  P_L  + Z_{ij, 12}  P_R 
\,, 
\end{eqnarray}
(also $\g^{(1)}_{c^{+} W^{*,-}_{\nu}}$ and
$\g^{(1)}_{c^{+} G^{*,-}}$ discussed in Section~\ref{sec:renorma}
should be taken into account) shows 
that the universal breaking terms, 
$\Psi^{{\cal S},(1)}_{c^{+} \bar{q}_i b}(p_q,p_b)$,
can be removed by 
adjusting the universal counterterms 
$\Xi^{(1)}_{W^{+}_{\mu} \bar{q}_i b}$ and $\Xi^{(1)}_{G^{+} \bar{q}_i b}$
(compare  with Eq.~(\ref{counterterms})).
Analogously $\Psi^{{\cal S},(1)}_{c^{-} \bar{s} q_i}(p_s,p_q)$ 
can be removed by $\Xi^{(1)}_{W^{-}_{\mu} \bar{s} q_i}$ 
and $\Xi^{(1)}_{G^{-} \bar{s} q_i}$.
Note that the Green functions $\g^{(1)}_{G^+\bar{q}_i b}(p_q,p_b)$ and
$\g^{(1)}_{G^- \bar{s} q_i }(p_s,p_q)$ enter the two-loop calculation
of $b\to s\gamma$ as
sub-diagrams which means that they have to be computed  anyway. 
Therefore, in this respect it is no extra effort to compute the relevant
counterterms of our subtraction scheme as compared to an invariant one.

However, in the calculation of the universal counterterms 
$\Xi^{(1)}_{W^{+}_{\mu} \bar{q}_i b}$, $\Xi^{(1)}_{G^{+} \bar{q}_i b}$,
$\Xi^{(1)}_{W^{-}_{\mu} \bar{s} q_i}$ and $\Xi^{(1)}_{G^{-} \bar{s} q_i}$
also the Green functions   
$\g_{c^+ \bar{q}^*_i b}$,
$\g_{c^+ \bar{q}_i b^*}$,
$\g_{c^- \bar{s} q^*_i}$
and 
$\g_{c^- \bar{s}^* q_i}$, 
appearing in~(\ref{re_sti_2.0}), are needed up to linear order in the
external momenta.
In Section~\ref{sub:bsg}, the complete set of identities for
$b\to s\gamma$ has been discussed and
Eqs.~(\ref{WTI_anti}) and~(\ref{WTI_anti.1}) have been derived
which constrain these Green functions occurring in our STIs.
The WTIs~(\ref{WTI_anti}) and~(\ref{WTI_anti.1})
are possibly broken by local terms like
$\Delta^{{\cal W}, (1)}_{\lambda_A c^+ \bar{q}^*_i  b}$.   

One finds  that within any mass-independent  regularization scheme  
the breaking terms  $\Delta^{{\cal W}, (1)}_{\lambda_A c^+ \bar{q}^*_i  b}$
and $\Delta^{{\cal W}, (1)}_{\lambda_Z c^+ \bar{q}^*_i  b}$  are zero
and no additional adjustment is needed.
In the following, we will explain the latter point in more detail.
In addition, we also cover the case where a general regularization
scheme is used.

The breaking terms  $\Delta^{{\cal W}, (1)}_{\lambda_A c^+ \bar{q}^*_i  b}$
and $\Delta^{{\cal W}, (1)}_{\lambda_Z c^+ \bar{q}^*_i  b}$
could again be removed by Taylor 
subtraction,  however, over-subtraction would generate
the following universal breaking terms 
(cf. Eqs.~(\ref{WTI_anti}) and~(\ref{WTI_anti.1}))
\begin{eqnarray}\label{fip_1}
  \Psi^{{\cal W}, (1)}_{\lambda_A c^+ \bar{q}^*_i  b}(-p_q-p_b, p_q,p_b)
  &=& 0
  \,,
  \nonumber \\
  \Psi^{{\cal W}, (1)}_{\lambda_Z c^+ \bar{q}^*_i  b}(-p_q-p_b, p_q,p_b)
  &=& - M_Z \g^{(1)}_{G_0 c^+ \bar{q}^*_i  b}(0,0,0) 
  \,,
\end{eqnarray}
where $M_Z$ is the $Z$ boson mass and $G_0$ is the corresponding
would-be Goldstone boson.
Note that $\g^{(1)}_{G_0 c^+ \bar{q}^*_i  b}(0,0,0)$ is finite and could
in principle be replaced by $\gg^{(1)}_{G_0 c^+ \bar{q}^*_i  b}(0,0,0)$.
For brevity the corresponding
equations involving $\lambda_+$ have been omitted.

As discussed in Section~\ref{sub:remove}, we will use our modified
subtraction technique to compute the finite part of the
remaining Green functions. In conjunction with the prescription
provided there
we found a very convenient and definitely  more effective way to compute the
breaking terms $\Delta^{{\cal W}, (1)}_{\lambda_A c^+ \bar{q}^*_i  b}$
and $\Delta^{{\cal W}, (1)}_{\lambda_Z c^+ \bar{q}^*_i  b}$ where
all external momenta are considered as large:
\begin{eqnarray}\label{fip_2}
\Delta^{{\cal W}, (1)}_{\lambda_A c^+ \bar{q}^*_i  b} &=& \lim_{\rho
  \rightarrow\infty} 
\left. \frac{\delta^4 {\cal W}^{(1)}_{(\lambda)}(\g)}
{\delta \lambda_A(0) \delta c^+(-\rho \,p_q -\rho \,p_b)  
\delta \bar{q}^*_i(\rho \, p_q)  \delta b(\rho \,p_b)}
\right|_{\phi=0}
\,,
\nonumber \\
\Delta^{{\cal W}, (1)}_{\lambda_Z c^+ \bar{q}^*_i  b} &=& \lim_{\rho
  \rightarrow\infty} 
\left. \frac{\delta^4 {\cal W}^{(1)}_{(\lambda)}(\g)}
{\delta \lambda_Z(0) \delta c^+(-\rho \,p_q -\rho \,p_b)  
\delta \bar{q}^*_i(\rho \, p_q)  \delta b(\rho \,p_b)}
\right|_{\phi=0}
\,.
\end{eqnarray}
For high values of the momenta the Green functions like 
$\g^{(1)}_{G_0 c^+ \bar{q}^*_i  b}$ in~(\ref{WTI_anti.1}) 
tend to zero because of Weinberg's theorem~\cite{wein}.
Also in the other WTIs we find that
$\Delta^{{\cal W}, (1)}_{\lambda_A c^+ \bar{q}^*_i  b}$
and 
$\Delta^{{\cal W}, (1)}_{\lambda_Z c^+ \bar{q}^*_i  b}$ 
are only combinations of the Green functions like
$\g_{c^+ \bar{q}^*_i b}$ and $\g_{c^+ \bar{q}_i b^*}$.
Note that the latter are simple to compute
at one-loop order. Furthermore only one diagram is involved.
Thus, the calculation of $\g_{c^+ \bar{q}^*_i b}$ and $\g_{c^+ \bar{q}_i
  b^*}$  at zero momentum which is needed for Eq.~(\ref{re_sti_2.0})
requires this simple additional step.
Using this procedure Eq.~(\ref{invariance_2}) can be applied in order to find
the breaking terms
$\Psi^{{\cal S},(1)}_{c^{+} \bar{q}_i b}(p_q,p_b)$
and
$\Psi^{{\cal S},(1)}_{c^{-} \bar{s} q_i}(p_s,p_q)$
where the symmetric Green functions
$\gg_{c^+ \bar{q}^*_i b}$ and $\gg_{c^+ \bar{q}_i b^*}$ can be identified with
the ``d-terms'' of~(\ref{invariance_2}).

Note that in a mass-independent schemes, like for instance Analytic
Regularization and Dimensional Regularization with 't~Hooft-Veltman
$\gamma_5$, one shows that these breaking terms are zero, as already stated
above. 

Unphysical renormalization constants as the wave function 
renormalization and the renormalization of gauge parameters can
be used to optimize further
the algebraic technique as explained in Section~\ref{sec:renorma}.

%%%%%%%%%%%%%%%%%%%%%%%%%%%%%%%%%%%%%%%%%%%%%%%%%%%%%%%%%%%%

\subsection{Imposing renormalization conditions}

As a last step, we have to fix the physical renormalization conditions
in order to obtain the correct amplitude for \bsg. 
We stress once more that the renormalization conditions must be imposed on
the renormalized Green functions $\gg$ by using 
the invariant counterterms ${\cal L}^{(1),INV}_{b\rightarrow s
  \gamma}$~\cite{STII,hollik_2,krau_ew}.

For the $W$ boson we choose the following condition\footnote{The superscript
  $T$ indicates the transversal part.}
\begin{eqnarray}\label{nor_con}
  \left[ \gg^{(1),T}_{W^{+} W^{-}}(p) \right]_{Re(p^{2}) = M^{2}_{W}} = 0
\end{eqnarray}  
on the quantum self energies
although it might not be convenient to fix the mass counterterms by using the
conventional quantum two-point functions $\gg^{(1)}_{W^{+}_{\mu} W^{-}_{\nu}}$
within the BFM. There, in fact,  the renormalization condition
on background two-point functions     
$\gg^{(1)}_{\hat{W}^{+}_{\mu} \hat{W}^{-}_{\nu}}$ are used. However, in the   
decay $b \rightarrow s \gamma$ none of the background functions must be 
computed and thus it is convenient to adopt~(\ref{nor_con}). Moreover we
have to notice that the zeros of background two-point functions and the
zeros of quantum two-point functions coincide. Thus the mass
renormalization is the same in both approaches.

For fermions we consider the corresponding two-point function 
\begin{eqnarray}\label{nor_con_fer}
\gg_{\bar{u} u'}(p) = 
\not\!p P_L \gg^L_{\bar{u} u'}(p) +
\not\!p P_R \gg^R_{\bar{u} u'}(p) +
P_L \gg^D_{\bar{u} u'}(p) +
P_R \gg^{D\dagger}_{\bar{u} u'}(p) 
\,,
\end{eqnarray}
where we have singled out the four different spinor structures entering in the 
amplitude $\gg_{\bar{u} u'}(p)$. An analogous equation holds for
the $d$ quark. 
We consider the zeros of the function~\cite{donoghue}
\begin{equation}\label{nor_con_fer.1}
{\rm Det}\left(  \gg^L -   (\gg^D)^{\dagger}  (\gg^R)^{-1} \gg^{D} \right)(p) 
\,,
\end{equation}
where $\gg^L,\gg^D$ and $\gg^R$ are matrices in the flavour space.
The matrices in~(\ref{nor_con_fer.1}) are hermitian.
The zeros $p^*_i$ of (\ref{nor_con_fer.1}) can be identified with 
the masses of the quark fields. Thus, this fixes the free parameters,
namely the ``diagonal values'' of the Yukawa matrices. 

Finally, the gauge parameters $\xi_A, \xi_Z$ and $\xi_W$ have to be fixed in
such a way that the restricted 't~Hooft gauge or the Feynman
gauge can be imposed. We also have to fix the charges $\alpha_{QED}, G_F$ and
$\alpha_s$. This is done in the usual way~\cite{hollik_2}.

For the renormalization of the CKM matrix, there are
two possible choices as was described at the end of Section~\ref{sec:sti}.

%%%%%%%%%%%%%%%%%%%%%%%%%%%%%%%%%%%%%%%%%%%%%%%%%%%%%%%%%%%%

\section*{Acknowledgments}
We thank Dieter Maison for many useful discussions regarding  
algebraic renormalization and for a careful reading of the manuscript.
We also thank  Carlo Becchi  and Peter Breitenlohner 
for illuminating comments and suggestions.
The work of M.S. has been supported by the {\it Schweizerischer 
Nationalfonds}.

%%%%%%%%%%%%%%%%%%%%%%%%%%%%%%%%%%%%%%%%%%%%%%%%%%%%%%%%%%%%

\renewcommand {\theequation}{\Alph{section}.\arabic{equation}}
\begin{appendix}

\setcounter{equation}{0} 
\section{BFM Lagrangian and notations} 
\label{app:sourceterms}

In this appendix we describe some ingredients useful 
for the calculation of radiative corrections in the SM within the BFM. 
In particular, we discuss how to derive the complete Lagrangian  
for the SM with background fields, we provide the BRST source (anti-fields)
terms  and the gauge fixing terms. 

\subsection*{Fields}

Here, the field content of the SM including background fields is listed.
Their Faddeev-Popov charges, $Q_{\fp}$, and
their UV and IR degrees\footnote{The UV and IR degrees are defined according
  to the BPHZL prescriptions given in~\cite{zimm}.}
are given in the form $(Q_{\fp})^{IR}_{UV}$.
The UV degree and the IR one for massless particles
coincides with the mass dimension of the corresponding field.
For massive fields the IR degree is set to two.
\begin{itemize}   
\item {\bf Quantum fields}   
  \begin{itemize}   
  \item Gauge fields\footnote{We use the convention $W^3_\mu = c_W Z_\mu - s_W
    A_\mu$.} 
    ${\bf V} \equiv \{A_{\mu},Z_{\mu},    
    W^{\pm}_{\mu}, G_{\mu}^{a} \}$ with   
    $\{ 0^{1}_{1}, 0^{2}_{1},  0^{2}_{1}, 0^{1}_{1} \}$
  \item Scalar fields ${\bf \Phi}  \equiv \{G^{0}, G^{\pm}, H \}$  with
    $\{0^{2}_{1}, 0^{2}_{1},  0^{2}_{1}\}$
  \item Ghost fields  ${\bf C} \equiv \{c^{A},c^{Z},c^{\pm}, c^{a} \}$
    with    
    $\{1^{0}_{0}, 1^{1}_{0},  1^{1}_{0}, 1^{0}_{0}\}$  
  \item Anti-ghost fields     
    ${\bf \bar{C}} \equiv \{\bar{c}^{A},\bar{c}^{Z},\bar{c}^{\pm},
    \bar{c}^{a} \}$    with    
    $\{(-1)^{2}_{2}, (-1)^{3}_{2},  (-1)^{3}_{2}, (-1)^{2}_{2}\}$
  \item Fermion fields\footnote{The colour index is   
      omitted. The index ``$i$'' denotes the    
      two components of the $SU(2)$ doublet, $L^{L}, Q^{L}$. In the
      case of the leptons
      the low components have the    
      quantum numbers $0^{\frac{3}{2}}_{\frac{3}{2}}$.
      This is because the neutrinos are massless.} 
    ${\bf \Psi}    
    \equiv \{L^{L}_{i}, Q^{L}_{i}, l^{R}, u^{R},  d^{R} \}$  with   
    $\{0^{2}_{\frac{3}{2}}, 0^{2}_{\frac{3}{2}},  0^{2}_{\frac{3}{2}},   
    0^{2}_{\frac{3}{2}}, 0^{2}_{\frac{3}{2}}\}$
  \item Nakanishi-Lautrup fields    
    ${\bf b} \equiv \{b^{A}, b^{Z}, b^{\pm}, b^{a} \}$ with   
    $\{0^{2}_{2}, 0^{3}_{2},  0^{3}_{2}, 0^{2}_{2}\}$
  \end{itemize}             
  
\item {\bf Background fields}   
  \begin{itemize}   
  \item Gauge fields  ${\bf \hat{V}} \equiv \{\hat{A}_{\mu},\hat{Z}_{\mu},    
    \hat{W}^{\pm}_{\mu}, \hat{G}_{\mu}^{a} \}$   with   
    $\{ 0^{1}_{1}, 0^{2}_{1},  0^{2}_{1}, 0^{1}_{1} \}$  
  \item Scalar fields ${\bf  \hat{\Phi}}  \equiv \{\hat{G}^{0},    
    \hat{G}^{\pm}, \hat{H} \}$   with   
    $\{0^{2}_{1}, 0^{2}_{1},  0^{2}_{1}\}$
  \end{itemize}   
  
\item {\bf External fields}   
  \begin{itemize}   
  \item BRST sources (or anti-fields) 
    
    ${\bf \phi^*} \equiv \{{W}^{*,3}_{\mu}$,     
    $W^{*,\pm}_{\mu}$, $G^{a,*}_{\mu}$, $c^{*,3}$, $c^{*,\pm}$, $c^{*,a}$,    
    ${H}^*$, $G^{*,\pm}$, $G^{*,0}$, ${L}^*_{L,i}$, ${L}^*_{Q,i}$,    
    ${l}^*$, ${u}^*$,  ${d}^* \}$ with   
    $\{ (-1)^{3}_{3}$, $(-1)^{3}_{3}$, $(-1)^{4}_{3}$,
    $(-2)^{4}_{4}$,    
    $(-2)^{4}_{4}$, $(-2)^{5}_{4}$, $(-1)^{3}_{3}$, $(-1)^{3}_{3}$,    
    $(-1)^{3}_{3}$,    
    $(-1)^{3}_{\frac{5}{2}}$, $(-1)^{3}_{\frac{5}{2}}$,   
    $(-1)^{3}_{\frac{5}{2}}$,   
    $(-1)^{3}_{\frac{5}{2}}$, $(-1)^{3}_{\frac{5}{2}}  
    \}$  
    
  \item Background sources

    ${\bf \Omega} \equiv    
    \{\Omega^{3}_{\mu}, \Omega^{\pm}_{\mu}, \Omega^{a}_{\mu},     
    \Omega^{H}, \Omega^{\pm}, \Omega^{0} \}$   with    
    $\left\{  (1)^{2}_{1}, (1)^{2}_{1}\right.$,$\left.(1)^{1}_{1},
      (1)^{2}_{1},    
      (1)^{2}_{1}, (1)^{2}_{1} \right\}$  
    
  \end{itemize}             
\end{itemize}

%%%%%%%%%%%%%%%%%%%%%%%%%%%%%%%%%%%%%%%%%%%%%%%%%%%%%%%%%%%%

\subsection*{Lagrangian and Feynman rules}
\def\gc{\g_0}

To derive the Lagrangian of the SM within the BFM we consider the 
invariant Lagrangian ${\cal L}_{INV}[V,\Phi,\Psi,\bar{\Psi}](x)$
presented in~\cite{aoki} and we replace the 
gauge and scalar fields by the sum
of the quantum and the corresponding background field: ${\cal
L}_{INV}[V,\Phi,\Psi,\bar{\Psi}](x)\longrightarrow 
{\cal L}_{INV}[V + \hat{V},\Phi+ \hat{\Phi},\Psi,\bar{\Psi}](x)
$. It is easy to check that this procedure provides the 
Feynman rules of~\cite{msbkg}. However, going beyond
the one-loop level, one needs the BRST-source terms to derive the
Feynman rules and also the WTIs and STIs.

For the BRST-source terms we have the following equation
\begin{eqnarray} \label{BRS_sources}
  {\cal L}^{BRST} &=& \sum_i \phi^*_i s \phi^i
  \nonumber \\
  &=&   
  {W}^{*,3}_{\mu} \left\{ \left( c_W \partial_{\mu} c^{Z} -   s_W
      \partial_{\mu} c^{A} \right) - 
    {i\, e \over s_{W}} \left[ \left(       
        W^{+}_{\mu} + \hat{W}^{+}_{\mu} \right) c^{-} - \left(       
        W^{-}_{\mu} + \hat{W}^{-}_{\mu} \right) c^{+} \right] \right\}
  \nonumber \\&&\mbox{}
  +   
  W^{*,\mp}_{\mu}  \left\{\!\partial_{\mu} c^{\pm} \mp i e \left(
      W^{\pm}_{\mu} + \hat{W}^{\pm}_{\mu} \right) \left( c^{A} -       
      \frac{c_{W}}{s_{W}} c^{Z} \right) 
  \right.\nonumber\\&&\left.\qquad\mbox{}
  \pm i e c^{\pm}\left[       
      \left( A_{\mu} + \hat{A}_{\mu} \right) - \!\frac{c_{W}}{s_{W}}      
      \left( Z_{\mu} + \hat{Z}_{\mu} \right) \right] \right\} 
  \nonumber \\&&\mbox{}
  +
  G^{*,a}_{\mu} \left\{ \partial_{\mu} c^{a} - g_{s} f^{abc}  
    \left( G^{b}_{\mu} + \hat{G}^{b}_{\mu} \right) c^{c} \right\} + 
  c^{*,3} \left\{ - {i\, e \over s_{W}} c^{-} c^{+}  \right\} 
  \nonumber \\&&\mbox{}
  +
  c^{*,\mp} \left\{ \mp \frac{i e}{2} c^{\pm} \left( c^{A} -       
      \frac{c_{W}}{s_{W}} c^{Z} \right)\right\} - 
  c^{*,a}  \left\{ \frac{1}{2} g_{s} f^{abc} c^{b} c^{c}  \right\}
  \nonumber\\&&\mbox{}
  +    {H}^* \left\{  \frac{i e}{2 s_{W}} \left[ \left(G^{+} +  
        \hat{G}^{+}\right) c^{-} -         
      \left(G^{-} + \hat{G}^{-}\right) c^{+} \right] +  
    \frac{e}{2 s_{W} c_{W}}       
    \left( G^{0} + \hat{G}^{0} \right) c^{Z} \right\}  
  \nonumber \\&&\mbox{}
  +   G^{*,\mp} \left\{ \pm \frac{i e}{2 s_{W}}  
    \left[ H + \hat{H} + v \pm i       
      \left( G^{0} + \hat{G}^{0} \right) \right] c^{\pm}  
  \right.\nonumber\\&&\left.\qquad\mbox{}
    \mp i e \left(G^{\pm} +       
      \hat{G}^{\pm}\right)\left(c^{A} -  
      \frac{c^{2}_{W} - s^{2}_{W}}{2 c_{W} s_{W}} c^{Z}      
    \right) \right\} 
  \nonumber \\&&\mbox{}
  + G^{*,0} \left\{  \frac{e}{2 s_{W}} \left[
      \left(G^{+} +   
        \hat{G}^{+}\right) c^{-} +         
      \left(G^{-} + \hat{G}^{-}\right) c^{+} \right] -  
    \frac{e}{2 s_{W} c_{W}}       
    \left( H + \hat{H} + v \right) c^{Z} \right\}
  \nonumber \\&&\mbox{} 
  +  \left( \bar{L}^*_{u} ,  \bar{L}^*_{d} \right)  
  \left(  
    \begin{array}{c}   
      \dms{\frac{ie}{\sqrt{2}s_{W}}}      
      L_{d} c^{+} -i e \left[ Q_{L_{u}} c^{A}      
        - \left(\dms{\frac{1}{2 s_{W} c_{W}}} -  
          Q_{L_{u}} \dms{\frac{s_{W}}{c_{W}}}      
        \right) c^{Z} \right] L_{u}  
      \vspace{.2cm}       \\  
      \dms{\frac{ie}{\sqrt{2}s_{W}}}  
      L_{u} c^{-} -i e      
      \left[ Q_{L_{d}} c^{A}      
        + \left(\dms{\frac{1}{2 s_{W} c_{W}}} +  
          Q_{L_{d}} \dms{\frac{s_{W}}{c_{W}}}      
        \right) c^{Z} \right] L_{d}   
    \end{array}  
  \right) 
  \nonumber \\&&\mbox{}
  +  \left( \bar{q}^*_{u} ,  \bar{q}^*_{d}
  \right)\!
  \left(\!\!
    \begin{array}{c}   
      \dms{\frac{ie V_{ud}}{\sqrt{2}s_{W}}}      
      q_{d}^L c^{+} \!\! - i e \left[ Q_{{u}} c^{A}      
        - \left(\dms{\frac{1}{2 s_{W} c_{W}}} \!-\!  
          Q_{{u}} \dms{\frac{s_{W}}{c_{W}}}      
        \right) c^{Z} \right] q_{u}^L  
      \!+\!  
      \frac{i}{2} g_s \left( \lambda_{a}  
        q_{u}^L \right) c_{a}   
      \vspace{.2cm} \nonumber \\  
      \dms{\frac{ie V^*_{ud}}{\sqrt{2}s_{W}}}  
      q_{u}^L c^{-} \!\! - i e      
      \left[ Q_{{d}} c^{A}      
        + \left(\dms{\frac{1}{2 s_{W} c_{W}}} \!+\!  
          Q_{{d}} \dms{\frac{s_{W}}{c_{W}}}      
        \right) c^{Z} \right] q_{d}^L   
      \!+\!   
      \frac{i}{2} g_s \left( \lambda_{a} q_{d}^L \right) c_{a} 
    \end{array}  
  \!\!\right) 
  \nonumber \\&&\mbox{}
  +  \bar{e}^*  \left\{  -i e Q_{e} \left( c^{A} +  
      \dms{\frac{s_{W}}{c_{W}}}      
      c^{Z} \right) e^{R}   \right\}  
  \nonumber \\&&\mbox{}  
  +   \bar{u}^*   \left\{  -i e Q_{u} \left( c^{A} +  
      \dms{\frac{s_{W}}{c_{W}}}      
      c^{Z} \right) u^{R}   +   
    \frac{i}{2} g_s \left( \lambda_{a} u^{R} \right) c_{a}
  \right\}
  \nonumber \\&&\mbox{}
  +  \bar{d}^*  \left\{  -i e Q_{d} \left( c^{A} +  
      \dms{\frac{s_{W}}{c_{W}}}      
      c^{Z} \right) d^{R}   +   
    \frac{i}{2} g_s \left( \lambda_{a}  
      d^{R} \right) c_{a}  \right\} + {\rm h.c.}
  \,.
\end{eqnarray} 

The gauge fixing terms, the Faddeev-Popov terms and 
the terms depending on $\Omega$ are easily obtained by 
\begin{equation}
{\cal L}_{GF} + {\cal L}_{\Phi\Pi} = 
s {\cal H}(x) 
\,,
\end{equation}
where ${\cal H}(x)$ is given by
\begin{eqnarray}
  \label{A13} 
  {\cal H}(x) &\equiv& \sum_{\alpha=A,Z,\pm,a} \bar{c}^{\alpha}
  {\cal F}^{\alpha}  \nonumber \\   
  & = &  \bar{c}^{A}  
  \left\{     
    \partial^{\lambda}A_{\lambda} - i\, e \,      
    \left( \hat{W}^{+}_{\lambda}\, W^{-,\lambda} -     
      \hat{W}^{-}_{\lambda}\, W^{+,\lambda}  \right)    
    - {i\, e \,\xi_A}
    \left({G^{-}} \,  {\hat{G}^{+}}  -      
      {G^{+}} \,  {\hat{G}^{-}}  \right)   
  \right\}
  \nonumber\\&&\mbox{}
  +   \bar{c}^{Z}   
  \left\{  
    \partial^{\lambda}Z_{\lambda} -  i\, e\, {c_W \over s_W}    
    \left( \hat{W}^{-}_{\lambda}\, W^{+,\lambda} -     
      \hat{W}^{+}_{\lambda}\, W^{-,\lambda} \right) 
  \right.  
  \nonumber\\&&\mbox{}
  \left.  
    +
    { i \, e \, \xi_Z } {c^2_W -s^2_W \over 2 c_W s_W} 
    \left( {G^{-}} \,  {\hat{G}^{+}}   
      -  {G^{+}} \,  {\hat{G}^{-}} \right) 
    -    e\, \xi_Z {1 \over 2 c_W s_W} 
    \left( {G^{0}} \,  {\hat{H}} -   {H} \,  {\hat{G}^{0}} +   
      v {G^{0}}  \right)   
  \right\}
  \nonumber\\&&\mbox{}  
  +   \bar{c}^{\mp}  
  \left\{     
    \partial^{\lambda}W^{\pm}_{\lambda} \mp \, i \, e\,     
    \left( A_{\lambda}\, \hat{W}^{\pm,\lambda}     
      - \hat{A}_{\lambda}\, W^{\pm,\lambda} \right)  \pm  
    i\, e\, {c_W \over s_W}     
    \left( \hat{W}^{\pm}_{\lambda}\, Z^{\lambda} 
      - \hat{Z}_{\lambda}\, W^{\pm,\lambda} \right) 
  \right.
  \nonumber\\&&\mbox{}     
  +   \left.  
    e\, \xi_W {1 \over 2 s_W} 
    \left( G^{\pm} \,  {\hat{G}^{0}} - G^{0} \,  {\hat{G}^{\pm}} \right) 
    \mp  i\, e\, \xi_W {1 \over 2 s_W} 
    \left( {G^{\pm}} \,  {\hat{H}}  -
      {H} \,  {\hat{G}^{\pm}}  +  v {G^{\pm}} \, \right)  \right\}  
  \nonumber\\&&\mbox{}
  + \bar{c}_{a} \hat{\nabla}^{ab} G^{b}_{\mu} +   
  \frac{\xi_{QCD}}{2} \bar{c}_{a} b^{a} +  
  \dms{\frac{\xi_A}{2}}{\bar{c}_{A}} b^{A} 
  + \dms{\frac{\xi_Z}{2}} {\bar{c}_{Z}} b^{Z}
  +  \frac{\xi_W}{2}
  \left( {\bar{c}_{+}}  b^{-} + {\bar{c}_{-}}  b^{+} \right) 
  \,.
\end{eqnarray}  
$s$ is the BRST transformation which can be read from the
Eqs.~(\ref{BRS_sources})  
and~(\ref{ST}). 
{Notice that by taking into account the extended BRST transformations for the 
background fields (cf Eqs. (\ref{new_1})) the action of $s$ on the gauge fermion ${\cal H}$ produce new terms. 
An analysis of these new Feynman rules has been presented in \cite{grassi}.}

%%%%%%%%%%%%%%%%%%%%%%%%%%%%%%%%%%%%%%%%%%%%%%%%%%%%%%%%%%%%

\subsection*{Background gauge transformations} 

Here we present the Background gauge transformations (BKG) for
the fields and the sources. 

\begin{itemize}   
\item {\bf BKG transformations for gauge fields}   
   
  In the next expression, $\lambda_{A}, \lambda_{Z},\lambda_{\pm}$ and
  $\lambda_{a}$ are the infinitesimal parameters of    
  the gauge group $SU_C(3)\times SU_I(2) \times U_Y(1)$.
  $\lambda_{A}$ generates the transformations of the subgroup $U_{Q}(1)$ of
  QED. 
  \begin{eqnarray}
    \label{BKG_tranf}
    \delta_{\lambda} W^{\pm}_{\mu} &=& \mp i e W^{\pm}_{\mu} \left(
      \lambda_{A} - 
      \frac{c_{W}}{s_{W}} \lambda_{Z} \right) \pm i e \lambda_{\pm}\left(
      A_{\mu} - \!\frac{c_{W}}{s_{W}}  Z_{\mu} \right)        
    \,,\nonumber\\
    \delta_{\lambda} \hat{W}^{\pm}_{\mu} &=& \partial_{\mu}\lambda_{\pm}    
    \mp i e \hat{W}^{\pm}_{\mu} \left( \lambda_{A} -       
      \frac{c_{W}}{s_{W}} \lambda_{Z} \right) \pm i e \lambda_{\pm}\left(
      \hat{A}_{\mu} - \!\frac{c_{W}}{s_{W}} \hat{Z}_{\mu} \right)        
    \,,\nonumber\\
    \delta_{\lambda}  Z_{\mu} &=& - i e \frac{c_{W}}{s_{W}} \left(    
      W^{+}_{\mu}  \lambda_{-} -      
      W^{-}_{\mu}  \lambda_{+} \right)           
    \,,\nonumber\\
    \delta_{\lambda}  \hat{Z}_{\mu} &=& \partial_{\mu} \lambda_{Z}- i e
    \frac{c_{W}}{s_{W}} \left(     
      \hat{W}^{+}_{\mu}  \lambda_{-} -      
      \hat{W}^{-}_{\mu}  \lambda_{+} \right)   
    \,,\nonumber\\
    \delta_{\lambda}  A_{\mu} &=& i e \left(   
      W^{+}_{\mu} \lambda_{-} -    
      W^{-}_{\mu} \lambda_{+} \right)              
    \,,\nonumber\\
    \delta_{\lambda}  \hat{A}_{\mu} &=& \partial_{\mu} \lambda_{A} + i e
    \left(     
      \hat{W}^{+}_{\mu} \lambda_{-} -    
      \hat{W}^{-}_{\mu} \lambda_{+} \right)       
    \,,\nonumber\\
    \delta_{\lambda}  G^{a}_{\mu} &=& - g_{s} f^{abc} G^{b}_{\mu} \lambda_{c}
    \,,\nonumber\\ \mbox{}              
    \delta_{\lambda}  \hat{G}^{a}_{\mu} &=& \partial_{\mu} \lambda_{a} -   
    g_{s} f^{abc} \hat{G}^{b}_{\mu} \lambda_{c}   
  \end{eqnarray}      
  
\item {\bf BKG transformations for scalar  fields}   
  \begin{eqnarray}
    \label{scal_tras_bkg}
    \delta_{\lambda} G^{\pm} &=& \pm \frac{i e}{2 s_{W}} \left( H \pm i G^{0}
    \right)     
    \lambda_{\pm} \mp i e G^{\pm}       
    \left(\lambda_{A} - \frac{c^{2}_{W} - s^{2}_{W}}{2 c_{W} s_{W}} \lambda_{Z}
    \right)       
    \,,\nonumber\\
    \delta_{\lambda} \hat{G}^{\pm} &=& \pm \frac{i e}{2 s_{W}} \left( \hat{H}
      + v 
      \pm i \hat{G}^{0} \right)     
    \lambda_{\pm} \mp i e \hat{G}^{\pm}       
    \left(\lambda_{A} - \frac{c^{2}_{W} - s^{2}_{W}}{2 c_{W} s_{W}} \lambda_{Z}
    \right)       
    \,,\nonumber\\
    \delta_{\lambda} H &=& \frac{i e}{2 s_{W}} \left( G^{+}  \lambda_{-} - 
      G^{-} \lambda_{+} \right) + \frac{e}{2 s_{W} c_{W}}       
    G^{0}  \lambda_{Z}       
    \,,\nonumber\\
    \delta_{\lambda} \hat{H} &=& \frac{i e}{2 s_{W}} \left( \hat{G}^{+}
      \lambda_{-} 
      -          
      \hat{G}^{-} \lambda_{+} \right) + \frac{e}{2 s_{W} c_{W}}       
    \hat{G}^{0} \lambda_{Z}       
    \,,\nonumber\\
    \delta_{\lambda} G^{0} &=& \frac{e}{2 s_{W}} \left( G^{+} \lambda_{-} +         
      G^{-} \lambda_{+} \right) - \frac{e}{2 s_{W} c_{W}} H \lambda_{Z}       
    \,,\nonumber\\
    \delta_{\lambda} \hat{G}^{0} &=& \frac{e}{2 s_{W}} \left( \hat{G}^{+}
      \lambda_{-} +          
      \hat{G}^{-} \lambda_{+} \right) - \frac{e}{2 s_{W} c_{W}}       
    \left( \hat{H} + v \right) \lambda_{Z}     
  \end{eqnarray}      
  Here one must notice that the quantum scalar fields undergo a
  homogeneous  transformation, while the background scalar fields
  transform according to the     
  inhomogeneous ones where the shift of the vacuum $v$ appears.    
\item {\bf BKG transformations for ghost fields, for anti-ghost and
    Nakanishi-Lautrup fields} 
  \begin{eqnarray}
    \label{BKG_tranf_ghost}
    \delta_{\lambda} c^{\pm} &=& \mp {i e} \left[ c^{\pm} \left(
        \lambda_{A} 
        -        
        \frac{c_{W}}{s_{W}} \lambda_{Z} \right) + \lambda_{\pm} \left(
        c^{A} -        
        \frac{c_{W}}{s_{W}} c^{Z} \right) \right]    
    \,,\nonumber\\
    \delta_{\lambda} c^{Z} &=& - {i e} \frac{c_{W}}{s_{W}} \left(
      c^{+} 
      \lambda_{-}  +  \lambda_{+} c^{-} \right)     
    \,,\nonumber\\
    \delta_{\lambda} c^{A} &=& i e \left( c^{+} \lambda_{-}  +  \lambda_{+}
      c^{-} 
    \right)      
    \,,\nonumber\\
    \delta_{\lambda} c^{a} &=& -g_{s} f^{abc} c^{b} \lambda_{c}   
  \end{eqnarray}      
  The BKG transformations for the Nakanishi-Lautrup
  fields and for the anti-ghost fields are
  identical with the     
  previous ones after substituting the ghost fields with the other fields.    
\item {\bf BKG transformations for the fermion fields}      
  \begin{eqnarray}\label{BKG_trans_ferm} 
    \delta_{\lambda} {L}_{u} &=& \dms{\frac{ie}{\sqrt{2}s_{W}}}
    {L}_{d} \lambda_{+} -i e \left[ Q_{{u}} \lambda_{A}      
      - \left(\dms{\frac{1}{2 s_{W} c_{W}}} - Q_{{u}}
        \dms{\frac{s_{W}}{c_{W}}} 
      \right) \lambda_{Z} \right] {L}_{u} 
    \,,\nonumber\\
    \delta_{\lambda} {L}_{d} &=& \dms{\frac{ie}{\sqrt{2}s_{W}}}
    {L}_{u} 
    \lambda_{-} -i e       
    \left[ Q_{{d}} \lambda_{A}      
      + \left(\dms{\frac{1}{2 s_{W} c_{W}}} + Q_{{d}}
    \dms{\frac{s_{W}}{c_{W}}} 
      \right) \lambda_{Z} \right] {L}_{d}  
    \,,\nonumber\\
    \delta_{\lambda} l^R  &=& -i e Q_{l} \left( \lambda_{A} +
      \dms{\frac{s_{W}}{c_{W}}}       
      \lambda_{Z} \right) l^{R}   
    \,,
\nonumber \\
 \delta_{\lambda} {q}_{u} &=& \dms{\frac{ie V_{ud}}{\sqrt{2}s_{W}}}
    {q}_{d} \lambda_{+} -i e \left[ Q_{{u}} \lambda_{A}      
      - \left(\dms{\frac{1}{2 s_{W} c_{W}}} - Q_{{u}}
        \dms{\frac{s_{W}}{c_{W}}} 
      \right) \lambda_{Z} \right] {q}_{u} + \frac{i}{2} g_s \left(
      \tau^{a} {q}_{u} \right)  
    \lambda_{a}   
    \,,\nonumber\\
    \delta_{\lambda} {q}_{d} &=& \dms{\frac{ie V^*_{ud}}{\sqrt{2}s_{W}}}
    {q}_{u}  
    \lambda_{-} -i e       
    \left[ Q_{{d}} \lambda_{A}      
      + \left(\dms{\frac{1}{2 s_{W} c_{W}}} + Q_{{d}}
    \dms{\frac{s_{W}}{c_{W}}} 
      \right) \lambda_{Z} \right] {q}_{d}  + \frac{i}{2} g_s \left(
      \tau^{a} q_{d} \right)  
    \lambda_{a}   
    \,,\nonumber\\
    \delta_{\lambda} f^{R}  &=& -i e Q_{f} \left( \lambda_{A} +
      \dms{\frac{s_{W}}{c_{W}}}       
      \lambda_{Z} \right) f^{R}   + \frac{i}{2} g_s
    \left( 
      \tau^{a} f^{R} \right)  
    \lambda_{a}
  \end{eqnarray}      
  $V_{ud}$ is the CKM matrix, $f=u,d$ and $L=(\nu,e)$.  

\item{\bf BKG transformations for the BRST sources}    
  
  The BKG transformations of BRST sources correspond to the gauge
  transformations of the corresponding quantum   
  gauge field according to their specific representations. 
\end{itemize}

\subsection*{Linearized Slavnov-Taylor operator and functional Taylor
  operator}  
The linearized Slavnov-Taylor operator for a generic functional 
${\cal F}$ is given by
\begin{eqnarray}
{\cal S}_{\gg} (\cf) & \equiv &  \int {\rm d}^4 x
\left( s_W \partial_\mu c_Z + c_W \partial_\mu c_A \right)  
\left( s_W \frac{\delta \cf}{\delta Z_\mu} +  
c_W \frac{\delta \cf}{\delta A_\mu} \right)
\nonumber \\&&\mbox{}
+ \sum_{\alp=A,Z,\pm,a} b_{\alp}  \frac{\delta \cf}{\delta \bar{c}^{\alp}}
+ (\gg, \cf) + (\cf,\gg)
\,,
\label{line_ST}
\end{eqnarray}
where 
\begin{eqnarray}\label{bracket}
\left( X, Y \right) &=& \int {\rm d}^4x \, \left[
\frac{\delta X}{\delta W^{*,3}_\mu}  \frac{\delta Y}{\delta W^3_\mu} +   
\frac{\delta X}{\delta W^{*,\pm}_\mu}  \frac{\delta Y}{\delta W^\mp_\mu} +   
\frac{\delta X}{\delta G^{*,a}_\mu}  \frac{\delta Y}{\delta G^a_\mu} +  
\frac{\delta X}{\delta c^{*,\pm}}  \frac{\delta Y}{\delta c^\mp} +  
\frac{\delta X}{\delta c^{*,3}} \frac{\delta Y}{\delta c^3} 
\nonumber\right.\\&&\left.\mbox{} 
+
\frac{\delta X}{\delta c^{*,a}}  \frac{\delta Y}{\delta c^a} + 
\frac{\delta X}{\delta c^{*,\pm}}  \frac{\delta Y}{\delta c^\mp} +  
\frac{\delta X}{\delta G^{*,\pm}}  \frac{\delta Y}{\delta G^\mp} +  
\frac{\delta X}{\delta G^{*,0}}  \frac{\delta Y}{\delta G^0} +  
\frac{\delta X}{\delta H^*}  \frac{\delta Y}{\delta H} 
\nonumber\right.\\&&\left.\mbox{} 
+
\sum_{I=L,Q,u,d,e} \left(  
\frac{\delta X}{\delta \bar{\psi}^{*I}} \frac{\delta Y}{\delta \psi^{I}} + 
{\rm h.c.} \right) \right] 
\,.
\end{eqnarray}
Since ${\cal S}(\gg) =0$, the
operator ${\cal S}_{\gg}$ is nilpotent. We also introduce the tree-level 
linearized operator ${\cal S}_0$ which is equivalent to ${\cal S}_{\g_0}$ 
where $\g_0$ is the tree level action. 

The Taylor operator $T^\delta$ on the functional $\Gamma$ is defined as
follows.
One first considers the relevant amplitude which results from functional
derivatives w.r.t. fields denoted by subscripts
$\Gamma_{\phi_1(p_1)\phi_2(p_2)...\phi_m(p_m)}$
with
$\sum_{j=1}^{m}p_j=0$.
Then  the Taylor expansion $T^{\delta}$
in the independent momenta up to degree $\delta$ acts formally
as
\begin{eqnarray}
T^\delta \Gamma = \sum_{m=1}^{\infty} \int \prod_{i=1}^{m} {\rm d}^4 p_i 
\phi_i (p_i) \delta^4 ({\scriptstyle \sum_{j=1}^{m}} p_j)
T^{\delta}_{p_1,\dots,p_m} 
\left .
\Gamma_{\phi_1(p_1)\phi_2(p_2)...\phi_m(p_m)}
\right |_{\sum_{j=1}^{m}p_j=0}
\label{definitionT2}
\,.
\end{eqnarray}
A remarkable property of $T^\delta$ is that $T^{\delta_1} T^{\delta_2} =
T^{\delta}$ with $\delta={\rm min}\{\delta_1,\delta_2\}$. 
Note that 
the Taylor operator is scale invariant, however, it does not commute
with spontaneous symmetry breaking. If massless fields are present
IR problems can occur.

%%%%%%%%%%%%%%%%%%%%%%%%%%%%%%%%%%%%%%%%%%%%%%%%%%%%%%%%%%%%

\setcounter{equation}{0}
\section{An example for the derivation of STIs}
\label{app:stiex}

In this appendix we explicitly derive the STIs for the amplitude
involving a scalar matter field $\Phi$ and two gauge fields $V^{\mu}_1$ and
$V^{\nu}_2$. Thereby we follow the general rules derived in
Section~\ref{sub:conventional}.
The amplitude for this process is computed in terms of the following
irreducible Green functions:
the two-point functions $\gg_{V^{\mu}_i V^{\nu}_j}$,
$\gg_{V^{\mu}_i \Phi}$ ($i,j=1,2$)
and $\gg_{\Phi \Phi}$ which represent corrections to external legs
and the three-point function
$\gg_{V^{\mu}_1 V^{\nu}_2 \Phi}$.
The renormalization program also requires the correct
definitions of other Green functions which are absent in the
physical amplitude which, however, arise in the analysis of
the STIs.

Let us in a first step discuss
the two point functions $\gg_{V^{\mu}_i V^{\nu}_j}$. 
The ghost field, associated with the gauge boson $V^{\mu}_i$,
is denoted by $c_i$.
According to rule~\ref{rule3} we
have to differentiate Eq.~(\ref{ST}) w.r.t. $c_i$ and $V^{\nu}_j$:
\begin{eqnarray}
\label{e_1}
\frac{\delta^2 {\cal S}(\gg)}{\delta c_i(-p) \delta V^{\nu}_j(p)
  }\Bigg|_{\phi=0} &=&
i p_{\mu} \left( s_W \delta_{iZ} + c_W \delta_{iA} \right)  
\left( s_W {\gg}_{Z^\mu V^{\nu}_j}(p) +  c_W {\gg}_{A^\mu V^{\nu}_j}(p) \right)
\nonumber \\&&\mbox{}
+\gg_{c_i V^{*}_{k, \rho}}(p) \gg_{V^{\rho}_k V^{\nu}_j}(p)
+\sum_{\Phi=G^\pm,G^0,H}\gg_{c_i \Phi^*}(p) \gg_{\Phi V^{\nu}_j}(p)
\nonumber\\
&=&
0
\,,
\end{eqnarray}
where in the second line a summation over $k$ is understood.
Notice that for simplicity
in the second line the gauge eigenstates notation is used. 
The subscript ``$\phi=0$'' means that after differentiation, 
all fields are set to zero.
$\Phi$ and $V$ represent all the scalar and gauge 
fields,
respectively. $H$ denotes the Higgs field and $\Phi^*$, $V^*$ and $H^*$
are the corresponding anti-fields.
In the following, the summation over the scalar fields is omitted.
If one does not use normal ordering in the perturbative expansion, 
one gets an additional tadpole term on the r.h.s. of (\ref{e_1}), namely
$\gg_{c_i H^* V_j^\nu}(0,p) \gg_H (0)$. 
However, this contribution is easily removed by using the
renormalization condition $\gg_H(0)=0$ which implements the spontaneous
symmetry breaking at the quantum level.
The linear terms in Eq.~(\ref{e_1})
take into account
the contributions arising from the abelian gauge field.
The quantity $\delta_{ij}$ represents the Kronecker delta carrying
gauge field indices. Notice that 
no charged gauge boson is involved in these terms. 

As mentioned below Eq.~(\ref{ST}), 
formula (\ref{e_1}) is valid to all orders.
We want to stress that at tree level, 
$\gg^{(0)}_{V^{\rho}_k V^{\nu}_j}$  
can be read off the invariant Lagrangian~\cite{STII,hollik_2,krau_ew}.

Let us now continue the construction of the closed set of
STIs for the example under consideration.
Eq.~(\ref{e_1}) contains new Green functions, namely the two-point function
for the mixing between the gauge and the scalar fields,
$\gg_{\Phi V^{\nu}_j}(p)$, and the ones involving one ghost 
and one anti-field, namely $\gg_{c_i \Phi^*}(p)$ and $\gg_{c_i V^*_{j,
    \mu}}(p)$. 
According to rule~\ref{rule1},  these 
Green functions are not vanishing and they are related to the two-point
functions of ghost and anti-ghost through the ghost equations
which reads in the `t~Hooft gauge fixing~\cite{brs,krau_ew,grassi}:
\begin{equation}
  \label{e_2}
  \gg_{c_i \bar{c}_j}(p) = -i p^{\mu}   \gg_{c_i V^{*}_{j, \mu}}(p) + 
  \xi_j M_{j,\Phi} \gg_{c_i \Phi^*}(p)
  \,.
\end{equation}
$M_j$ is defined below~(\ref{bfm_2}).
This equation is independent from the STIs and 
it implements the
equation of motion for the ghost fields. It is a consequence of
the choice of the gauge fixing. 

In order to get an identity involving $\gg_{\Phi V^{\nu}_j}(p)$
we can derive a new independent STI by differentiating
(\ref{ST}) w.r.t. $c_j$ and $\Phi$. 
\begin{eqnarray}
  \label{e_3}
  \frac{\delta^2 {\cal S}(\gg)}{\delta c_j(-p) \delta \Phi(p)}\Bigg|_{\phi=0}
  &=&
  i p_{\mu} \left( s_W \delta_{jZ} + c_W \delta_{jA} \right)  
  \left( s_W {\gg}_{Z^\mu  \Phi}(p) +  c_W {\gg}_{A^\mu  \Phi}(p) \right) 
  \nonumber \\ && \mbox{}
  +\gg_{c_j V^{*}_{k, \rho}}(p) \gg_{V^{\rho}_k \Phi}(p) + 
  \gg_{c_j \Phi^{'*}}(p) \gg_{\Phi' \Phi}(p)
  \nonumber \\
  &=& 0
  \,.  
\end{eqnarray}
In this equation, only one new Green function emerges, namely
$ \gg_{\Phi^\prime \Phi}(p)$.
As it does not contain any gauge or ghost fields
the process of finding a closed set of STIs is completed.
There is no independent STI which can give us new
information on the Green functions.
Notice that also the Green
functions involving ghost fields, $\gg_{c_j \Phi^*}(p)$ and
$\gg_{c_j V^*_{k, \mu}}(p)$, are the same as before.

Let us now consider the three-point function $\gg_{V^{\mu}_1 V^{\nu}_2 \Phi}$.
Replacing one of the gauge bosons
by its corresponding ghost field leads to (omitting the summation over the
Goldstone fields)  
\begin{eqnarray}
  \label{e_4}
  \lefteqn{\frac{\delta^3 {\cal S}(\gg)}{\delta c_i(p_3) 
      \delta V^j_{\nu}(p_1) \delta \Phi (p_2)}\Bigg|_{\phi=0} =}
\nonumber\\&&\mbox{}
- i p_3^{\mu} \left( s_W \delta_{iZ} + c_W \delta_{iA} \right)  
\left( s_W {\gg}_{Z^\mu  V^j_{\nu} \Phi}(p_1,p_2) +  c_W {\gg}_{A^\mu
      V^j_{\nu} 
    \Phi} (p_1,p_2)\right)  
\nonumber\\&&\mbox{}
+ \gg_{c_i V^{*}_{k, \rho}}(-p_3) \gg_{V_{k, \rho}   V^j_{\nu} \Phi}(p_1,
p_2)   
+  \gg_{c_i V^{*}_{k, \rho} V^j_{\nu}}(p_2, p_1) \gg_{V_{k, \rho} 
  \Phi}(p_2) 
\nonumber\\&&\mbox{}
+  \gg_{c_i V^{*}_{k, \rho} \Phi }(p_1, p_2) \gg_{V_{k,  \rho}
  V^{\nu}_j}(p_1)   
+ \gg_{c_i \Phi^{'*}}(-p_3) \gg_{\Phi'  V^j_{\nu} \Phi}(p_1, p_2) 
\nonumber\\&&\mbox{}
+ \gg_{c_i \Phi^{'*} V^j_{\nu}}(p_2, p_1) \gg_{\Phi' \Phi}(p_2)
+ \gg_{c_i \Phi^{'*} \Phi}(p_1, p_2) \gg_{\Phi' V^j_{\nu}}(p_1)
\nonumber\\
&=& 0
\,.  
\end{eqnarray}
Besides the two-point functions already discussed above, new Green
functions arise:
$ \gg_{\Phi'  V^j_{\nu} \Phi}$,
$\gg_{c_i \Phi^{'*} V^j_{\nu}}$ and
$\gg_{c_i V^{*}_{k, \rho} V^j_{\nu}}$. The first one satisfies a new
independent STI because of the presence of only one gauge field. 
In fact, differentiation w.r.t.
$\Phi'$, $c^j$ and $\Phi$ leads to the identity 
\begin{eqnarray}
  \label{e_5}
  \lefteqn{\frac{\delta^3 {\cal S}(\gg)}{\delta c_i(p_3) 
      \delta \Phi'(p_1) \delta \Phi (p_2)}\Bigg|_{\phi=0} =}
  \nonumber\\&&\mbox{}
- i p_3^{\mu} \left( s_W \delta_{iZ} + c_W \delta_{iA} \right)  
  \left( s_W {\gg}_{Z^\mu  \Phi' \Phi}(p_1,p_2) +  c_W {\gg}_{A^\mu  \Phi'
      \Phi}(p_1,p_2) \right)  
  \nonumber\\&&\mbox{}
  + \gg_{c_i V^{*}_{k, \rho}}(-p_3) \gg_{V_{k, \rho}   \Phi' \Phi}(p_1,
  p_2)   
  + \gg_{c_i V^{*}_{k, \rho} \Phi'}(p_2, p_1) \gg_{V_{k, \rho} \Phi}(p_2)
  \nonumber\\&&\mbox{}
  + \gg_{c_i V^{*}_{k, \rho} \Phi}(p_1, p_2) \gg_{V_{k,\rho} \Phi'}(p_1)  
  + \gg_{c_i \Phi^{''*}}(-p_3) \gg_{\Phi'' \Phi' \Phi}(p_1, p_2) 
  \nonumber\\&&\mbox{}
  + \gg_{c_i \Phi^{''*} \Phi'}(p_2, p_1) \gg_{\Phi'' \Phi}(p_2)
  + \gg_{c_i \Phi^{''*} \Phi}(p_1, p_2) \gg_{\Phi'' \Phi'}(p_1)
  \nonumber\\
  &=& 0.  
\end{eqnarray}
The summation over the Goldstone fields $\Phi''$ is understood. 
In this identity, only the new Green function $ \gg_{\Phi''
  \Phi' \Phi}$ occurs which is not constrained by an independent
STI. Concerning the Green functions with only external quantum
fields the system is now complete.
However, we are left with the ones involving also ghost fields.
In the case of $\gg_{c_i V^{*}_{k, \rho} V^j_{\nu}}$, 
we differentiate w.r.t. $c_i$, $c_j$ and $V^{*}_{k, \rho}$ and get   
\begin{eqnarray}   
  \label{e_6}
  \lefteqn{ \frac{\delta^3 {\cal S}(\gg)}{\delta c_i(p_3)  
      \delta c_j(p_1)  \delta V^{*}_{k, \rho}(p_2)}\Bigg|_{\phi=0} =}
  \nonumber\\&&\mbox{}
  - i p_3^{\mu} \left( s_W \delta_{iZ} + c_W \delta_{iA} \right)  
  \left( s_W {\gg}_{Z^\mu   c_j V^{*}_{k, \rho}}(p_1, p_2) 
    +  c_W {\gg}_{A^\mu   c_j V^{*}_{k, \rho}}(p_1, p_2)  \right) 
  \nonumber\\&&\mbox{}
  + i p_1^{\mu} \left( s_W \delta_{jZ} + c_W \delta_{jA} \right)  
  \left( s_W {\gg}_{Z^\mu  c_i V^{*}_{k, \rho}}(p_3, p_2) +  
    c_W {\gg}_{A^\mu  c_i V^{*}_{k, \rho} }(p_3, p_2) \right) 
  \nonumber\\&&\mbox{}
  +  \gg_{c_j V^{*}_{l, \sigma}}(-p_1) \gg_{V_{l, \sigma} c_i  V^{*}_{k,
      \rho}}(p_3,p_2)  
  +  \gg_{c_i \Phi^{*}}(-p_3) \gg_{\Phi c_j  V^{*}_{k, \rho}}(p_1,p_2)
  \nonumber\\&&\mbox{}
  +  \gg_{c_j \Phi^{*}}(-p_1)  \gg_{\Phi c_i  V^{*}_{k, \rho}}(p_3,p_2) 
  +  \gg_{c_i c_j c^{*}_l}(p_1,p_2) \gg_{c_l  V^{*}_{k, \rho}}(p_2)
  \nonumber\\&&\mbox{}
  + \gg_{c_i V^{*}_{l, \sigma}}(-p_3) \gg_{V_{l, \sigma} c_j  V^{*}_{k,
      \rho}}(p_1,p_2)  
\nonumber \\
&=& 0
\,.
\end{eqnarray}
An analogous equation is obtained for $\gg_{c_V \Phi^{'*} V^j_{\nu}}$.
In~(\ref{e_6}) $c^*_i$ is the anti-field of $c_i$.
At this point a short comment
is in order.
As is well known, the BRST
transformations are non-linear and local. At the quantum level, 
they receive radiative corrections.
Moreover, if the regularization breaks the symmetries, the correct
non-linear transformation rules at the quantum level can be obtained only by
adjusting the finite counterterms which are needed for the 
STI~(\ref{e_6}).
Note that they --- in contrast
to~(\ref{e_4}) and~(\ref{e_5}) --- contain derivatives w.r.t. anti-fields.

In~(\ref{e_6}) the new function $\gg_{c_i c_j c^*_l}$ which does not contain
gauge fields occurs. According to rule~\ref{rule4}, we get the following
identity: 
\begin{eqnarray}
  \label{e_8}
  \lefteqn{\frac{\delta^3 {\cal S}(\gg)}{\delta c_i  (p_4)
      \delta c_j(p_1)  \delta c_k(p_2) \delta c^*_m(p_3)}\Bigg|_{\phi=0} =}
  \nonumber\\&&\mbox{}
  - i p_4^{\mu} \left( s_W \delta_{iZ} + c_W \delta_{iA} \right)  
  \left( s_W {\gg}_{Z^\mu   c_j c_k c^{*}_m}(p_1, p_2,p_3) 
    +  c_W {\gg}_{A^\mu   c_j  c_k c^{*}_m}(p_1, p_2,p_3) \right) 
  \nonumber\\&&\mbox{}
  +  i p_1^{\mu} \left( s_W \delta_{jZ} + c_W \delta_{jA} \right)  
  \left( s_W {\gg}_{Z^\mu   c_i c_k c^{*}_m}(p_4,p_2,p_3) 
    +  c_W {\gg}_{A^\mu   c_i  c_k c^{*}_m}(p_4,p_2,p_3)   \right) 
  \nonumber\\&&\mbox{}
  -  i p_2^{\mu} \left( s_W \delta_{k,Z} + c_W \delta_{k,A} \right)  
  \left( s_W {\gg}_{Z^\mu   c_i c_j c^{*}_m}(p_4,p_1,p_3)
    +  c_W {\gg}_{A^\mu   c_i  c_j c^{*}_m}(p_4,p_1,p_3)\right) 
   \nonumber\\&&\mbox{}
  +  \gg_{c_k V^*_{l,\mu}}(-p_2) {\gg}_{V^{l,\mu}  c_i c_j
      c^{*}_m}(p_4,p_1,p_3) 
  +  \gg_{c_k \Phi^*}(-p_2) {\gg}_{\Phi  c_i c_j c^{*}_m}(p_4,p_1,p_3) 
  \nonumber\\&&\mbox{}
  +  \gg_{c_j V^*_{l,\mu}}(-p_1) {\gg}_{V^{l,\mu}  c_k c_i
      c^{*}_m}(p_2,p_4,p_3)  
  +  \gg_{c_j \Phi^*}(-p_1) {\gg}_{\Phi  c_k c_i c^{*}_m}(p_2,p_4,p_3) 
 \nonumber\\&&\mbox{}
  +  \gg_{c_i V^*_{l,\mu}}(-p_4) {\gg}_{V^{l,\mu}  c_j c_k
      c^{*}_m}(p_1,p_2,p_3) 
  +  \gg_{c_i \Phi^*}(-p_4) {\gg}_{\Phi  c_j c_k c^{*}_m}(p_1,p_2,p_3)
  \nonumber\\&&\mbox{}
  + \gg_{c_i c_j c^{*}_l}(p_1, p_2 +p_3) \gg_{c_l  c_k c^{*}_m}(p_2,p_3) 
  \nonumber\\&&\mbox{}
  + \gg_{c_j c_k c^{*}_l}(p_3, p_2 +p_1) \gg_{c_l  c_i c^{*}_m}(p_4,p_3) 
  \nonumber\\&&\mbox{}
  +\gg_{c_k c_i c^{*}_l}(p_4, p_1 +p_3) \gg_{c_l  c_j c^{*}_m}(p_1,p_3)
  \nonumber\\
  &=&0
  \,,
\end{eqnarray}
which can be considered as the extension of the Jacobi identity
to the quantum level.

From rule~\ref{rule4} it also follows that derivatives w.r.t. 
$c_i c_j V^k_\mu V^{*l}_\nu, c_i c_j \Phi \Phi^{*}$, $c_i c_j V^k_\mu \Phi^*$
have to be considered.
The corresponding identities do not involve any new Green function
which requires a subtraction,
however, they provide new constraints\footnote{Here we would like to underline
  that the problem of the  
  complete set of identities involved is a pure algebraic problem. 
  This is equivalent to derive the well-known {\it descent
    equations}
  in the space of local functionals~\cite{libro}.
  If anti-fields are required, the relevant constraints 
  can easily be derived from the descent equation formulated in the 
  Batalin-Vilkovisky anti-field formalism~\cite{BV}. As an
  example, consider the amplitude for $VV \rightarrow cV$. According to
  rule~1 the STI for $c\, V$ produces $c\,  V^*$. The latter requires 
  the identity for $ c\,  V^* \rightarrow c^2 \,  V^*$
  which finally produces $c^3 c^*$. }.
For brevity we will not list them explicitly.

Now all symmetry constraints for the amplitude and the
related Green functions have been derived.
In a second step, the same procedure has to be applied to the sub-divergences.
Again the independent irreducible contributions have to be extracted and
a closed set of independent STIs has to be derived.
At that point, one is able to study the counterterms
and the breaking terms  involved in the set of identities.

%%%%%%%%%%%%%%%%%%%%%%%%%%%%%%%%%%%%%%%%%%%%%%%%%%%%%%%%%%%%

\setcounter{equation}{0}
\section{Triangular organization of counterterms}
\label{app:triangula}

As is clear from the applications discussed in 
Sections~\ref{sec:hgg_coun} and~\ref{sec:bsg_coun}  
and from the theoretical analysis of Section~\ref{sub:consi}, 
the triangular organization of functional identities and of
counterterms is very important. With triangular structure of functional
identities 
we mean the possibility to organize the set of functional identities
into a hierarchical structure in such a way that we can restore the
identities one after the other without spoiling those which  are already
recovered.  
As an example, we recall the \bsg~calculation. We have seen that we can 
organize the counterterms in such a way that in a first step the
WTIs and in a second one the STIs can be restored.
This can only be achieved if the fixing of the STIs
does not destroy the already restored WTIs. This means that the
counterterms needed to restore the STIs must be invariant under the
action of the WTIs, i.e. they must be background gauge invariant. Clearly
this is only possible   
if the breaking terms to the STIs are background gauge invariant.  
In  Section~\ref{sub:consi}, by using the consistency conditions and repairing 
the WTI by suitable non-invariant counterterms, we have shown that the new
breaking terms 
\begin{equation}\label{new}
\hat{\Delta'}_S = \Delta'_S - {\cal S}'_0\left(\int {\rm d}^4x {\cal
  L}^{WTI}_{b\rightarrow s\gamma} \right) 
\end{equation}
are indeed background gauge invariant ${\cal W'}_{(\lambda)}\left(\hat{\Delta'}_S \right)=0 $. 

We mention here that this fact can be extended to the complete set of functional
identities. We recall that the SM in the BFM~\cite{grassi} is 
completely  defined in terms of the following functional identities (up to
free arbitrary constant parameters)
\begin{itemize}
\item The Nakanishi-Lautrup identities (\ref{eq:gau_fix}), which implement the
  gauge fixing conditions to all orders,
\item the Faddeev-Popov equations (see \cite{grassi,krau_ew}), 
\item the Abelian Anti-ghost Equation (in the case of BFM \cite{grassi}), 
\item the WTI  given by Eq.~(\ref{WTI}) for the background gauge invariance
  and  
\item the STI given by Eq.~(\ref{ST}) for the BRST symmetry.
\end{itemize}

We specialize now the argument to each subspace of
counterterms and we show that it is indeed possible to restore the WTIs,
respectively, the STIs by starting from the identities for two-point
functions,
then for three-point functions and, finally, to fix the counterterms
for four-point functions. Clearly, this is a considerable  simplification 
for practical
calculations. It helps us to restrict the set of identities
involved into a specific calculation and, as a consequence, it restricts
the number of counterterms which must be effectively restored.

We first focus on the WTI. The main problem in dealing with the WTI 
is that 
the 
operator ${\cal W}'_{(\lambda)}$ is not homogeneous in the fields. 
Since this operator implements the local background gauge symmetry, the 
transformations of the gauge fields contain an inhomogeneous term which
does not depend on the gauge fields itself.
Furthermore, the transformations
of the scalar fields contain the shift of the Higgs fields in order to
take into account the spontaneous symmetry breaking mechanism. 
Following the extended anti-field formalism~\cite{BV} we promote the
infinitesimal parameter $\lambda$ to a local dimensionless Grassman 
parameters $\omega$ in the adjoint representation of the gauge
group  
which transforms as $\omega \rightarrow \omega + \omega\wedge\omega$ and we 
decompose  ${\cal W}'_{(\omega)} + (\omega\wedge\omega)_a \frac{\delta
  }{\delta \omega^a}$ into 
\begin{eqnarray}
\label{eq:app_D_9}
{\cal W}'_{(\omega)} +  (\omega\wedge\omega)_a \frac{\delta
  }{\delta \omega^a} &=& {\cal W}'^0_{(\omega)} + {\cal W}'^1_{(\omega)},
\end{eqnarray}
where ${\cal W}'^1_{(\omega)}$ contains the transformations of
$\omega$. This definition implies that 
\begin{eqnarray}
\label{eq:app_D_10}
\left( {\cal W}'^0_{(\omega)} \right)^2=0,~~~ 
\left( {\cal W}'^1_{(\omega)} \right)^2=0,~~~
\left\{ {\cal W}'^0_{(\omega)},  {\cal W}'^1_{(\omega)} \right\}=0. 
\end{eqnarray}
On the other side, the action $\g$ and the breaking term $\Delta'_W(\omega) $
can be  decomposed into amplitudes with $n$ external legs\footnote{This
means that $\g_n$ is generically an n-point function and $\Delta'_n$ 
are coefficients of monomials with $n$ fields (comprehensive of $\omega$).} 
$\g = \sum_{n\geq1}  \g_{n, CT}^{ WTI}$ respectively
$\Delta'_W(\omega) = \sum^{N}_{n=1} \Delta'_{n,W}(\omega)$ where $N$ is
the upper bound. In the case of renormalizable theories we have $N=4$. 
Therefore, the contribution with the highest content of fields, namely
$n=N-1,N$, to Eqs.~(\ref{cc.2}) 
are 
\begin{eqnarray}
  \widetilde{\cal W}'^1 \left( \g_{N} \right) &=&
  \Delta'_{N-1,W} - \widetilde{\cal W}'^0 \left(  \g_N \right)
  \nonumber\\
  \widetilde{\cal W}'^0 \left( \g_{N+1} \right) &=&
  \Delta'_{N,W} - \widetilde{\cal W}'^1 \left(   \g_{N+1} \right)
  \label{eq:app_D_12}
  \,,
\end{eqnarray}
where $\widetilde{\cal W}'^{0,1}$ are the matrices acting on each single 
monomials of $\g_n$. The breaking terms satisfy
\begin{eqnarray}
  \widetilde{\cal W}'^1 \left( \Delta'_{N,W} \right) &=& 0
  \nonumber\\
  \widetilde{\cal W}'^0 \left( \Delta'_{N,W} \right) + 
  \widetilde{\cal W}'^1 \left( \Delta'_{N-1,W} \right) &=&0
  \,.
\label{eq:app_D_14}
\end{eqnarray} 
Notice that $\widetilde{\cal W}'^0 \left( \g_{N+1} \right)$ and
$\widetilde{\cal W}'^1 \left( \g_{N} \right)$ are
combinations of amplitudes with $N$ external fields whose finite parts 
can be adjusted by counterterms with monomials with $N$ fields. 
On the other side, $\widetilde{\cal W}'^0 \left(  \g_N \right)$
are combinations of amplitudes with $N-1$ external fields whose finite
parts are already fixed by the WTI for breaking terms with $0,\dots,N-2$
external fields. Finally, $\widetilde{\cal W}'^1 \left(   \g_{N+1} \right)$
is expressed in terms of $N+1$ external fields amplitudes which cannot be
modified by adjusting overall local counterterms. 
 
Therefore the two systems of Eqs.~(\ref{eq:app_D_12}) 
must be solved simultaneously and this can be done only if the system is
redundant. To show that we apply the operator $\widetilde{\cal W}'^{0}$
from the right on the first equation and the operator  $\widetilde{\cal
  W}^{1}$  
on the second and by using the consistency conditions (\ref{eq:app_D_14}),
we immediately find that a non-trivial combination of the two system of 
equations exists which implies the redundancy of the two system. 
The consistency conditions help us to disentangle the independent
equations. 

For the STI the problem is a little more involved. The functional
operator ${\cal S}'_0$ is not linear, and decomposing it w.r.t.
the power of the fields we get a mixing between terms coming from the
functional operator and those coming from the action $\g$. 

We introduce the notation ${\cal S}'_{0,N}$ ($N=1,2,3,4$) to
denote the contribution linear, quadratic, cubic and quartic to the
non-linear Slavnov-Taylor operator and, as above, we indicate 
with $\Delta'_{N,STI}$ ($N=1,2,3,4,5$) the breaking terms of the STI. By
expanding the equation  ${\cal S}'_0(\g) = \Delta'_{STI}$ and the 
consistency conditions ${\cal S}'_0(\Delta'_{STI})=0$ in powers of
fields, we have the following equations 
\begin{eqnarray}\label{app_D_1}
  &&\left\{\begin{array}{c}
      {\cal S}'_{0,2} \g_2 = \Delta'_{2,STI} \,, \\
      {\cal S}'_{0,2} \Delta'_{2,STI} = 0\,,
    \end{array} \right.
  \nonumber \\
  &&\left\{\begin{array}{c}
      {\cal S}'_{0,3} \g_2 + {\cal S}'_{0,2} \g_3 = \Delta'_{3,STI} \,,\\
      {\cal S}'_{0,3} \Delta'_{2,STI} + {\cal S}'_{0,2} \Delta'_{3,STI} = 0\,,
    \end{array} \right.
  \nonumber \\
  &&\left\{\begin{array}{c}
      {\cal S}'_{0,4} \g_2 + {\cal S}'_{0,3} \g_3 +  {\cal S}'_{0,3} \g_4 =
      \Delta'_{4,STI} \,,\\ 
      {\cal S}'_{0,4} \Delta'_{2,STI} + {\cal S}'_{0,3} \Delta'_{3,STI} 
      + {\cal S}'_{0,2} \Delta'_{4,STI}= 0\,,
    \end{array} \right.
  \nonumber \\
  &&\left\{\begin{array}{c}
      {\cal S}'_{0,4} \g_3 + {\cal S}'_{0,3} \g_4 +  {\cal S}'_{0,2} \g_5 =
      \Delta'_{5,STI} \,,\\ 
      {\cal S}'_{0,4} \Delta'_{3,STI} + {\cal S}'_{0,3} \Delta'_{4,STI} 
      + {\cal S}'_{0,2} \Delta'_{5,STI} = 0\,.
    \end{array} \right.
\end{eqnarray}
Rearranging the last two equations, we can immediately see that they
must compatible by means of the consistency conditions otherwise they
cannot be solved in terms of the remaining free parameter $\g_4$. Notice
that the Green functions $\g_5$ are superficially convergent and
therefore they cannot be fixed by the previous equations. We
apply ${\cal S}'_0$ from the left on both equations and use the
commutation properties 
\begin{eqnarray}\label{app_D_2}
  \left( S'_{0,2} \right)^2 = 0\,,
  &
  \left( S'_{0,3} \right)^2 = 0\,,
  &
  \left(S'_{0,4}\right)^2 = 0\,,
  \nonumber \\ 
  \left\{ S'_{0,2}, S'_{0,3} \right\} = 0\,, 
  &
  \left\{ S'_{0,2}, S'_{0,4} \right\}  = 0\,,
  &
  \left\{ S'_{0,4},  S'_{0,3} \right\} = 0\,,
  \nonumber
\end{eqnarray}
in combination with
the consistency conditions. Then one checks the compatibility
between the two equations. This completes the proof of the triangular
structure of the counterterms for the STIs.

%%%%%%%%%%%%%%%%%%%%%%%%%%%%%%%%%%%%%%%%%%%%%%%%%%

\end {appendix}

%%%%%%%%%%%%%%%%%%%%%%%%%%%%%%%%%%%%%%%%%%%%%%%%%%

\def\ap#1#2#3{{\it Ann. Phys. (NY)} {\bf #1} (#2) #3}  
\def\jmp#1#2#3{{\it  J. Math. Phys.} {\bf #1} (#2) #3}  
\def\rmp#1#2#3{{\it Rev. Mod. Phys.} {\bf #1} (#2) #3}  

\def\app#1#2#3{{\it Act.~Phys.~Pol.~}{\bf B #1} (#2) #3}
\def\apa#1#2#3{{\it Act.~Phys.~Austr.~}{\bf#1} (#2) #3}
\def\cmp#1#2#3{{\it Comm.~Math. Phys.~}{\bf #1} (#2) #3}
\def\cpc#1#2#3{{\it Comp.~Phys.~Commun.~}{\bf #1} (#2) #3}
\def\epjc#1#2#3{{\it Eur.\ Phys.\ J.\ }{\bf C #1} (#2) #3}
\def\fortp#1#2#3{{\it Fortschr.~Phys.~}{\bf#1} (#2) #3}
\def\ijmpc#1#2#3{{\it Int.~J.~Mod.~Phys.~}{\bf C #1} (#2) #3}
\def\ijmpa#1#2#3{{\it Int.~J.~Mod.~Phys.~}{\bf A #1} (#2) #3}
\def\jcp#1#2#3{{\it J.~Comp.~Phys.~}{\bf #1} (#2) #3}
\def\jetp#1#2#3{{\it JETP~Lett.~}{\bf #1} (#2) #3}
\def\mpl#1#2#3{{\it Mod.~Phys.~Lett.~}{\bf A #1} (#2) #3}
\def\nima#1#2#3{{\it Nucl.~Inst.~Meth.~}{\bf A #1} (#2) #3}
\def\npb#1#2#3{{\it Nucl.~Phys.~}{\bf B #1} (#2) #3}
\def\nca#1#2#3{{\it Nuovo~Cim.~}{\bf #1A} (#2) #3}
\def\plb#1#2#3{{\it Phys.~Lett.~}{\bf B #1} (#2) #3}
\def\prc#1#2#3{{\it Phys.~Reports }{\bf #1} (#2) #3}
\def\prd#1#2#3{{\it Phys.~Rev.~}{\bf D #1} (#2) #3}
\def\pR#1#2#3{{\it Phys.~Rev.~}{\bf #1} (#2) #3}
\def\prl#1#2#3{{\it Phys.~Rev.~Lett.~}{\bf #1} (#2) #3}
\def\pr#1#2#3{{\it Phys.~Reports }{\bf #1} (#2) #3}
\def\ptp#1#2#3{{\it Prog.~Theor.~Phys.~}{\bf #1} (#2) #3}
\def\sovnp#1#2#3{{\it Sov.~J. Nucl. Phys.~}{\bf #1} (#2) #3}
\def\tmf#1#2#3{{\it Teor.~Mat.~Fiz.~}{\bf #1} (#2) #3}
\def\yadfiz#1#2#3{{\it Yad.~Fiz.~}{\bf #1} (#2) #3}
\def\zpc#1#2#3{{\it Z.~Phys.~}{\bf C #1} (#2) #3}
\def\ppnp#1#2#3{{\it Prog.~Part.~Nucl.~Phys.~}{\bf #1} (#2) #3}
\def\ibid#1#2#3{{ibid.~}{\bf #1} (#2) #3}

%  
% references  
%  

\end{document}